\documentclass[review]{elsarticle}

\usepackage{lineno,hyperref,pgfplots}
\usepackage{color}
\usepackage[norelsize,ruled,linesnumbered]{algorithm2e}
\usepackage{amsmath,amsfonts,amssymb,amsthm,mathrsfs,graphicx,multirow}
\usepackage{wrapfig}
\usepackage{booktabs}
\usepackage{todonotes}
\usepackage{libertine}
\usepackage{etex}
\usepackage{xargs}

\usepackage{xspace}

\usepackage{xstring}

\usepackage{boolexpr}

\usepackage[cmex10]{mathtools}
\usepackage{amssymb}
\usepackage{amsfonts}
\usepackage{mathrsfs}
\usepackage{latexsym}
\usepackage{textcomp}
\usepackage{pifont}

\newcommand{\argemp}[2]{\if&#1&\else#2\fi}

\newcommand{\argdef}[2]{\if&#1&#2\else#1\fi}

\newcommand{\argint}[3]{\if&#2&\else#1#2#3\fi}

\newcommand{\argext}[3]{\if&#1&#3\else#1\if&#3&\else#2#3\fi\fi}

\newcommandx{\mthfnt}[3][1=, 2=0]{{
	\IfStrEqCase{#1}
	{%
		{}%
		{#3}%
		{Name}%
		{%
			\IfStrEqCase{#2}
			{%
				{0}{\mathcal{#3}}%
				{1}{\mathscr{#3}}%
				{2}{\mathfrak{#3}}%
				{3}{\mathbb{#3}}%
			}
			[\ensuremath{\clubsuit}]%
		}%
		{Set}%
		{%
			\IfStrEqCase{#2}
			{%
				{0}{\mathrm{#3}}%
				{1}{\mathsf{#3}}%
				{2}{\mathbb{#3}}%
				{3}{\mathbf{#3}}%
			}
			[\ensuremath{\clubsuit}]%
		}%
		{Fun}%
		{%
			\IfStrEqCase{#2}
			{%
				{0}{\mathsf{#3}}%
				{1}{\mathrm{#3}}%
			}
			[\ensuremath{\clubsuit}]%
		}%
		{Rel}%
		{%
			\IfStrEqCase{#2}
			{%
				{0}{\mathit{#3}}%
				{1}{\mathtt{#3}}%
			}
			[\ensuremath{\clubsuit}]%
		}%
		{Sym}%
		{%
			\IfStrEqCase{#2}
			{%
				{0}{\mathtt{#3}}%
				{1}{\mathbf{#3}}%
			}
			[\ensuremath{\clubsuit}]%
		}%
		{Elm}%
		{\mathnormal{#3}}
	}
[\ensuremath{\clubsuit}]%
}}

\newcommand{\mthsub}[1]{\argemp{#1}{\ensuremath{_{\mathnormal{#1}}}}}

\newcommand{\mthsup}[1]{\argemp{#1}{\ensuremath{^{\mathnormal{#1}}}}}

\newcommandx{\mth}[5][1=, 2=0, 4=, 5=]{{\ensuremath{\mthfnt[#1][#2]{#3}\mthsub{#4}\mthsup{#5}}}}

\newcommandx{\mtharg}[6][1=, 2=0, 4=, 5=]{{\mth[#1][#2]{#3}[#4][#5]\ensuremath{\argint{(}{#6}{)}}}}

\newcommand{\mthempty}{\mth[][]}

\newcommand{\mthstyname}{0}
\newcommand{\mthname}[1][]{\mth[Name][\argdef{#1}{\mthstyname}]}

\newcommand{\mthstyset}{0}
\newcommand{\mthset}[1][]{\mth[Set][\argdef{#1}{\mthstyset}]}

\newcommand{\mthstyfun}{0}
\newcommand{\mthfun}[1][]{\mth[Fun][\argdef{#1}{\mthstyfun}]}

\newcommand{\mthstyrel}{0}
\newcommand{\mthrel}[1][]{\mth[Rel][\argdef{#1}{\mthstyrel}]}

\newcommand{\mthstysym}{0}
\newcommand{\mthsym}[1][]{\mth[Sym][\argdef{#1}{\mthstysym}]}

\newcommand{\tuple}[1]
{\ensuremath{\!\argint{\langle}{#1}{\rangle}}}

\theoremstyle{theorem}
\newtheorem{theorem}{Theorem}

\theoremstyle{definition}
\newtheorem{definition}{Definition}

\theoremstyle{lemma}
\newtheorem{lemma}{Lemma}

\theoremstyle{corollary}
\newtheorem{corollary}{Corollary}

\newcommand{\SetN}{\mathbb{N}}
\def\Nat{\mathbb{N}}

\newcommand{\set}[2]{\ensuremath{\argint{\{}{\argext{#1}{\allowbreak:\allowbreak}{#2}}{\}}}}
\newcommand{\pow}[1]{\ensuremath{2^{#1}}}
\newcommand{\card}[1]{\mthempty{\argint{\vert}{#1}{\vert}}}

\newcommand{\Trans}{\mthfun{T}}

\newcommand{\LTL}{\mthfun{LTL}\xspace}
\def\TEMPORAL#1{\mbox{\small\boldmath$\mathbf{#1}$}}
\def\ltlnext{\TEMPORAL{X}}
\def\sometime{\TEMPORAL{F}} 
\def\always{\TEMPORAL{G}}
\def\until{\,\TEMPORAL{U}\,}
\def\Nat{\mathbb{N}} 

\newcommand{\CTL}{\mthfun{CTL}\xspace}

\newcommand{\Ag}{\mthset{N}}
\newcommand{\Ac}{\mthset{Ac}}
\newcommand{\AcProf}{\vec{\Ac}}
\newcommand{\St}{\mthset{St}}
\newcommand{\AP}{\mthset{AP}}
\newcommand{\APSet}{\AP}
\newcommand{\Pun}{\text{Pun}}

\newcommand{\Par}{\mthset{PAR}}

\newcommand{\CGModel}{\mthname{M}}
\renewcommand{\Game}{\mthname{G}}
\newcommand{\LTLGame}{\Game_{\LTL}}
\newcommand{\ParGame}{\Game_{\Par}}
\renewcommand{\Pun}{\mthset{Pun}}
\newcommand{\pun}{\mthsym{pun}}

\newcommand{\Automaton}{\mthname{A}}
\newcommand{\Language}{\mthname{L}}

\newcommand{\ERel}{\mthrel{E}}

\newcommand{\labFun}{\lambda}

\newcommand{\WinSet}{\mthset{Win}}

\newcommand{\trnFun}{\mthfun{tr}}

\newcommand{\NE}{\mthset{NE}}

\newcommand{\winsym}{\mthset{Win}}
\newcommandx{\Win}[3][1=, 2=, 3=]
{\mthset{\winsym#3}[#1][#2]}

\newcommand{\presym}{\mthfun{Pre}}
\newcommandx{\Pre}[3][1=, 2=, 3=]
{\mthset{\presym#3}[#1][#2]}

\newcommand{\eqsym}{\mthfun{Eq}}
\newcommandx{\Eq}[3][1=, 2=, 3=]
{\mthset{\eqsym#3}[#1][#2]}

\newcommandx{\AFW}[5][1=, 2=, 3=, 4=, 5=]
{\txtargname{AFW#5{\small\argint{$[$}{#1}{$]$}}}[#2][#3]{#4}\xspace}

\def\MOCHA{\textsf{MOCHA}\xspace}
\def\PRISM{\textsf{PRISM\xspace}}
\def\PRALINE{\textsf{PRALINE}\xspace}
\def\MCMAS{\textsf{MCMAS}\xspace}
\def\EVE{\textsf{EVE}\xspace}
\SetCommentSty{xCommentSty}
\newlength{\wordlength}

\newcommand{\Bisim}{R}
\newcommand{\Bisimilar}{\sim}

\newcommand{\valf}{\lambda}
\newcommand{\direction}{\vec{a}}

\usepackage{tikz}
\usetikzlibrary{arrows,shapes,snakes}

\usepackage{wrapfig}
\tikzstyle{every node} =
	[draw = none, fill = white, thin]
\tikzstyle{every edge} +=
	[thin]

\tikzstyle{noall} =
	[draw = none, fill = none]
\tikzstyle{nodraw} =
	[draw = none, fill = white]
\tikzstyle{nofill} =
	[draw = black, fill = none]

\tikzstyle{cnode} =
	[circle, draw = black]
\tikzstyle{snode} =
	[regular polygon, regular polygon sides = 4, draw = gray!50]
\tikzstyle{lnode} =
	[diamond, draw = gray!75]
\tikzstyle{pnode} =
	[regular polygon, regular polygon sides = 5, draw = gray]
\tikzstyle{rnode} =
[rectangle, draw = black, thin]

	{

	}

		\newcommand{\figseqgam}
		{
	\begin{figure}[t]
		\vspace{-10pt}
		\begin{center}
			\footnotesize
			\mbox{
				{
					\label{fig:seqgame:left}
					\scalebox{1.00}[1.00]{
						\begin{tikzpicture}
						[node distance = 3cm]
						\node [cnode]
						(S0)
						{$s_{1}$};
						\node [cnode]
						(S1)
						[right of = S0]
						{$s_{2}$};
						\path[->]
						(S0)	edge	[]
						node [] {\textcolor{black}{$(\vec{a}_{-j}, a_{j})$}}
						(S1)
						;
						\end{tikzpicture}
					}}
					\qquad %\qquad
					{
						\label{fig:seqgame:right}
						\scalebox{1.00}[1.00]{
						\begin{tikzpicture}
						[node distance = 1.5cm]
						\node [cnode]
						(S0)
						{$s_{1}$};
						\node [rnode]
						(S1)
						[right of = S0]
						{$(s_{1}, \vec{a}_{-j})$};
						\node [cnode]
						(S2)
						[right of = S1]
						{$s_{2}$};
						\path[->]
						(S0)	edge	[]
						(S1)
						(S1)	edge	[]
						(S2)
						;
						\end{tikzpicture}
						}}
					}
					\caption{\label{fig:seqgame} Sequentialisation of a game.
						On the left, a representation of a transition from $s_{1}$ to $s_{2}$ using action profile $(\vec{a}_{-j}, a_{j})$.
						On the right, the two states $s_{1}$ and $s_{2}$ are assigned to Player~$0$ in the parity game, which are interleaved with a state of Player~$1$ corresponding to the choice of $\vec{a}_{-j}$ by coalition $-j$ in the original game.} 
				\end{center}
				\vspace{-15pt}
			\end{figure}
		}
		
		\newcommand{\figNEpun}
		{
			\begin{figure*}%[htbp]
				\vspace{-10pt}
				\begin{center}
					\footnotesize
					\mbox{
							\scalebox{1.00}{
								\begin{tikzpicture}
								[node distance = 2cm]
								\node
								(S0)
								{$s_{0}$};
								\node
								(S1)
								[right of = S0]
								{$s_{1}$};
								\node
								(Sdots)
								[right of = S1]
								{$\ldots$};
								\node
								(Sk)
								[right of = Sdots]
								{$s_{k}$};
								\node
								(Sk+1)
								[right of = Sk]
								{$s_{k + 1}$};
								\node
								(S')
								[above right of = Sk]
								{$s'$};
								\node
								(Sdots')
								[right of = Sk+1]
								{$\ldots$};
								\node
								(Spun)
								[right of = S']
								{$\ldots$};
								\path[->]
								(S0)	edge	[]
								node [] {$\vec{a}_{0}$}
								(S1)
								(S1)	edge	[]
								node [] {$\vec{a}_{1}$}
								(Sdots)
								(Sdots)	edge	[]
								node [] {$\vec{a}_{k - 1}$}
								(Sk)
								(Sk)	edge	[]
								node [] {$\vec{a}_{k}$}
								(Sk+1)
											edge	[]
								node [] {$((\vec{a}_{k})_{-j}, a_{j}')$}
								(S')
								(S')	edge	[]
								node [] {$\sigma_{i}^{\pun j}$}
								(Spun)
								(Sk+1)	edge	[]
								node [] {$\vec{a}_{k + 1}$}
								(Sdots')
								;
								\end{tikzpicture}
							}
							}
							\caption{ \small \label{fig:NEPun} Representation of the strategy $\sigma_{i}$.
							At the beginning, player $i$ follows the transducer $\Trans_{\eta}$ that generates the action profile run $\eta$.
							The strategy adheres to it until a unilateral deviation from player $j$ occurs, here represented at the $k$-th step of the play.
							Once the deviation has occurred, and the game entered a state $s'$, player~$i$ starts executing the strategy $\sigma_{i}^{\pun j}$, to employ the punishment strategy against player~$j$.}
								
						\end{center}
						\vspace{-15pt}
					\end{figure*}
				}

		\newcommand{\Compute}{\textbf{Compute}\xspace}
		
		\newcommand{\algparsolv}
		{
			\begin{algorithm}
				\textbf{Input: }{An \LTL game $\LTLGame = (\Ag, (\Ac_{i})_{i \in \Ag}, \St, s_{0}, \trnFun, \labFun, (\gamma_i)_{i \in \Ag})$.}\\
				\textbf{Output: }{``Yes'' if $\NE(\LTLGame) \neq \emptyset$; ``No'' otherwise. }\\
				$\ParGame \Longleftarrow \LTLGame$ \tcc*{from Section \ref{secn:ltltoparity} (Theorem \ref{thm:NEinvariance})}
				\ForEach{$ W \subseteq \Ag $}{
					\ForEach{$j \in L = \Ag \setminus W$}{
					\Compute $\Pun_{j}(\ParGame)$ \tcc*{from Section \ref{secn:characterisation} (Theorem \ref{thm:punchar})}
					}
					\Compute $\ParGame^{-L}$ \\ 
					\ForEach{$i \in W$}{
						\Compute $\Automaton_{i}$ and $\mthname{S}_{i}$ from $\ParGame^{-L}$
					}
					\If{$\Language(\bigtimes_{i \in W}(\mthname{S}_{i})) \neq \emptyset$ \tcc*{from Section \ref{secn:characterisation} (Theorem \ref{thm:NEchar})}}{\Return ``Yes''}
				}
				\Return ``No''
				\caption{Nash equilibrium via Parity games}
				\label{algo:NEviaParity}
			\end{algorithm}
		}

		\newcommand{\GridEx}
		{
			\begin{figure}
				\begin{center}
					\vspace{-15pt}
					\begin{tikzpicture}[every node/.style={minimum size=1cm-\pgflinewidth, outer sep=0pt}]
					\draw[step=1cm,color=black] (-2,-2) grid (2,2);
					\node[fill=gray] at (-0.5,-0.5) {};
					\node[fill=gray] at (-0.5,0.5) {};
					\node[fill=gray] at (-1.5,0.5) {};
					\node[fill=gray] at (-1.5,-0.5) {};
					\node[fill=gray] at (1.5,0.5) {};
					\node[fill=gray] at (1.5,-0.5) {};
					\draw [->](-1.5,1.5) -- (-0.5,1.5){};
					\draw [->](1.5,-1.5) -- (0.5,-1.5){};
					\node[circle, thick, fill=white, minimum size=0.5cm,draw] at (-1.5,1.5){};
					\node[rectangle, fill=black, minimum size=0.5cm] at (1.5,-1.5){};
					\draw (-2,2) -- (-2.5,2.5) node[pos=0.75,xshift=0.25cm] {x} node[pos=0.75,yshift=-0.25cm] {y};
					\node[] at (-1.5,2.5) {0};
					\node[] at (-0.5,2.5) {1};
					\node[] at (0.5,2.5) {2};
					\node[] at (1.5,2.5) {3};
					
					\node[] at (-2.5,1.5) {0};
					\node[] at (-2.5,0.5) {1};
					\node[] at (-2.5,-0.5) {2};
					\node[] at (-2.5,-1.5) {3};

					\end{tikzpicture}		
					\vspace{-15pt}
				\end{center}
				\caption{Example of a $ 4 \times 4 $ grid world.}
				\label{fig:GridEx}
			\end{figure}
			
		}
	
	\newcommand{\GridExNE}
	{
		\begin{figure}
			\begin{center}
				\vspace{-15pt}
				\begin{tikzpicture}[every node/.style={minimum size=1cm-\pgflinewidth, outer sep=0pt}]
				\draw[step=1cm,color=black] (-2,-2) grid (2,2);
				\node[fill=gray] at (-0.5,-0.5) {};
				\node[fill=gray] at (-0.5,0.5) {};
				\node[fill=gray] at (-1.5,0.5) {};
				\node[fill=gray] at (-1.5,-0.5) {};
				\node[fill=gray] at (1.5,0.5) {};
				\node[fill=gray] at (1.5,-0.5) {};
				\draw [->](-1.5,1.5) -- (-0.5,1.5){};
				\draw [->](-1.5,-1.5) -- (-0.5,-1.5){};
				\node[circle, thick, fill=white, minimum size=0.5cm,draw] at (-1.5,1.5){};
				\node[rectangle, fill=black, minimum size=0.5cm] at (-1.5,-1.5){};
				\draw (-2,2) -- (-2.5,2.5) node[pos=0.75,xshift=0.25cm] {x} node[pos=0.75,yshift=-0.25cm] {y};
				\node[] at (-1.5,2.5) {0};
				\node[] at (-0.5,2.5) {1};
				\node[] at (0.5,2.5) {2};
				\node[] at (1.5,2.5) {3};
				
				\node[] at (-2.5,1.5) {0};
				\node[] at (-2.5,0.5) {1};
				\node[] at (-2.5,-0.5) {2};
				\node[] at (-2.5,-1.5) {3};
				
				\end{tikzpicture}		
				\vspace{-15pt}
			\end{center}
			\caption{A $ 4 \times 4 $ grid world with safe Nash equilibrium.}
			\label{fig:GridExNE}
		\end{figure}
		
	}
	
	\newcommand{\GridPlot}
	{
		\begin{figure}
			\begin{center}
				\vspace{-15pt}
				\resizebox{0.8\textwidth}{!}{
					\begin{tikzpicture}
					\begin{axis}[
					xlabel={size},
					xmin=3, xmax=10,
					ymin=0, ymax=100000,
					xtick={3,4,5,6,7,8,9,10},
					ymode=log,
					ytick={0,1,10,100,1000,10000,100000},
					legend pos=outer north east,
					ymajorgrids=true,
					grid style=dashed,
					]
					
					\addplot[
					solid, every mark/.append style={solid, fill=white}, mark=*
					]
					coordinates {
						(3,15)(4,40)(5,94)(6,155)(7,228)(8,491)(9,564)(10,916)
					};
					\addlegendentry{KS}
					
					\addplot[
					dashed, every mark/.append style={solid, fill=black}, mark=square*
					]
					coordinates {
						(3,44)(4,150)(5,398)(6,655)(7,994)(8,2297)(9,2687)(10,4780)
					};
					\addlegendentry{KE}
					
					\addplot[
					densely dotted, every mark/.append style={solid, fill=white}, mark=otimes*
					]
					coordinates {
						(3,60)(4,156)(5,376)(6,619)(7,909)(8,1963)(9,2256)(10,3657)
					};
					\addlegendentry{GS}
					
					\addplot[
					solid, every mark/.append style={solid, fill=white}, mark=triangle*
					]
					coordinates {
						(3,173)(4,595)(5,1591)(6,2622)(7,3969)(8,9190)(9,10748)(10,19102)
					};
					\addlegendentry{GE}
					
					\addplot[
					dashed, every mark/.append style={solid, fill=black},mark=diamond*
					]
					coordinates {
						(3,0.44)(4,0.98)(5,4.73)(6,9.53)(7,17.69)(8,50.91)(9,100.94)(10,211.30)
					};
					\addlegendentry{$ \nu $}
					
					\addplot[
					dashed, every mark/.append style={solid, fill=black}, mark=x, mark size=4pt
					]
					coordinates {
						(3,1.21)(4,1.57)(5,22.51)(6,32.32)(7,48.90)(8,121.33)(9,6002.80)(10,6871.16)
					};
					\addlegendentry{$ \epsilon $}
					
					\end{axis}
					\end{tikzpicture}
				}\vspace{-20pt}
			\end{center}
			\caption{Plots from Table \ref{tab:grid}. Y-axis is in logarithmic scale.}
			\label{fig:GridPlot}
		\end{figure}
	}

\graphicspath{{./graphics/}}
\let\phi=\varphi
\usepackage{changebar}

\begin{document}

\begin{frontmatter}

\title{Automated Temporal Equilibrium Analysis: \\ Verification and Synthesis of Multi-Player Games}
%\tnotetext[mytitlenote]{Fully documented templates are available in the elsarticle package on \href{http://www.ctan.org/tex-archive/macros/latex/contrib/elsarticle}{CTAN}.}

%% Group authors per affiliation:
\author[monash]{Julian Gutierrez}
\ead{julian.gutierrez@monash.edu}
\author[TUK]{Muhammad Najib}
\ead{najib@cs.uni-kl.de}
\author[rome]{Giuseppe Perelli}
\ead{perelli@diag.uniroma1.it}
\author[oxford]{Michael Wooldridge}
\ead{michael.wooldridge@cs.ox.ac.uk}
%\fntext[myfootnote]{Since 1880.}

%
%%% or include affiliations in footnotes:
%\author[mymainaddress,mysecondaryaddress]{Elsevier Inc}
%\ead[url]{www.elsevier.com}
%
\cortext[corresponding]{Corresponding author: Julian Gutierrez.}
%\ead{julian.gutierrez@cs.ox.ac.uk}
%
\address[monash]{Faculty of Information Technology, Monash University}
\address[TUK]{Department of Computer Science, University of Kaiserslautern}
\address[rome]{Department of Computer, Automatic, and Management Engineering, Sapienza University of Rome}
\address[oxford]{Department of Computer Science, University of Oxford}

\begin{abstract}
  In the context of multi-agent systems, the \textit{rational
    verification} problem is concerned with checking which temporal
  logic properties will hold in a system when its constituent agents
  are assumed to behave rationally and strategically in pursuit of
  individual objectives.  Typically, those objectives are expressed as
  temporal logic formulae which the relevant agent desires to see
  satisfied.  Unfortunately, rational verification is computationally
  complex, and requires specialised techniques in order to obtain
  practically useable implementations.  In this paper, we present such
  a technique.  This technique relies on a reduction of the rational
  verification problem to the solution of a collection of parity
  games.
 % One key aspect of our approach is that it preserves equilibria across bisimilar systems, which other existing approaches to rational verification do not.  
  Our approach has been implemented in the \emph{E}quilibrium
  \emph{V}erification \emph{E}nvironment (\EVE) system.  The \EVE\
  system takes as input a model of a concurrent/multi-agent system
  represented using the Simple Reactive Modules Language (SRML), where
  agent goals are represented as Linear Temporal Logic (\LTL) formulae,
  together with a claim about the equilibrium behaviour of the system,
  also expressed as an \LTL\ formula.  \EVE\ can then check whether
  the \LTL\ claim holds on some (or every) computation of the system
  that could arise through agents choosing Nash equilibrium
  strategies; it can also check whether a system has a Nash
  equilibrium, and synthesise individual strategies for players in the
  multi-player game.  After presenting our basic framework, we
  describe our new technique and prove its correctness.  We then
  describe our implementation in the \EVE\ system, and present
  experimental results which show that \EVE\ performs favourably in
  comparison to other existing tools that support rational
  verification.
\end{abstract}

\end{frontmatter}
%\linenumbers
%%%%%%%%%%%%%%%%%%%%%%%%%%%%%%%%%%%%%%%%%%%%%%%%%%%%%%%%%%%%%%%%%%%%%%
\section{Introduction}\label{secn:intro}
The deployment of AI technologies in a wide range of application areas
over the past decade has brought the problem of \emph{verifying} such
systems into sharp focus. Verification is the problem of ensuring that
a particular system is correct with respect to some specification. The
most successful approach to automated formal verification is that of
\emph{model checking}~\cite{CGP02}. With this approach, we first
derive a finite state abstract model of the system $\mathcal{S}$ being
studied; a common approach involves representing the system as a
directed graph in which vertices correspond to states of
the system, and edges correspond to the execution of program
instructions, or the performance of actions; branching in the graph
represents either input from the environment, or choices available to
components of the system. With this approach, the directed graph is
typically referred to as a labelled transition system, or Kripke
structure: each path through the transition system represents a
possible execution or computation of the system
$\mathcal{S}$. Correctness properties of interest are expressed as
formulae $\phi$ of propositional temporal logic---the most popular
such logics for this purpose are Linear Temporal Logic (\LTL) and the
Computation Tree Logic (\CTL). In the case of properties $\phi$
expressed as \LTL\ formulae, we typically want to check whether $\phi$
is satisfied on some or all possible computations of $\mathcal{S}$,
that is, on some or all possible paths through the transition
system/Kripke structure representing $\mathcal{S}$.

Great advances have been made in model checking since the approach was
first proposed in the early 1980s, and the technique is now widely
used in industry. Nevertheless, the verification of practical software
systems is by no means a solved problem, and remains the subject of
intense ongoing research.  The verification of AI systems, however,
raises a distinctive new set of challenges.  The present paper is
concerned with the problem of verifying {\em multi-agent systems},
which are AI systems consisting of multiple interacting
semi-autonomous software components known as {\em
  agents}~\cite{Woo01,shoham:2008a}.  Software agents were originally
proposed in the late 1980s, but it is only over the past decade that
the software agent paradigm has been widely adopted.  At the time of
writing, software agents are ubiquitous: we have software agents in
our phone ({\em e.g.}, Siri), processing requests online,
automatically trading in global markets, controlling complex
navigation systems ({\em e.g.}, those in self-driving cars), and even
carrying out tasks on our behalf in our homes ({\em e.g.},
Alexa). Typically, these agents do not work in isolation: they may
interact with humans or with other software agents. The field of
multi-agent systems is concerned with understanding and engineering
systems that have these characteristics.

We typically assume that agents are acting in pursuit of goals or
preferences that are delegated to them by their users. However,
whether an agent is able to achieve its goal, or the extent to which
it can bring about its preferences, will be directly influenced by the
behaviour of other agents. Thus, to act optimally, an agent must
reason \emph{strategically}, taking into account the goals/preferences
of other agents, and the fact that they too will be acting
strategically in the pursuit of these, taking into account the
goals/preferences of other agents and their own strategic
behaviour. \emph{Game theory} is the mathematical theory of strategic
interaction, and as such, it provides a natural set of tools for
reasoning about multi-agent systems~\cite{OR94}.

With respect to the problem of verifying multi-agent systems, the
relevance of game theory is as follows. Suppose we are interested in
whether a multi-agent system $\mathcal{S}$, populated by
self-interested agents, might exhibit some property represented by an
\LTL\ formula $\phi$. We can, of course, directly apply standard model
checking techniques, to determine whether $\phi$ holds on some or all
computations of $\mathcal{S}$. However, given that our agents are
assumed to act rationally, whether $\phi$ holds on some or all
computations is not relevant if the computations in question involve
irrational choices on behalf of some agents in the system. A much more
relevant question, therefore, is whether $\phi$ holds on some or all
computations \emph{that could result from agents in the system making
  rational choices}. This raises the question of what counts as a
rational choice by the agents in the system, and for this game theory
provides a number of answers, in the form of \emph{solution concepts}
such as Nash equilibrium~\cite{OR94,shoham:2008a}.  Thus, from the point of
view of game theory, {\em correct behaviour} would correspond to {\em
  rational behaviour} according to some game theoretic solution
concept, which is another way of saying that agents in the system will
act {\em optimally} with respect to their preferences/goals, under the
assumption that other agents do the same.

This approach to reasoning about the behaviour of multi-agent AI
systems establishes a natural connection between multi-agent systems
and multi-player games: agents correspond to players, computations of
the multi-agent system correspond to plays of the game, individual
agent behaviours correspond to player strategies (which define how
players make choices in the system over time), and correct behaviour
would correspond to rational behaviour---in our case, player
behaviour that is consistent with the set of Nash equilibria of the
multi-player game, whenever such a set is non-empty.  Our main
interest in this paper is the development of the theory, algorithms,
and tools for the automated game theoretic analysis of concurrent and
multi-agent systems, and in particular, the analysis of temporal logic
properties that will hold in a multi-agent system under the assumption
that players choose strategies which form a Nash
equilibrium\footnote{Although in this work we focus on Nash
  equilibrium, a similar methodology may be applied using refinements
  of Nash equilibrium and other solution concepts.}.

The connection between AI systems (modelled as multi-agent systems)
and multi-player games is well-established, but one may still wonder
why {\em correct behaviour} for the AI system should correspond to
{\em rational behaviour} in the multi-player game. This is a
legitimate question, especially, because game theory offers very many
different notions of rationality, and therefore of optimal behaviour
in the system/game. For instance, solution concepts such as
subgame-perfect Nash equilibrium (SPNE) and strong Nash equilibrium
(SNE) are refinements of Nash equilibrium where the notion of
rationality needs to satisfy stronger requirements. Consequently,
there may be executions of a multi-agent system that would correspond
to a Nash equilibrium of the associated multi-player game (thus,
regarded as correct behaviours of the multi-agent system), but which
do not correspond to a subgame-perfect Nash equilibrium or to a strong
Nash equilibrium of the associated multi-player game. We do not argue
that Nash equilibrium is the only solution concept of relevance in the
game theoretic analysis of multi-agent systems, but we believe (as do
many others~\cite{shoham:2008a,Halpern08,AbrahamAH11}) that Nash
equilibrium is a natural and appropriate starting point for such an
analysis.
Taking Nash equilibrium as our baseline notion of rationality in
multi-player games, and therefore of correctness in multi-agent
systems, we focus our study on two problems related to the
temporal equilibrium analysis of multi-agent
systems~\cite{GHW17-aij,WooldridgeGHMPT16}, as we now explain.

\vspace*{1ex}\paragraph*{Synthesis and Rational Verification} The two
main problems of interest to us are the \emph{rational verification}
and \emph{automated synthesis} problems for concurrent and multi-agent
systems modelled as multi-player games. In the \emph{rational
  verification} problem, we desire to check which temporal logic
properties are satisfied by the system/game \emph{in equilibrium},
that is, temporal logic properties satisfied by executions of the
multi-agent system generated by strategies that form a Nash
equilibrium.  A little more formally, let $P_1,\ldots,P_n$ be the
agents in our concurrent and multi-agent system, and let
$\NE(P_1,\ldots,P_n)$ denote the set of all executions, hereafter
called runs, of the system that could be generated by agents selecting
strategies that form a Nash equilibrium. Finally, let $\phi$ be an
\LTL\ formula.  Then, in the rational verification problem, we want to
know whether for some/every run $\pi\in \NE(P_1,\ldots,P_n)$ we have
$\pi\models\phi$.

In the automated \emph{synthesis} problem, on the other hand, we additionally
desire to \emph{construct} a profile of strategies for players so that
the resulting profile is an equilibrium of the multi-player game, and induces a
run that satisfies a given property of interest, again
expressed as a temporal logic formula.  That is, we are given the
system $P_1, \ldots, P_n$, and a temporal logic property $\phi$, and
we are asked to compute Nash equilibrium strategies
$\vec{\sigma} = (\sigma_1,\ldots,\sigma_n)$, one for each player in the game, 
that would result in
$\phi$ being satisfied in the run $\pi(\vec{\sigma})$ that would be
generated when these strategies are enacted. 

\vspace*{1ex}\paragraph*{Our Approach} In this paper, we present a new  
approach to the rational verification and automated synthesis problems
for concurrent and multi-agent systems. 
%, using a
%model of strategies that is bisimulation invariant---that is, in which
%individual strategies for system components are valid across all
%bisimilar systems, and which satisfy the same temporal logic
%properties in equilibrium. 
In particular, we develop a novel technique that can be used for both
rational verification and automated synthesis using a reduction to the
solution of a collection of \emph{parity games}. The technique can be
efficiently implemented making use of powerful techniques for parity
games and temporal logic synthesis and verification, and has been
deployed in the \emph{E}quilibrium \emph{V}erification
\emph{E}nvironment (\EVE~\cite{eve18}), which supports high-level
descriptions of systems/games using the {\em Simple Reactive Modules
  Language} (SRML~\cite{HoekLW06,GHW17-aij}) and temporal logic
specifications given by Linear Temporal Logic
formulae~\cite{pnueli:77a}.

The central decision problem that we consider is that of
\textsc{Non-Emptiness}, the problem of checking if the set of Nash
equilibria in a multi-player game is empty; as we will later show,
rational verification and synthesis can be reduced to this problem.
% Thus, the key problem of interest to us may
% be stated as follows:
% %
% \begin{quote}
% 	\textbf{Given}: A game $G$, representing a concurrent/multi-agent system.\\
% 	\textbf{Question}: Is it the case that $G$ has some Nash equilibrium?
% \end{quote}
%
If we consider concurrent and multi-player games in which players have
goals expressed as temporal logic formulae, this problem is known to
be 2EXPTIME-complete for a wide range of system representations and
temporal logic languages. For instance, for games with perfect
information played on labelled graphs, the problem is
2EXPTIME-complete when goals are given as \LTL\
formulae~\cite{MogaveroMPV14}, and 2EXPTIME-hard when goals are given
in \CTL~\cite{GHW17-apal}.  The problem is 2EXPTIME-complete even if
succinct representations~\cite{FismanKL10,GutierrezHW15} or only
two-player games~\cite{ChatterjeeHP10} are considered, and becomes
undecidable if imperfect information and more than two players are
allowed~\cite{GutierrezPW18}, showing the very high complexity of
solving this problem, from both practical and theoretical viewpoints.

A common feature of the results above mentioned is that---modulo
minor variations---their solutions are, in the end, reduced to the
construction of an alternating parity automaton over \emph{infinite
  trees} (APT~\cite{Loding12}) which are then checked for
non-emptiness.  Here, we present a novel, simpler, and more direct
technique for checking the existence of Nash equilibria in games where
players have goals expressed in \LTL. In particular, our technique
does not rely on the solution of an APT. Instead, we reduce the
problem to the solution of (a collection of) {parity
  games~\cite{EmersonJ91}, which are widely used for synthesis and
  verification problems.

  Formally, a parity game is a two-player zero-sum turn-based game
  given by a labelled finite graph~$H = (V_0,V_1,E,\alpha)$ such that
  $V = V_0 \cup V_1$ is a set of states partitioned into Player $0$
  ($V_0$) and Player $1$ ($V_1$) states, respectively,
  $E\subseteq V\times V$ is a set of edges/transitions, and
  $\alpha : V \to \mathbb{N}$ is a labelling priority function. Player
  0 wins if the smallest priority that occurs infinitely often in the
  infinite play is even. Otherwise, player 1 wins. It is known that
  solving a parity game (checking which player has a winning strategy)
  is in NP$\ \cap \ $coNP~\cite{Jurdzinski98}, and can be solved in
  quasi-polynomial time~\cite{CaludeJKLS17}~\footnote{Despite more than 30 years of research, and promising practical performance for algorithms to solve them, it remains unknown whether parity games can be solved in polynomial time.}.

  Our technique uses parity games in the following way. We take as
  input a game~$G$ (representing a concurrent and multi-agent system)
  and build a parity game~$H$ whose sets of states and transitions are
  doubly exponential in the size of the input but with priority
  function only exponential in the size of the input game. Using a
  deterministic Streett automaton on \emph{infinite words}
  (DSW~\cite{Kupferman18}), we then solve the parity game, leading to
  a decision procedure that is, overall, in 2EXPTIME, and, therefore,
  given the hardness results we mentioned above, essentially optimal.

  \vspace*{1ex}\paragraph*{Context} Games have several dimensions: for
  example, they may be cooperative or non-cooperative; have perfect or
  imperfect information; have perfect or imperfect recall; be
  stochastic or not; amongst many other features. Each of these
  aspects will have a modelling and computational impact on the work
  to be developed, and so it is important to be precise about the
  nature of the games we are studying, and therefore the assumptions
  underpinning our approach.

  Our framework considers non-cooperative multi-player general-sum
  games with perfect information, with Nash equilibrium as the main
  game-theoretic solution concept. The games are played on finite
  structures (state-transition structures induced by high-level SRML
  descriptions), with players having goals (preferences over plays)
  given by \LTL\ formulae and deterministic strategies represented by
  finite-state machines with output (Moore machines, sometimes
  referred to as transducers).  Because of the features of our
  framework -- chiefly, the fact that players have \LTL\ goals and
  games are played on finite structures -- considering deterministic
  strategies modelled as finite-state machines does not represent a
  restriction: in our framework, anything that a player can achieve
  with a perfect-recall strategy can also be achieved with a
  finite-state machine strategy (see,
  \emph{e.g.,}~\cite{GutierrezHW15} for the formal results).

  Finally, we note that our games have equilibria that are
  \emph{bisimulation invariant}: that is, bisimilar structures have
  the same set of Nash equilibria. This is a highly desirable
  property, and to the best of our knowledge, in this respect our work
  is unique in the computer science and multi-agent systems
  literatures.

\vspace*{1ex}\paragraph*{The \EVE\ System} The technique outlined
above and described in detail in this paper has been successfully
implemented in the \emph{E}quilibrium \emph{V}erification
\emph{E}nvironment (\EVE) system~\cite{GNPW18}.  \EVE\ takes as input a model of a
concurrent and multi-agent system, in which agents are specified using the
Simple Reactive Modules Language
(SRML)~\cite{HoekLW06,GHW17-aij}, and preferences for agents
are defined by associating with each agent a goal, represented as a
formula of \LTL~\cite{pnueli:77a}. 
Note that we believe our choice of the Reactive Modules language is a
very natural one~\cite{AlurH99b}: The language is both widely used in
practical model checking systems, such as
\textsf{MOCHA}~\cite{AlurHMQRT98}
and~\textsf{PRISM}~\cite{KwiatkowskaNP09}, and close to real-world
(declarative) programming models and specification languages.

Now, given a specification of a multi-agent system and player
preferences, the \EVE\ system can: (i) check for the existence of a
Nash equilibrium in a multi-player game; (ii) check whether a given
\LTL formula is satisfied on some or every Nash equilibrium of the
system; and (iii) synthesise individual player strategies in the game.
As we will show in the paper, \EVE\ performs favourably compared with
other existing tools that support rational verification. Moreover,
\EVE\ is the first and only tool for automated temporal equilibrium
analysis for a model of multi-player games where Nash equilibria are
preserved under bisimilarity\footnote{Other tools to compute Nash
  equilibria exist, but they do not use our model of strategies. A
  comparison with those other techniques for equilibrium analysis are
  discussed later.}.

Note that our approach may be used to model a wide range of
multi-agent systems. For example, as shown in \cite{GHW17-aij}, it is
easy to capture multi-agent STRIPS systems~\cite{BrafmanD08}.

\vspace*{1ex}\paragraph*{Structure of the paper} The remainder of this
article is structured as follows.
\begin{itemize}
\item Section~\ref{secn:prelim} presents the
relevant background on games, logic, and automata. 
\item In
Section~\ref{secn:preface}, we formalise the main problem of interest
and give a high-level description of the core decision procedure for
temporal equilibrium analysis developed in this paper. 
\item In Sections~\ref{secn:ltltoparity}, \ref{secn:characterisation},
  and~\ref{secn:computation}, we describe in detail our main decision
  procedure for temporal equilibrium analysis, prove its correctness,
  and show that it is essentially optimal with respect to
  computational complexity.
\item In Section~\ref{secn:synvef}, we show how to use our main
  decision procedure to do rational verification and automated
  synthesis of logic-based multi-player games.
\item In Section~\ref{sec:implementation}, we describe the \EVE\ system, and
  give detailed experimental results which demonstrate that \EVE\
  performs favourably in comparison with other tools that support
  rational verification.
\item In Section~\ref{secn:conc}, we
  conclude, discuss relevant related work, and propose some avenues
  for future work.    
\end{itemize}
The source code for \EVE\ is available online\footnote{See
  \url{https://github.com/eve-mas/eve-parity}}, and the system can
also be accessed via the web\footnote{See
  \url{http://eve.cs.ox.ac.uk/}}.
\section{Preliminaries}\label{secn:prelim}
\paragraph{\bf Games}
A \emph{concurrent (multi-player) game structure} (CGS) is a tuple 
$$\CGModel = (\Ag, (\Ac_{i})_{i \in \Ag}, \St, s_{0}, \trnFun)$$ where $\Ag = \{1, \dots, n \}$ is a set of \emph{players}, each $ \Ac_i $ is a set of \emph{actions}, $\St$ is a set of \emph{states}, with a designated \emph{initial} state $ s_{0} $. With each player $ i \in \Ag $ and each state $ s \in \St $, we associate a non-empty set $ \Ac_{i}(s) $ of \textit{available} actions that, intuitively, $ i $ can perform when in state $ s $.
We refer to a profile of actions
$\vec{a} = (a_{1}, \dots, a_{n}) \in \AcProf = \Ac_{1} \times \dots \times \Ac_{n}$ as a \emph{direction}.
%We also consider \emph{partial} directions.
A direction $ \vec{a} $ is available in state $ s $ if for all $ i $ we have $ a_{i} \in \Ac_{i}(s) $. Write $ \AcProf(s) $ for the set of available directions in state $ s $. For a given set of players
$A \subseteq \Ag$ and an action profile $\vec{a}$, we let
$\vec{a}_{A}$ and $\vec{a}_{-A}$ be two tuples of actions,
respectively, one for each player in $A$ and one for each player in
$\Ag \setminus A$.  We also write $\vec{a}_{i}$ for $\vec{a}_{\{i\}} $
and $ \vec{a}_{-i} $ for $ \vec{a}_{\Ag \setminus \{i\}} $.  Furthermore,
for two directions $\vec{a}$ and $\vec{a}'$, we write
$(\vec{a}_{A}, \vec{a}_{-A}')$ to denote the direction where the
actions for players in $ A $ are taken from $\vec{a}$ and the actions
for players in $ \Ag \setminus A $ are taken from $\vec{a}'$. Finally, $ \trnFun $ is a \textit{deterministic transition function}, which associate each state $ s $ and every available direction $ \vec{a} $ in $ s $ a state $ s' \in \St $.

Whenever there is $ \vec{a}$ such that $\trnFun(s, \vec{a}) = s'$, we
say that $s'$ is \emph{accessible} from $s$. A \emph{path}
$\pi = s_{0}, s_{1}, \ldots \in \St^\omega$ is an infinite sequence of
states such that, for every $k \in \SetN$, $s_{k + 1}$ is accessible
from $s_{k}$.  By $\pi_{k}$ we refer to the $(k + 1)$-th state in
$\pi$ and by $\pi_{\leq k}$ to the (finite) prefix of $\pi$ up to the
$(k + 1)$-th element.  An \emph{action profile run} is an infinite
sequence $\eta = \vec{a}_{0}, \vec{a}_{1}, \ldots$ of action profiles.
Note that, since $\CGModel$ is deterministic (\emph{i.e.}, the
transition function $\trnFun$ is deterministic), for a given state
$s_{0}$, an action profile run uniquely determines the path $\pi$ in
which, for every $k \in \SetN$,
$\pi_{k + 1} = \trnFun(\pi_{k}, \vec{a}_{k})$.

%$$\CGModel = (\Ag, (\Ac_{i})_{i \in \Ag}, \St, s_{0}, \trnFun)$$ 

A CGS is a type of concurrent system. As such, behaviourally equivalent CGSs should give rise to strategically equivalent games. However, that is not always the case. A comprehensive study of this issue can be found in \cite{GHPW17,GHPW19} where the strategic power of games is compared using one of the most important behavioural (also called observational) equivalences in concurrency, namely bisimilarity, which is usually defined over Kripke structures or labelled transition systems (see, {\em e.g.}, \cite{milner:89a,HM85}). However, the equivalence can be uniformly defined for general CGSs, where directions play the role of, for instance, actions in transition systems. 
Formally, 
let $M=(\Ag,(\Ac_{i})_{i \in \Ag},\St,s_0,\trnFun)$ and $M'=(\Ag,(\Ac_{i})_{i \in \Ag},\St',s'_0,\trnFun')$ 
be two CGSs, and $\valf:St\to\AP$ and $\valf':\St'\to\AP$ be two labelling functions over a set of propositional variables $\AP$. 
A \emph{bisimulation}, denoted by $\Bisimilar$, 
%$\Bisimilar$
between states $s^*\in\St$ and $t^*\in\St'$ is a non-empty binary 
relation $\Bisim \subseteq \St \times \St'$, such that $s^* \mathrel{\Bisim} t^*$ and for all $s,s'\in\St$, $t,t'\in\St'$, and $\direction\in\AcProf$: %each of the following conditions holds:
\begin{itemize}
	\item $s\mathrel{\Bisim} t$ implies $\valf(s)=\valf'(t)$,
	\item $s\mathrel{\Bisim} t$ and 
	%$s\transarrow{\direction}s'$ 
	$\trnFun(s,\direction)=s'$
	implies 
	%$t\transarrow{\direction}t''$ 
	$\trnFun(t,\direction)=t''$
	for some $t''\in\St'$ with $s'\mathrel{\Bisim}t''$,
	\item $s\mathrel{\Bisim} t$ and 
	%$t\transarrow{\direction}t'$ 
	$\trnFun(t,\direction)=t'$
	implies 
	%$s\transarrow{\direction}s''$ 
	$\trnFun(s,\direction)=s''$
	for some $s''\in\St$ with $s''\mathrel{\Bisim}t'$.
\end{itemize}
Then, if there is a bisimulation between two states~$s^*$ and~$t^*$, we say that they are {\em bisimilar} and write $s^*\Bisimilar t^*$ in such a case. We also say that CGSs~$M$ and~$M'$ are \emph{bisimilar} (in symbols $M\Bisimilar M'$) if $s_0 \Bisimilar s'_0$. Bisimilar structures satisfy the same set of temporal logic properties, a desirable property that will be relevant later. 

A CGS defines the dynamic structure of a game,
but lacks a central aspect of games in the sense of game theory:
preferences, which give games their strategic structure.  A
\emph{multi-player game} is obtained from a structure $\CGModel$ by
associating each player with a goal.  In this paper, we consider
multi-player games with parity and Linear Temporal Logic (\LTL) goals.

\LTL~\cite{pnueli:77a} extends classical propositional logic with two
operators, $\ltlnext$ (``next'') and $\until$ (``until''), that can be
used to express properties of paths.  The syntax of \LTL is defined
with respect to a set $\APSet$ of propositional variables as follows:
$$ \phi ::= 
\mathop\top \mid
p \mid
\neg \phi \mid
\phi \vee \phi \mid
\ltlnext \phi \mid
\phi \until \phi 
$$
where $p \in \APSet$. The remaining classical logical connectives are defined in terms of $ \lnot $ and $ \vee $ in the usual way. Two key derived \LTL operators are $ \sometime $ (``eventually'') and $ \always $ (``always''), which are defined in terms of $ \until $ as follows: $ \sometime \phi = \top \until \phi $ and $ \always \phi = \lnot \sometime \lnot \phi $.

We interpret formulae of \LTL with respect to tuples $(\pi,t,\labFun)$, where
$\pi$ is a path over some multi-player game, $t \in \Nat$ is a
temporal index into $\pi$, and $\labFun: \St \to \pow{\APSet}$ is a
labelling function, that indicates which propositional variables are
true in every state.  Formally, the semantics of \LTL is
given by the following rules:
$$
\begin{array}{lcl}
(\pi,t,\labFun)\models\mathop\top	\\
(\pi,t,\labFun)\models p 				&\text{ iff }&	p\in\labFun(\pi_t)\\
(\pi,t,\labFun)\models\neg \phi			&\text{ iff }&   \text{it is not the case that $(\pi,t,\labFun) \models \phi$}\\
(\pi,t,\labFun)\models\phi \vee \psi		&\text{ iff }&	\text{$(\pi,t,\labFun) \models \phi$  or $(\pi,t,\labFun) \models \psi$}\\
(\pi,t,\labFun)\models\ltlnext\phi			&\text{ iff }&	\text{$(\pi,t+1,\labFun) \models \phi$}\\
(\pi,t,\labFun)\models\phi\until\psi	&\text{ iff }&   \text{for some $t' \geq t: \ \big((\pi,t',\labFun) \models \psi$  and }\\
&&\quad\text{for all $t \leq t'' < t': \ (\pi,t'',\labFun) \models \phi \big)$.}\\
\end{array}
$$
If $(\pi,0,\labFun)\models\phi$, we write $\pi\models\phi$ and say that
\emph{$\pi$ satisfies~$\phi$}. 

\begin{definition}
	\label{dfn:ltlgame}
	A \emph{(concurrent multi-player) \LTL game} is a tuple $$\LTLGame = (\CGModel, \labFun, (\gamma_i)_{i \in \Ag})$$ where $\labFun: \St \to \pow{\APSet}$ is a labelling function on the set of states~$\St$ of~$\CGModel$, and each $\gamma_i$ is the goal of player~$i$, given as an \LTL formula over~$\APSet$.
\end{definition}

To define multi-player games with parity goals we consider priority functions. 
Let~$\alpha: \St \to \SetN$ be a priority function. 
A path~$\pi$ satisfies~$\alpha: \St \to \SetN$, and write $\pi \models \alpha$ in that case, if the minimum number occurring infinitely often in the infinite sequence~$\alpha(\pi_0),\alpha(\pi_1),\alpha(\pi_2),\ldots$ is even.

%\cbstart
Observe that parity conditions are \emph{prefix-independent}, that is, for every path $\pi$ and a finite sequence $h \in \St^{*}$, it holds that $h \cdot \pi \models \alpha$ if and only if $\pi \models \alpha$.
%\cbend

\begin{definition}
	\label{dfn:pargame}
	A \emph{(concurrent multi-player) Parity game} is a tuple $$\ParGame = (\CGModel, (\alpha_i)_{i \in \Ag})$$ where $\alpha_i: \St \to \SetN$ is the goal of player~$i$, given as a priority function over $\St$.
\end{definition}

Hereafter, for statements regarding either \LTL or Parity games\footnote{To simplify notations, note that , hereafter, by ``Parity game'' we denote the concurrent and multi-player extension defined here of the well-known two-player turn-based parity games in the literature.}, we will simply denote the underlying structure as~$\Game$.
Games are played by each player~$i$ selecting a \emph{strategy}~$\sigma_i$
that will define how to make choices over time.
Formally, for a given game $\Game$, a strategy~$\sigma_{i} = (S_{i}, s_{i}^{0}, \delta_i, \tau_i) $ for player~$i$ is a finite state machine with output (a transducer), where $S_{i}$ is a finite and non-empty set of \emph{internal states}, $ s_{i}^{0} $ is the \emph{initial state}, $\delta_i: S_{i} \times \AcProf \rightarrow S_{i} $ is a deterministic \emph{internal transition function}, and $\tau_i: S_{i} \rightarrow \Ac_i$ an \emph{action function}.
Note that strategies are required to output actions that are available to the agent in the current state.
To enforce this, we assume that the current state $s \in \St$ in the arena is encoded in the internal state $s_i$ in $S_i$ of agent $i$ and that the action $\tau_{i}(s_i)$ taken by the action function belongs to $\Ac_{i}(s)$.
Let $\Sigma_i$ be the set of strategies for player~$i$.
A strategy is \emph{memoryless} in $\Game$ from $s$ if $S_{i} = \St$, $s_{i}^{0} = s$, and $\delta_{i} = \trnFun$.
Once every player~$i$ has selected a strategy~$\sigma_i$, a 
\emph{strategy profile}~$\vec{\sigma} = (\sigma_1, \dots, \sigma_n)$ results and the game
has an \emph{outcome}, a path in~$\CGModel$, which we will denote by $\pi(\vec{\sigma})$. 
Because strategies are deterministic, $\pi(\vec{\sigma})$ is 
the unique path induced by~$\vec{\sigma}$, 
that is, the infinite sequence $s_0,s_1,s_2,\ldots$ such that 
\begin{itemize}
\item $s_{k+1} = \trnFun (s_k,(\tau_1(s^k_1),\cdots,\tau_n(s^k_n)))$, and 
\item $s^{k+1}_i = \delta_i(s^k_i,(\tau_1(s^k_1),\cdots,\tau_n(s^k_n)))$, 
for all $k\geq 0$. 
\end{itemize}

%\cbstart

Note that the path induced by the strategy profile $\vec{\sigma} (\sigma_1, \ldots, \sigma_{n})$ from state $s_0$ corresponds to the one generated by the finite transducer $\Trans_{\vec{\sigma}}$ obtained from the composition of the strategies $\sigma_{i}$'s in $\vec{\sigma}$, with input set $\St$ and output set $\vec{\Ac}$, where the initial input is $s_{0}$.
Since such transducer is finite, the generated path $\pi$ is \emph{ultimately periodic}, that is, there exists $p, r \in \SetN$ such that $\pi_k = \pi_{k + r}$ for every $p \leq k$.
This means that, after the prefix $\pi_{\leq p}$, the path loops indefinitely over the sequence $\pi_{p + 1} \ldots \pi_{p + r}$.

%\cbend

%\vspace{0.3cm}
%\noindent
%{\bf Nash equilibrium.}
\paragraph{\bf Nash equilibrium}
Since the outcome of a game determines if a player goal is satisfied,  we can define a 
preference relation~$\succeq_i$ over outcomes for each player~$i$.
Let~$w_i$ be $\gamma_i$ if $\Game$ is an \LTL game, and be $\alpha_i$ if $\Game$ is a Parity game. 		
%	Let~$w_i=\gamma_i$ and $w_i=\alpha_i$ if $\Game$ is an \LTL game or if $\Game$ is a Parity game, respectively. 
Then, for two strategy profiles~$\vec\sigma$ and $\vec\sigma'$ in $\Game$, we have 
$$
\text{$\pi(\vec\sigma)\succeq_i\pi(\vec\sigma')$\ if and only if \ 
	$\pi(\vec\sigma')\models w_i$ implies $\pi(\vec\sigma)\models w_i$.}
$$
On this basis, we can define the concept of Nash equilibrium~\cite{OR94} for a multi-player game with \LTL or parity goals: given a game $\Game$, a strategy profile $\vec{\sigma}$ is a \emph{Nash equilibrium} of~$\Game$ if, for every player~$i$ and strategy $\sigma'_i\in\Sigma_i$, we have 
$$
\pi(\vec{\sigma})	\succeq_i	\pi((\vec{\sigma}_{-i},\sigma'_i))
$$
where $(\vec{\sigma}_{-i},\sigma'_i)$ denotes $(\sigma_1, \dots, \sigma_{i - 1}, \sigma'_i, \sigma_{i + 1}, \dots, \sigma_n)$, the strategy profile where the strategy of player~$i$ in $\vec{\sigma}$ is replaced by $\sigma'_i$. 
Let $\NE(\Game)$ denote the set of Nash equilibria of~$\Game$. 
In~\cite{GHPW17,GHPW19} we showed that, using the model of strategies defined above, the existence of Nash equilibria is preserved across bisimilar systems.
This is in contrast to other models of strategies considered in the concurrent games literature, which do not preserve Nash equilibria.
Because of this, hereafter, we say that $\{\Sigma_i\}_{i\in\Ag}$ is a set of \emph{bisimulation-invariant strategies} and that $\NE(\Game)$ is the set of bisimulation-invariant Nash equilibrium profiles of~$\Game$.

\paragraph{\bf Automata}
A \emph{deterministic automaton on infinite words} is a tuple
$$ \Automaton = (\AP, Q, q^0, \rho, \mathcal{F}) $$ where $ Q $ is a
finite set of states, $ \rho: Q \times \AP \rightarrow Q $ is a
transition function, $ q^0 $ is an initial state, and $\mathcal{F}$ is an
acceptance condition.  We mainly use \emph{parity} and \emph{Streett}
acceptance conditions. A parity condition $ \mathcal{F} $ is a partition
$ \{F_1,\dots,F_n\} $ of $ Q $, where $ n $ is the \emph{index} of the
parity condition and any $ [1,n] \ni k $ is a \emph{priority}. We use
a \emph{priority function} $ \alpha: Q \rightarrow \mathbb{N} $ that maps
states to priorities such that $ \alpha(q)=k $ if and only if
$ q \in F_k $. For a run $ \pi = q^0,q^1,q^2\dots $, let
$ \mathit{inf}(\pi)$ denote the set of states occurring infinitely often
in the run: 
\[ \mathit{inf}(\pi) = \{q \in Q \,|\, q = q^i \text{ for infinitely
    many \emph{i}'s}\} \] A run $\pi$ is accepted by a deterministic
parity word (DPW) automaton with condition $ \mathcal{F} $ if the
minimum priority that occurs infinitely often is even, {\em i.e.}, if the
following condition is satisfied:
\[ \left(\min_{k \in [1,n]}(\mathit{inf}(\pi) \cap F_k \neq \varnothing)\right)
  \bmod 2 = 0.\] 
A Streett condition $ \mathcal{F} $ is a set of pairs
$ \{(E_1,C_1),\dots,(E_n,C_n)\} $ where $ E_k \subseteq Q$ and
$ C_k \subseteq Q $ for all $ k \in [1,n] $. A run $ \pi $ is
accepted by a deterministic Streett word (DSW) automaton~$\mthname{S}$
with condition $ \mathcal{F} $ if  $\pi$
either visits $ E_k $ finitely many times or visits $ C_k $ infinitely
often, {\em i.e.}, if for every $ k $ either
$ \mathit{inf}(\pi) \cap E_k = \varnothing $ or
$ \mathit{inf}(\pi) \cap C_k \neq \varnothing $.

\GridEx

\paragraph{\bf Example} In order to illustrate the usage of our framework, consider the following example. Suppose we have two robots/agents inhabiting a grid world (an abstraction of some environment, \textit{e.g.,} a warehouse) with dimensions $ n \times n $. Initially, the agents are located at some corners of the grid; 
%specifically, agent 1 is located at the  top-left corner (coordinate $ (0,0) $) and agent 2 at the bottom-right corner $ (n-1,n-1) $. 
The agents are each able to move around the grid in directions \textit{north}, \textit{south}, \textit{east}, and \textit{west}. The goal of each agent is to reach the opposite corner. For instance, if agent \textit{i}'s initial position is $ (0,0) $, then the goal is to reach position~$ (n-1,n-1) $. A number of obstacles may also appear on the grid. The agents are not allowed to move into a coordinate occupied by an obstacle or outside the grid world. To make it clearer, consider the configuration shown in Figure~\ref{fig:GridEx}; a (grey) filled square depicts an obstacle. Agent 1, depicted by $ \blacksquare $, can only move west to $ (2,3) $, whereas agent 2, depicted by $ \bigcirc $, can only move east to $ (1,0) $.

In this example we make the following assumptions: (1) at each timestep, each agent has to make a move, that is, it cannot stay at the same position for two consecutive timesteps, and it can only move at most one step; (2) the goal of each agent is, as stated previously, to eventually reach the opposite corner of her initial position. From system design point of view, the question that may be asked is: can we synthesise a strategy profile such that it induces a stable (Nash equilibrium) run and at the same time ensures that the agents never crash into each other?

\GridExNE

Checking the existence of such strategy profile is not trivial. For instance, the configuration in Figure \ref{fig:GridEx} does not admit any safe Nash equilibrium runs, that is, where all agents get their goals achieved without crashing into each other. Player $ \bigcirc $ can reach $ (3,3) $ without crashing into $ \blacksquare $, since $ \blacksquare $ can safely ``wait'' by moving back and forth between $ (0,3) $ and $ (1,3) $ until $ \bigcirc $ reaches $ (3,3) $. However, there is no similar safe ``waiting zone'' for $ \bigcirc $ to get out of $ \blacksquare $'s way. On the other hand, the configuration in Figure \ref{fig:GridExNE}, admits safe Nash equilibrium; $ \bigcirc $ and $ \blacksquare $ have safe waiting zones $ (0,0) $ and $ (1,0) $, and $ (0,3) $ and $ (1,3) $, respectively. Clearly, such a reasoning is not always straightforward, especially when the setting is more complex, and therefore, having a tool to verify and synthesise such scenario is desirable. Later in Section \ref{sec:exp_grid} we will discuss how to encode and check such systems using our tool.

\section{A Decision Procedure using Parity Games}\label{secn:preface}
We are now in a position to formally state the \textsc{Non-Emptiness} problem:
\begin{quote}
	\emph{Given}: An \LTL Game $\LTLGame$.\\
	\emph{Question}: Is it the case that $\NE(\LTLGame)\neq\emptyset$?
\end{quote}
%
%It is known~\cite{Gao0W17} that using only polynomial translations, this decision problem, called \textsc{Non-Emptiness}, can be used to do both rational synthesis~\cite{FismanKL10} and rational verification~\cite{WooldridgeGHMPT16}. 

%~

%\noindent{\bf The decision procedure.} 
As indicated before, we solve both verification and synthesis through
a reduction to the above problem.  The technique we develop consists
of three steps. First, we build a Parity game~$\ParGame$ from an input
\LTL game~$\LTLGame$. Then---using a characterisation of Nash
equilibrium (presented later) that separates players in the game into
those that achieve their goals in a Nash equilibrium (the ``winners'',
$W$) and those that do not achieve their goals (the ``losers'',
$L$)---for each set of players in the game, we eliminate nodes and
paths in~$\ParGame$ which cannot be a part of a Nash equilibrium, thus
producing a modified Parity game, $\ParGame^{-L}$. Finally, in the
third step, we use Streett automata on infinite words to check if the
obtained Parity game witnesses the existence of a Nash
equilibrium. The overall algorithm is presented in
Algorithm~\ref{algo:NEviaParity} which also includes some comments pointing to the relevant Sections/Theorems. The first step is contained in
line~3, while the third step is in lines~12--14. The rest of the
algorithm is concerned with the second step. In the sections that
follow, we will describe each step of the algorithm and, in
particular, what are and how to compute $\Pun_{j}(\ParGame)$ and
$\ParGame^{-L}$, two key constructions used in our decision procedure.

\algparsolv	

%It should be noted that even though 
%We note that although we use automata in our algorithm (specifically, Streett automata to check for the existence of a Nash equilibrium in line~12 of the algorithm, and parity automata to build a parity game in line~3 of the algorithm), most reasoning is done at the level of parity games, \emph{i.e.}, in the second step of the decision procedure. 

%\noindent {\bf Complexity.} 
%Regarding the complexity, 
\paragraph{\bf Complexity}
The procedure presented above
runs in doubly exponential time, matching the \emph{optimal} upper bound of
the problem. In the first step we obtain a doubly exponential
blowup. The underlying structure~$\CGModel$ of the obtained Parity
game~$\ParGame$ is doubly exponential in the size of the goals of the
input~\LTL game~$\LTLGame$, but the priority functions
set~$(\alpha_i)_{i\in\Ag}$ is only (singly) exponential. Then, in the
second step, reasoning takes only polynomial time in the size of the
underlying concurrent game structure of~$\ParGame$, but exponential
time in both the number of players and the size of the priority
functions set. Finally, the third step takes only polynomial time,
leading to an overall 2EXPTIME complexity.

\section{From LTL to Parity}\label{secn:ltltoparity}
We now describe how to realise line~3 of
Algorithm~\ref{algo:NEviaParity}, and in doing so we prove a strong
correspondence between the set of Nash equilibria of the input \LTL
game~$\LTLGame$ and the set of Nash equilibria of its associated
Parity game~$\ParGame$. This result allows us to shift reasoning
on the set of Nash equilibria of~$\LTLGame$ into reasoning on the set
of Nash equilibria of~$\ParGame$. The basic idea behind this step of
the decision procedure is to transform all \LTL
goals~$(\gamma_i)_{i\in\Ag}$ in~$\LTLGame$ into a collection of DPWs,
denoted by~$(\Automaton[\gamma_i])_{i\in\Ag}$, that will be used to
build the underlying CGS of~$\ParGame$. We construct~$\ParGame$ as
follows.

In general, using the results in~\cite{SistlaVW87,Piterman07}, 
from any \LTL formula~$ \varphi $ over $\AP$ one can build 
a DPW~$\Automaton[\varphi] = \tuple{\pow{\AP}, Q, q^{0}, \rho, \alpha}$ such that, $\Language(\Automaton[\varphi]) = \set{\pi \in (\pow{\AP})^{\omega}}{\pi \models \varphi}$, 
that is, 
the language accepted by~$\Automaton[\varphi]$ is exactly the set of words over~$\pow{\AP}$ that are models of~$\varphi$.
The size of $Q$ is doubly exponential in~$\card{\varphi}$ and the size of the range of~$\alpha$ is singly exponential in~$\card{\varphi}$.
Using this construction we can define, for each \LTL goal~$\gamma_i$, a DPW~$\Automaton[\gamma_i]$.

\begin{definition}
	\label{def:ltltopar}
	Let $\LTLGame = (\CGModel, \labFun, (\gamma_i)_{i \in \Ag})$ be an \LTL game whose underlying CGS is 
	$\CGModel = (\Ag, (\Ac_{i})_{i \in \Ag}, \St, s_{0}, \trnFun)$, and 
	%Moreover, 
	let $\Automaton[\gamma_{i}] = \tuple{\pow{\AP}, Q_{i}, q_{i}^{0}, \rho_{i}, \alpha_{i}}$ be the DPW corresponding to player $i$'s goal $\gamma_{i}$ in $\LTLGame$.
	The \emph{Parity game~$\ParGame$ associated to~$\LTLGame$}~is $\ParGame = (\CGModel', (\alpha_i')_{i \in \Ag})$, where $\CGModel' = (\Ag, (\Ac_{i})_{i \in \Ag}, \St', s_{0}', \trnFun[]['])$ and $(\alpha_i')_{i\in\Ag}$ are as follows:
	
	\begin{itemize}
		\item
		$\St' = \St \times \bigtimes_{i \in \Ag} Q_{i}$ and $s_{0}' = (s_{0}, q^{0}_{1}, \ldots , q^{0}_{n})$;
		
		%		\item
		%		$s_{0}' = (s_{0}, q^{0}_{1}, \ldots , q^{0}_{n})$;
		
		\item
		for each state $(s, q_{1}, \ldots , q_{n})\in \St'$ and action profile $\vec{a}$, \\ $\trnFun'((s, q_{1}, \ldots , q_{n}), \vec{a}) =(\trnFun(s, \vec{a}), \rho_{1}(q_{1}, \labFun(s)), \ldots, \rho_{n}(q_{n}, \labFun(s))$;			
		
		\item $\alpha_i'(s, q_{1}, \ldots q_{n}) = \alpha_{i}(q_{i})$.
	\end{itemize}
	
\end{definition}

Intuitively, the game $\ParGame$ is the product of the \LTL game $\LTLGame$ and the collection of parity (word) automata $\Automaton[\gamma_{i}]$ that recognise the models of each player's goal.
Informally, the game executes in parallel the original \LTL game together with the automata built on top of the \LTL goals.
At every step of the game, the first component of the product state follows the transition function of the original game~$\LTLGame$, while the ``automata'' components are updated according to the labelling of the current state of $\LTLGame$.
As a result, the execution in $\ParGame$ is made, component by component, by the original execution, say $\pi$, in the \LTL game~$\LTLGame$, paired with the unique runs of the DPWs $\Automaton[\gamma_{i}]$ generated when reading the word $\labFun(\pi)$.

%This construction is reported in Algorithm \ref{algo:cmlg2cmpg}.
%\todo[color=white]{Adjust the notation in the algorithm according to the new one. Not sure we need this, though.}
%
%\begin{algorithm}[H]
%	\textbf{Input: }{CMLG $ \mathcal{G}=(\AP,\Ag,\Ac,\St,s^0,\lambda,\tau,(\gamma_i)_{i \in \Ag}) $.}\\
%	\textbf{Output: }{CMPG $ \mathcal{G}^{\star} = ( \Ag,\St^\star,\Ac,s^{\star 0},\tau^\star,(\alpha_i)_{i \in \Ag} ) $. }\\
%	\ForEach{$ i \in \textup{Ag} $}{
%		build DPW $ \mathcal{A}_i=(\AP,Q_i,\delta_i,q^0_i,\alpha_i) $ from $ \gamma_i $;
%		}
%	$ s^{\star 0} \leftarrow \{ (s^0,(q^0_i)_{i \in \Ag}) \} $\;
%	\ForEach{$ (s,q_1,\dots,q_n) \in \textup{St} \times Q_1 \times \dots \times Q_n $}{
%		$ \St^\star \leftarrow \St^\star \cup \{(s,q_1,\dots,q_n)\} $\;
%		}
%	\ForEach{$ (s^\star,s^{\star\prime}) \in \textup{\St}^\star \times \textup{St}^\star $}{
%		\If{$ \tau(s,d)=s' \wedge \bigwedge_{i \in \Ag} \delta_i(q_i,\lambda(s)) = q_i', (s,q_i) \in s^\star, (s',q_i') \in s^{\star\prime} $}
%		{$ \tau^\star \leftarrow \tau^\star \cup \{ (s^\star,d,s^{\star\prime}) \} $ \;}
%		}
%\caption{CMLG to CMPG}
%\label{algo:cmlg2cmpg}
%\end{algorithm}

%Since the size of $ \mathcal{A}_i $ is doubly exponential to the size of $ \gamma_i $, it is easy to see that the construction produces a CMPG $ \mathcal{G}^{\star} $ with the size of doubly exponential with respect to the size of formulas in CMLG $ \mathcal{G} $.

%\begin{theorem}
%	Given a \textup{CMLG} $ \mathcal{G} $, there is a \textup{CMGP} $ \mathcal{G}^\star $ with $ 2^{2^{O|\gamma|}} $ states.
%\end{theorem}

%\newpage 
Observe that in the translation from~$\LTLGame$ to its associated~$\ParGame$ the set of actions for each player is unchanged. This, in turn, means that the set of strategies in both $\LTLGame$ and $\ParGame$ is the same, since for every state~$s\in\St$ and action profile~$\vec{a}$, it follows that $\vec{a}$ is available in $s$ if and only if it is available in $(s,q_1,\ldots,q_n)\in\St'$, for all $(q_1,\ldots,q_n)\in\bigtimes_{i \in \Ag} Q_{i}$. Using this correspondence between strategies in~$\LTLGame$ and strategies in~$\ParGame$, we can prove the following Lemma, which states an invariance result between $\LTLGame$ and $\ParGame$ with respect to the satisfaction of players' goals. 

\begin{lemma}[Goals satisfaction invariance]
	\label{lmm:satinvariance}
	Let $\LTLGame$ be an \LTL game and $\ParGame$ its associated Parity game.
	Then, for every strategy profile $\vec{\sigma}$ and player~$i$, it is the case that $\pi(\vec{\sigma}) \models \gamma_{i}$ in $\LTLGame$ if and only if $\pi(\vec{\sigma}) \models \alpha_{i}$ in $\ParGame$.
\end{lemma}

\begin{proof}
	We prove the statement by double implication.
	To show the left to right implication, assume that $\pi(\vec{\sigma}) \models \gamma_{i}$ in $\LTLGame$, for any player~$i \in \Ag$, and let $\pi$ denote the infinite path generated by $\vec{\sigma}$ in $\LTLGame$; 
	thus, we have that~$\labFun(\pi) \models \gamma_{i}$.
	On the other hand, let $\pi'$ denote the infinite path generated in~$\ParGame$ by the same strategy profile~$\vec{\sigma}$.
	Observe that the first component of~$\pi'$ is exactly~$\pi$.
	Moreover, consider the~$(i + 1)$-th component $\rho_{i}$ of $\pi'$.
	By the definition of $\ParGame$, it holds that $\rho_{i}$ is the run executed by the automaton $\Automaton[\gamma_{i}]$ when the word $\labFun(\pi)$ is read.
	By the definition of the labelling function of $\ParGame$, it holds that the parity of~$\pi'$ according to $\alpha'_{i}$ corresponds to the one recognised by $\Automaton[\gamma_{i}]$ in $\rho_{i}$.
	Thus, since we know that~$\labFun(\pi) \models \gamma_{i}$, it follows that $\rho_{i}$ is accepting in $\Automaton[\gamma_{i}]$ and therefore $\pi' \models \alpha_{i}$, which implies that~$\pi(\vec{\sigma}) \models \alpha_{i}$ in~$\ParGame$.
	For the %right to left 
	other direction, observe that all implications used above are equivalences. %Thus, 
	Using those equivalences one can reason backwards to prove the statement. 
	%\todo[color=white]{GP: Maybe we can improve the right-to-left.}
\end{proof}

Using Lemma~\ref{lmm:satinvariance} we can then show that the set of Nash Equilibria for any \LTL game exactly corresponds to the set of Nash equilibria of its associated Parity game. Formally, we have the following invariance result between games. 

\begin{theorem}[Nash equilibrium invariance]
	\label{thm:NEinvariance}
	Let $\LTLGame$ be an \LTL game and $\ParGame$ its associated Parity game.
	Then, $\NE(\LTLGame) = \NE(\ParGame)$.
\end{theorem}

\begin{proof}
	The proof proceeds by double inclusion.
	First, assume that a strategy profile~$\vec{\sigma} \in \NE(\LTLGame)$ is a Nash Equilibrium in $\LTLGame$ and, by contradiction, it is not a Nash Equilibrium in $\ParGame$.
	Observe that, due to Lemma~\ref{lmm:satinvariance}, we known that the set of players that get their goals satisfied by~$\pi(\vec{\sigma})$ in~$\LTLGame$ (the ``winners'', $W$) is the same set of players that get their goals satisfied by~$\pi(\vec{\sigma})$ in~$\ParGame$. Then, there is player~$j \in L = \Ag \setminus W$ and a strategy $\sigma_{j}'$ such that $\pi((\vec{\sigma}_{-j}, \sigma_{j}')) \models \alpha_{j}$ in~$\ParGame$.
	Then, due to Lemma~\ref{lmm:satinvariance}, we have that $\pi((\vec{\sigma}_{-j}, \sigma_{j}')) \models \gamma_{j}$ in~$\LTLGame$ and so $\sigma_{j}'$ would be a beneficial deviation for player~$j$ in $\LTLGame$ too---a contradiction.
	On the other hand, for every $\vec{\sigma} \in \NE(\ParGame)$, we can reason in a symmetric way and conclude that $\vec{\sigma} \in \NE(\LTLGame)$.
\end{proof}

\section{Characterising Nash Equilibria}\label{secn:characterisation}
Thanks to Theorem~\ref{thm:NEinvariance}, we can focus our attention
on Parity games, since a technique for solving such games will also
provide a technique for solving their associated \LTL games.  To do
this we characterise the set of Nash equilibria in the Parity
game construction~$\ParGame$ in our algorithm.
The existence of Nash Equilibria in \LTL games can be characterised in
terms of punishment strategies and memoryful
reasoning~\cite{GutierrezHW15-concur}.  We will show that a similar
characterisation holds here in a parity games framework, where
only memoryless reasoning is required. To do this, we first
introduce the notion of punishment strategies and regions formally, as
well as some useful definitions and notations.
%
%{\bf Notation:} 
In what follows, given a (memoryless) strategy profile $\vec{\sigma}=(\sigma_1,\ldots,\sigma_n)$ defined on a state~$s\in \St$ of a Parity game~$\ParGame$, that is, such that $s^0_i=s$ for every $i\in\Ag$, we write $\ParGame, \vec{\sigma}, s \models \alpha_{i}$ if $\pi(\vec{\sigma}) \models \alpha_{i}$ in~$\ParGame$. Moreover, if $s = s_{0}$ is the initial state of the game, we omit it and simply write $\ParGame, \vec{\sigma} \models \alpha_{i}$ in such a case. 

\begin{definition}[Punishment strategies and regions]
	\label{def:punish}
	For a Parity game $\ParGame$ and a player~$i \in \Ag$, we say that $\vec{\sigma}_{-i}$ is a \emph{punishment (partial) strategy profile} against $i$ in a state~$s$ if, for all strategies $\sigma_{i}'\in\Sigma_i$, it is the case that $\ParGame, (\vec{\sigma}_{-i}, \sigma_{i}'), s \not\models \alpha_{i}$.
	A state $s$ is \emph{punishing} for~$i$ if there exists a punishment (partial) strategy profile against~$i$ in~$s$.
	By $\Pun_{i}(\ParGame)$ we denote the set of punishing states, the \emph{punishment region}, for $i$ in $\ParGame$.
\end{definition}

To %better 
understand the meaning of a punishment (partial) strategy profile, it is useful to think of a modification of the game $\ParGame$, in which player $i$ still has its goal~$\alpha_{i}$, while the rest of the players are collectively playing in an adversarial mode, \emph{i.e.}, trying to make sure that~$i$ does not achieve~$\alpha_i$. 
This scenario is %formally 
represented by a two-player zero-sum game in which the winning strategies of the (coalition) player, denoted by~$-i$, correspond (one-to-one) to the punishment strategies in the original game~$\ParGame$.
As described in~\cite{GutierrezHW15-concur}, knowing the set of punishment (partial) strategy profiles in a given game is important to compute its set of Nash Equilibria.
For this reason, it is useful to compute the set $\Pun_{i}(\ParGame)$, that is, the set of states in the game from which a given player~$i$ can be punished. 
(\emph{e.g.}, to deter undesirable unilateral player deviations). 
To do this, we reduce the problem to computing a winning strategy in a %suitable 
turn-based two-player zero-sum parity game, whose definition is as follows.

%\begin{figure}_{h}%{r}{0.500\textwidth}
%	\missingfigure[figwidth=1\textwidth]{Figure about the sequentialised transition}
%\end{figure}

\begin{definition}
	\label{dfn:seqgame}
	For a (concurrent multi-player) Parity game 
	$$\ParGame = (\Ag, \St, (\Ac_{i})_{i \in \Ag}, s_{0}, \trnFun, (\alpha_{i})_{i \in \Ag})$$ and player~$j \in \Ag$, the \emph{sequentialisation} of $\ParGame$ with respect to player~$j$ is the (turn-based two-player) parity game $\ParGame^{j} = \tuple{V_{0}, V_{1}, \ERel, \alpha}$ where
	
	\begin{itemize}
		\item
		$V_{0} = \St$ and $V_{1} = \St \times \AcProf_{-j}$;
		
		\item
%		$\ERel = \St \times (\St \times \AcProf_{-j}) \cup \{((s, \vec{a}_{-j}), s') \in (\St \times \AcProf_{-j}) \times \St \ :$
		$ \ERel = \{(s,(s, \vec{a}_{-j})) \in \St 
%		\wedge \vec{a}_{-j} \in \AcProf_{-j}(s)
		\times (\St \times \AcProf_{-j})
		 \} \cup \{((s, \vec{a}_{-j}), s') \in (\St \times \AcProf_{-j}) \times \St \ :$
				\begin{flushright}
		$  \ \exists a_{j}' \in \Ac_{j}. \ s' = \trnFun(s, (\vec{a}_{-j}), a_{j}')\}$;		
				\end{flushright}
		
		\item
		$\alpha: V_{0} \cup V_{1} \to \SetN$ is such that \\
		$\alpha(s) = \alpha_{j}(s) + 1$ and $\alpha(s, \vec{a}_{-j}) = \alpha_{j}(s) + 1$.
	\end{itemize}
\end{definition}

\figseqgam

The formal connection between the notion of punishment in $\ParGame$ and the
set of winning strategies in $\ParGame^{j}$ is established in the
following theorem, where by $\WinSet[0](\ParGame^{j})$ we denote the
winning region of Player~$0$ in~$\ParGame^{j}$, that is, the %set of
states from which Player~$0$, representing the set of
players~$-j = \Ag\setminus \{j\}$ (the coalition of players not including~$j$), has a memoryless winning strategy
against player~$j$ in the two-player zero-sum parity game~$\ParGame^{j}$.

\begin{theorem}
	\label{thm:punchar}
	For all states~$s \in \St$, it is the case that $s \in \Pun_{j}(\ParGame)$ if and only if $s \in \WinSet[0](\ParGame^{j})$.
	In other words, it holds that $\Pun_{j}(\ParGame) = \WinSet[0](\ParGame^{j}) \cap \St$.
\end{theorem}

\newcommand{\odd}{\mthfun{odd}}
\newcommand{\even}{\mthfun{even}}
\newcommand{\InfSet}{\mthset{Inf}}

\begin{proof}
	The proof goes by double inclusion.
	From left to right, assume $s \in \Pun_{j}(\ParGame)$ and let $\vec{\sigma}_{-j}$ be a punishment  strategy profile against player~$j$ in $s$, \emph{i.e.}, such that $\ParGame, (\vec{\sigma}_{-j}, \sigma_{j}'),s \not\models \alpha_{j}$, for every strategy $\sigma_{j}'\in\Sigma_j$ of player~$j$.
	We now define a strategy $\sigma_{0}$ for player $0$ in $\ParGame^{j}$ that is winning in $s$.
	In order to do this, first observe that, for every finite path $\pi_{\leq k}' \in V^{*} \cdot V_{0}$ in $\ParGame^{j}$ starting from $s$, there is a unique finite sequence of action profiles $\vec{a}_{-j}^{0}, \ldots, \vec{a}_{-j}^{k}$ and a sequence $\pi_{\leq k} = s^{0}, \ldots, s^{k + 1}$ of states in $\St^*$ such that
	$$\pi_{\leq k}' = s^{0}, (s^{0}, \vec{a}_{- j}^{0}), \ldots, s^{k}, (s^{k}, \vec{a}_{- j}^{k}) , \ldots,  s^{k + 1} \ .$$ %.
	Now, for every path $\pi_{\leq k}'$ of this form that is consistent with $\vec{\sigma}_{-j}$, \emph{i.e.}, the sequence $\vec{a}_{-j}^{0}, \ldots, \vec{a}_{-j}^{k - 1}$ is generated by $\vec{\sigma}_{-j}$, define $\sigma_{0}(\pi_{\leq k}') = (s^{k + 1}, \vec{a}_{-j}^{k + 1})$, where $\vec{a}_{-j}^{k + 1}$ is the action profile selected by $\vec{\sigma}_{-j}$.
	To prove that $\sigma_{0}$ is winning, consider a strategy $\sigma_{1}$ for Player~$1$ and the infinite path $\pi'=\pi((\sigma_{0}, \sigma_{1}))$ generated by $(\sigma_{0}, \sigma_{1})$.
	It is not hard to see that the sequence $\pi_{\odd}'$ of odd positions in $\pi'$ belongs to a path $\pi$ in $\ParGame$ and it is consistent with $\vec{\sigma}_{-j}$.
	Thus, since $\vec{\sigma}_{-j}$ is a punishment strategy, $\pi_{\odd}'$ does not satisfy $\alpha_{j}$.
	Moreover, observe that the parity of the sequence $\pi_{\even}'$ of even positions equals that of~$\pi_{\odd}'$.
	Thus, we have that $\InfSet(\labFun'(\pi')) + 1 = \InfSet(\labFun'(\pi_{\odd}')) + 1 \cup \InfSet(\labFun'(\pi_{\even}')) + 1 = \InfSet(\labFun(\pi))$ and so $\pi'$ is winning for player $0$ in $\ParGame^{j}$ and $\sigma_{0}$ is a winning strategy.
	
	From right to left, let $s \in \St \cap \Win[0](\ParGame^{j})$ and let $\sigma_{0}$ be a winning strategy for Player~$0$ in~$\ParGame^{j}$, and 
	%In particular, we can 
	assume $\sigma_{0}$ %to be 
	is memoryless.
	Now, for every player~$i$, with~$i \neq j$, define the memoryless strategy $\sigma_{i}$ in $\ParGame$ such that, for every $s' \in \St$, if $\sigma_{0}(s') = (s', \vec{a}_{-j})$, then $\sigma_{i}(s') = (\vec{a}_{-j})_{i}$~\footnote{By an abuse of notation, we let~$\sigma_{i}(s')$ be the value of~$\tau_{i}(s')$.}, \emph{i.e.}, the action that player~$i$ takes in $\sigma_{0}$ at $s'$.
	Now, consider the (memoryless) strategy profile $\vec{\sigma}_{-j}$ given by the composition of all strategies~$\sigma_{i}$, and consider a play $\pi$ in $\ParGame$, starting from $s$, that is consistent with $\vec{\sigma}_{-j}$.
	Thus, there exists a play $\pi'$ in $\ParGame^{i}$, consistent with $\sigma_{0}$, such that $\pi = \pi_{\odd}'$.
	Moreover, since $\pi_{\odd}' = \pi_{\even}'$, we have that $\InfSet(\labFun'(\pi')) = \InfSet(\labFun'(\pi_{\odd}')) \cup \InfSet(\labFun'(\pi_{\even}')) = \InfSet(\labFun(\pi)) - 1$.
	Since $\pi'$ is winning for Player $0$, we know that $\pi\not\models\alpha_j$ and so $\vec{\sigma}_{-j}$ is a punishment strategy against Player~$j$ in $s$.
\end{proof}

Definition~\ref{dfn:seqgame} and Theorem~\ref{thm:punchar} not only
make a bridge from the notion of punishment strategy to the notion of
winning strategy for two-player zero-sum games, but also provide a way
to understand how to compute punishment regions as well as how to
synthesise an actual punishment strategy in Parity games.
In this way, by computing winning regions and winning strategies in these games
we can solve the \emph{synthesis} problem for individual players in the original game with \LTL goals, 
one of the problems we are interested in. Thus, from Definition \ref{dfn:seqgame} and Theorem \ref{thm:punchar}, we have the following corollary.

\begin{corollary}
	\label{cor:punchar}
	Computing~$\Pun_{i}(\ParGame)$ can be done in polynomial time with respect to the size of the underlying graph of the game~$\ParGame$ and exponential in the size of the priority function~$\alpha_{i}$, that is, to the size of the range of~$\alpha_{i}$. Moreover, there is a memoryless strategy $\vec{\sigma}_{i}$ that is a punishment against player~$i$ in every state~$s \in \Pun_{i}(\ParGame)$.
\end{corollary}

As described in~\cite{GutierrezHW15-concur}, in any (infinite) run \emph{sustained} by a Nash equilibrium~$\vec{\sigma}$ in deterministic and pure strategies, that is, in~$\pi(\vec{\sigma})$, it is the case that all players that do not get their goals achieved in $\pi(\vec{\sigma})$ can deviate from such a (Nash equilibrium) run only to states where they can be punished by the coalition consisting of all other players in the game. To formalise this idea in the present setting, we need one more concept about punishments, defined next. 

\begin{definition}
	\label{def:punsec}
	An action profile run $\eta = \vec{a}_{0}, \vec{a}_{1}, \ldots\in\AcProf^\omega$ is \emph{punishing-secure} in $s$ for player~$j$ if, for all $k \in \SetN$ and $a_{j}'$, we have $\trnFun(\pi_{j},((\vec{a}_{k})_{-j} , a_{j}')) \in \Pun_{j}(\ParGame)$, where $\pi$ is the only play in $\ParGame$ starting from $s$ and generated by $\eta$.
\end{definition}

\figNEpun

Using the above definition, we can characterise the set of Nash equilibria of a given game.
Recall that strategies are formalised as transducers, \emph{i.e.}, as finite state machines with output, so such Nash equilibria strategy profiles produce runs which are \emph{ultimately periodic}.
Moreover, since in every run~$\pi$ there are players who get their goals achieved in $\pi$ (and therefore do not have an incentive to deviate from~$\pi$) and players who do not get their goals achieve in $\pi$ (and therefore may have an incentive to deviate from~$\pi$), we will also want to explicitly refer to such players.
To do that, the following notation will be useful: Let $W(\ParGame, \vec{\sigma}) = \set{i \in \Ag}{\ParGame , \vec{\sigma} \models \alpha_i}$ denote the set of player that get their goals achieved in $\pi(\vec{\sigma})$.
%that is, the ``winners''~$W$ of the game when playing the strategy profile~$\vec{\sigma}$. 
We also write $W(\ParGame, \pi) = \set{i \in \Ag}{\ParGame , \pi \models \alpha_i}$. 

\begin{theorem}[Nash equilibrium characterisation]
	\label{thm:NEchar}
	For a Parity game $\ParGame$, there is a Nash Equilibrium strategy profile~$\vec{\sigma} \in \NE(\ParGame)$ if and only if there is an ultimately periodic action profile run $\eta$ such that, for every player~$j \in L = \Ag \setminus W(\ParGame, \pi)$, the run~$\eta$ is punishing-secure for~$j$ in state~$s_{0}$, where $\pi$ is the unique path generated by $\eta$ from $s_{0}$. 
\end{theorem}

\begin{proof}
	The proof is by double implication.
	From left to right, for $\vec{\sigma} \in \NE(\ParGame)$, let $\eta$ be the ultimately periodic sequence of action profiles generated by $\vec{\sigma}$.
	Moreover, assume for a contradiction that $\eta$ is not punishing-secure for some $j \in L$.
	By the definition of punishment-secure, there is $k \in \SetN$ and action $a_{j}'\in\Ac_j$ for player~$j$ such that $s' = \trnFun(\pi_{k},((\vec{a}_{k})_{-j} , a_{j}') \notin \Pun_{j}(\ParGame)$.
	Now, consider the strategy $\sigma_{j}'$ that follows $\eta$ up to the $(k - 1)$-th step, executes action $a_{j}'$ on step $k$ to get into state~$s'$, and applies a strategy that achieves~$\alpha_j$ from that point onwards.
	Note that such a strategy is guaranteed to exist since $s' \notin \Pun_{j}(\ParGame)$.
	Therefore, $\ParGame, (\vec{\sigma}_{-j}, \sigma_{j}') \models \alpha_{j}$ and so $\sigma_{j}'$ is a beneficial deviation for player~$j$, a contradiction to $\vec{\sigma}$ being a Nash equilibrium.
	
	From right to left, we need to define a Nash equilibrium $\vec{\sigma}$ assuming only the existence of $\eta$.
	First, recall that $\eta$ can be generated by a finite transducer $\Trans_{\eta} = (Q_{\eta}, q_{\eta}^{0}, \delta_{\eta}, \tau_{\eta})$ where $\delta_{\eta}: Q_{\eta} \to Q_{\eta}$ and $\tau_{\eta}: Q_{\eta} \to \AcProf$.
	Moreover, for every player~$i$ and deviating player~$j$, with $i\neq j$, there is a (memoryless) strategy $\sigma_{i}^{\pun j}$ to punish player~$j$ in every state in~$\Pun_{j}(\ParGame)$.
	By suitably combining the transducer with the punishment strategies, we define the following strategy
	$\sigma_{i} = (Q_{i}, q_{i}^{0}, \delta_{i}, \tau_{i})$ for player~$i$
	where
	\begin{itemize}
		\item
		$Q_{i} = \St \times Q_{\eta} \times (L \cup \{\top\})$ 
		and 		
		$q_{i}^{0} = (s^{0}, q_{\eta}^{0}, \top)$;
		
		%		\item
		%			$\delta_{i} = Q_{i} \times \AcProf \to Q_{i}$ is such that
		%			
		%			$\delta_{i}((s, q, \top), \vec{a}) =
		%			\begin{cases}
		%				(\trnFun(s, \vec{a}), \delta_{\eta}(q), \top), & \text{if } a = \tau_{\eta}(q) \\
		%				(\trnFun(s, \vec{a}), \delta_{\eta}(q), j), & \text{if } a_{-j} = (\tau_{\eta}(q))_{-j} \text{ and } \vec{a}_{j} \neq (\tau_{\eta}(q))_{j}
		%			\end{cases}$
		\item
		$\delta_{i} = Q_{i} \times \AcProf \to Q_{i}$ is defined as 
		
		$\delta_{i}((s, q, \top), \vec{a}) =
		\begin{cases}
		(\trnFun(s, \vec{a}), \delta_{\eta}(q), \top), & \text{ if } a = \tau_{\eta}(q) \\
		(\trnFun(s, \vec{a}), \delta_{\eta}(q), j), & \text{} a_{-j} = (\tau_{\eta}(q))_{-j} \text{ and } \vec{a}_{j} \neq (\tau_{\eta}(q))_{j} \\
		\bot, & \text{ otherwise }
		\end{cases}
		$~\footnote{For completeness, the function $\delta_{i}$ is assumed to take an available action.
		However, this is not important, as it is clear from the proof we never use this case.}
		%
		
%		
%		\begin{itemize}
%			\item $\delta_{i}((s, q, \top), \vec{a}) = (\trnFun(s, \vec{a}), \delta_{\eta}(q), \top)$, if $a = \tau_{\eta}(q)$, and 
%			\item $\delta_{i}((s, q, \top), \vec{a}) = (\trnFun(s, \vec{a}), \delta_{\eta}(q), j)$, if both 
%						\begin{flushright}
%			$a_{-j} = (\tau_{\eta}(q))_{-j}$ and $\vec{a}_{j} \neq (\tau_{\eta}(q))_{j}$; 
%						\end{flushright}
%		\end{itemize}			
		%
		\item
		$\tau_{i}: Q_{i} \to \Ac_{i}$ is such that 
		\begin{itemize}
			\item
			$\tau_{i}(s, q, \top) = (\tau_{\eta}(q))_{i}$, and
			
			\item
			$\tau_{i}(s, q, j) = \sigma_{i}^{\pun j}(s)$.
		\end{itemize}

	\end{itemize}
	
	%	\figNEpun
	
	To understand how strategy $\sigma_{i}$ works, observe that its set of internal states is given by the following triple.
	The first component is a state of the game, remembering the position of the execution.
	The second component is a state of the transducer $\Trans_{\eta}$, which is used to employ the execution of the action profile run $\eta$.
	The third component is either the symbol~$\top$, used to flag that no deviation has occurred, or the name of a losing player~$j$, used to remember that such a player has deviated from~$\eta$.
	At the beginning of the play, strategy $\sigma_{i}$ starts executing the actions prescribed by the transducer $\Trans_{\eta}$.
	It sticks to it until some losing player~$j$ performs a deviation.
	In such a case, the third component of the internal state of~$\sigma_{i}$ switches to remember the deviating player.
	Moreover, from that point on, it starts executing the punishment strategy $\sigma_{i}^{\pun j}$.
%	\cbstart
	Recall that parity conditions are prefix-independent.
	Therefore, no matter the result of the execution, if all the players start playing according to the punishment strategy $\sigma_{i}^{\pun j}$, the resulting path will not satisfy the parity condition $\alpha_{j}$.
%	Moreover, note that the punishment strategy is \emph{memoryless}.
%	Thus, the application of the punishment strategy takes effect no matter when it is triggered along the execution.
%	\cbend
	Now, define $\sigma$ to be the collection of all $\sigma_{i}$.
	It remains to prove that $\vec{\sigma}$  is a Nash Equilibrium.
	
	First, observe that since~$\vec{\sigma}$ produces exactly $\eta$, we have $W(\ParGame, \vec{\sigma}) = W(\ParGame, \eta)$, that is, the players that get their goals achieved in~$\pi(\vec{\sigma})$ and $\eta$ are the same. Thus, only players in~$L$ could have a beneficial deviation.
	Now, consider a player~$j \in L$ and a strategy $\sigma_{j}'$ and let $k \in \SetN$ be the minimum (first) step where $\sigma_{j}'$ produces an outcome that differs from $\sigma_{j}$ when executed along with~$\vec{\sigma}_{-j}$. We write $ \pi' $ for $ \pi((\vec{\sigma}_{-j},\sigma'_{j})) $.
	Thus, we have $\pi_{h} = \pi_{h}'$ for all $h \leq k$ and $\pi_{k + 1} \neq \pi_{k + 1}'$.
	Hence $\pi_{k + 1}' = \trnFun(\pi_{k}', (\eta_{k})_{-j}, a_{j}') = \trnFun(\pi_{k}, (\eta_{k})_{-j}, a_{j}') \in \Pun_{j}(\ParGame)$ and $\ParGame, (\vec{\sigma}_{-j}, \sigma_{j}') \not\models \alpha_{j}$, since $\sigma_{-j}$ is a punishment strategy from $\pi_{k + 1}'$.
	Thus, there is no beneficial deviation for~$j$ and $\vec{\sigma}$ is a Nash equilibrium.
\end{proof}

\section{Computing Nash Equilibria}\label{secn:computation}
Theorem~\ref{thm:NEchar}~allows us to reduce the problem of finding a Nash equilibrium to finding a path in the game satisfying certain properties, which we will show how to check using DPW and DSW automata. 
To do this, let us fix a given set~$W\subseteq\Ag$ of players in a given game~$\ParGame$, which are assumed to get their goals achieved. 
Now, due to Theorem~\ref{thm:NEchar}, we have that an action profile run~$\eta$ corresponds to a Nash equilibrium with~$W$ being the set of ``winners'' in the game if, and only if, the following two properties are satisfied:
\begin{itemize}
	\item
	$\eta$ is punishment-secure for $j$ in $s^{0}$, for all~$j \in L = \Ag \setminus W$;
	
	\item
	$\ParGame, \pi \models \alpha_i$, for every $i \in W$;
\end{itemize}
where $\pi$ is, as usual, the path generated by $\eta$ from $s^{0}$.

To check the existence of such~$\eta$, we have to check these two properties.
First, note that, for~$\eta$ to be punishment-secure for every losing player~$j \in L$, the game has to remain in the punishment region of each~$j$.
This means that an acceptable action profile run needs to generate a path that is, at every step, contained in the intersection $\bigcap_{j \in L} \Pun_{j}(\ParGame)$. Thus, to find a Nash equilibrium, we can remove all states not in such an intersection. 
We also need to remove some edges from the game. 
Indeed, consider a state~$s$ and a partial action profile~$\vec{a}_{-j}$.
It might be the case that $\trnFun(s, (\vec{a}_{-j}, a_{j}')) \notin \Pun_{j}(\ParGame)$, for some $a_{j}' \in \Ac_{j}$.
Therefore, an action profile run that executes the partial profile $\vec{a}_{-j}$ over $s$ cannot be punishment-secure, and so all outgoing edges from $(s,\vec{a}_{-j})$, can also be removed.
After doing this for every~$j\in L$, we obtain~$\ParGame^{-L}$, the game resulting from $\ParGame$ after the removal of the states and edges just described. As a consequence, $\ParGame^{-L}$ has all and only the paths that can be generated by an action profile run that is punishment-secure for every~$j\in L$.

The only thing that remains to be done is to check whether there exists a path in $\ParGame^{-L}$ that satisfies all players~in $W$.
To do this, we use DPW and DSW automata.
Since players goals are parity conditions, a path satisfying player~$i$ is an accepting run of the DPW $\Automaton[][i]$ where the set of states and transitions are exactly those of~$\ParGame^{-L}$ and the acceptance condition is given by~$\alpha_i$.
Then, in order to find a path satisfying the goals of all players in~$W$, we can solve the emptiness problem of the automaton intersection $\bigtimes_{i \in W} \Automaton[][i]$.
%\cbstart
%In general the product intersection of DPW provides a further exponential blow-up in the size.
However, observe that each $\Automaton[i]$ differs from each other only in its acceptance condition $\alpha_{i}$.
Moreover, each parity condition $\alpha = (F_{1}, \ldots, F_{n})$ can be regarded as a Street condition of the form $((E_1, C_1), \ldots, (E_m, C_m))$ with $m = \lceil \frac{n}{2} \rceil$ and $(E_i, C_i) = (F_{2i + 1}, \bigcup_{j \leq i} F_{2j})$, for every $0 \leq i < m$.
Therefore, the intersection language of $\bigtimes_{i \in W} \Automaton[][i]$ can be recognized by a Street automaton over the same set of states and transitions and the concatenation of all the Streett conditions determined by the parity conditions of the players in $W$.
%\cbend
%To do this, we can see each~$\Automaton[][i]$ as a DSW~$\mthname{S}_{i}$ in the usual way (parity conditions are a special case of Streett~\cite{Kupferman15}).
%Since Streett automata are closed under conjunctions of Streett conditions, $\bigtimes_{i \in W} \Automaton[][i]$
The overall translation is a DSW automaton with a number of Streett pairs being logarithmic in the number of its states, whose emptiness can be solved in polynomial time~\cite{PP04}.
Finally, as we fixed~$W$ at the \emph{beginning}, all we need to do is to use the procedure just described for each~$W\subseteq\Ag$, if needed (see Algorithm~\ref{algo:NEviaParity}).
%, obtaining an \emph{optimal} decision procedure that has only exponential time and polynomial space complexity in~$|\Ag|$, the number of agents in the system.%
%\footnote{Instead, the `suspect game' has exponential space in $|\Ag|$.}
\footnote{Some previous techniques, \emph{e.g.}~\cite{BouyerBMU15}, to the computation of pure Nash equilibria are not optimal as they have exponential space complexity in the number of players~$|\Ag|$.}

%\cbstart
Concerning the complexity analysis, consider again Algorithm~\ref{algo:NEviaParity} and denote by $n$ the number of agents and $\card{\St_{\LTL}}$ the number of states.
Observe that Line~3 of the algorithm builds a Parity game $\ParGame$ by making the product construction between $\LTLGame$ and all the DPW automata $\Automaton[\gamma_{i}]$, whose state space is $\pow{\pow{\card{\gamma_{i}}}}$, and the number of priorities is $\pow{\card{\gamma_{i}}}$.
Thus, the number of states of $\ParGame$ is $\card{\St_{\Par}} = \card{\St_{\LTL}} \cdot \pow{\pow{\card{\gamma_{1}}}} \cdot \ldots \cdot \pow{\pow{\card{\gamma_{n}}}}$.
Now, on the one hand, Line~6 requires to solve a parity game on the state-graph of $\ParGame$ with $\pow{\gamma_{i}}$ priorities.
This is solved by applying Zielonka's algorithm~\cite{Zie98}, that works in time $(\card{\St_{\Par}})^{2} \cdot (\card{\St_{\Par}})^{\pow{\gamma_{i}}}$, thus polynomial in the state space of $\ParGame$ and doubly exponential in the size of objectives $\gamma_{i}$'s.
On the other hand, Line~12 calls for the Non-Emptiness procedure of a DSW whose number of Street pairs is linear in the sum of priorities of the automata $\Automaton[\gamma_{1}], \ldots, \Automaton[\gamma_{n}]$ and so logarithmic in its state-space (that is doubly exponential in the size of the objectives).
Such procedure is polynomial in the state space of the automaton~\cite[Corollary 10.8]{PP04} and therefore polynomial in $\card{\St_{\Par}}$.
Finally, consider the consider the loops of Line~4 and Line~5, respectively.
The first is on all the possible subsets of agents, and thus of length $\pow{n}$.
The second is on all the possible agents, and thus of length $n$.
This sums up to an overall complexity for Algorithm~\ref{algo:NEviaParity} of:

$$
\pow{n} \cdot n \cdot ((\card{\St_{\Par}})^{2} \cdot (\card{\St_{\Par}})^{\sum_{i \in \Ag} \pow{\gamma_{i}}} + \card{\St_{\Par}})\text{.}
$$

Recall that $\card{\St_{\Par}}$ is linear in the set of states of the $\LTLGame$ and doubly exponential in every objective $\gamma_i$'s of the agents.
Thus, the procedure is \emph{polynomial} in $\card{\St_{\LTL}}$, exponential in $N$, and doubly exponential in the size of the formulas $\card{\gamma_{1}}, \ldots, \card{\gamma_{N}}$.˘

\section{Synthesis and Verification}\label{secn:synvef}
We now show how to solve the synthesis and verification problems using
\textsc{Non-Emptiness}. For \emph{synthesis}, the solution is already
contained in the proof of Theorem~\ref{thm:NEchar}, so we only need to
sketch out the approach here. Note that, in the computation of
punishing regions, the algorithm builds, for every player~$i$ and
potential deviator~$j$, a (memoryless) strategy that player~$i$ can
play in the collective strategy profile~$\vec{\sigma}_-j$ in order to
punish player~$j$, should player~$j$ wishes to deviate. If a Nash
equilibrium exists, the algorithm also computes a (ultimately
periodic) witness of it, that is, a computation~$\pi$ in~$G$, that, in
particular, satisfies the goals of players in~$W$. At this point,
using this information, we are able to define a strategy~$\sigma_i$
for each player~$i\in \Ag$ in the game (\emph{i.e.}, including those
not in~$W$), as follows: while no deviation occurs, play the action
that contributes to generate~$\pi$, and if a deviation of player~$j$
occurs, then play the (memoryless) strategy~$\sigma_{i}^{pun j}$ that
is defined in the game to punish player~$j$ in case~$j$ were to
deviate. Notice, in addition, that because of
Lemma~\ref{lmm:satinvariance} and Theorem~\ref{thm:NEinvariance},
every strategy for player~$i$ in the game with parity goals is also a
valid strategy for player~$i$ in the game with \LTL goals, and that
such a strategy, being bisimulation-invariant, is also a strategy for
every possible bisimilar representation of player~$i$. In this way,
our technique can also solve the synthesis problem for every player,
that is, can compute individual bisimulation-invariant strategies for
every player (system component) in the original multi-player game
(concurrent system).

%Now, for \emph{verification}, we can use a reduction of the following two problems, which we call \textsc{E-Nash} and \textsc{A-Nash}, to \textsc{Non-Emptiness}. 
For \emph{verification}, one can use a reduction of the following two problems, called \textsc{E-Nash} and \textsc{A-Nash} in~\cite{GutierrezHW15,WooldridgeGHMPT16,GHW17-aij}, to \textsc{Non-Emptiness}. 

%\begin{quote}
%%\textsc{E-Nash}: \\
%\emph{Given}: \LTL game $\LTLGame$, LTL formula $\phi$.\\
%\textsc{E-Nash}: Does 
%$\exists \vec{\sigma} \in \NE(\LTLGame).\ \rho(\vec{\sigma})\models \phi$ hold?
%\end{quote}
%\begin{quote}
%%\textsc{A-Nash}: \\
%\emph{Given}: \LTL game $\LTLGame$, LTL formula $\phi$.\\
%\textsc{A-Nash}: Does 
%$\forall\vec{\sigma} \in \NE(\LTLGame).\ \rho(\vec{\sigma})\models \phi$ hold?
%\end{quote}

\begin{quote}
	%\textsc{E-Nash}: \\
	\emph{Given}: Game $\LTLGame$, \LTL formula $\phi$.\\
	\textsc{E-Nash}: Is it the case that $\pi(\vec{\sigma})\models \phi$, for some $\vec{\sigma} \in \NE(\LTLGame)$~?\\
	\textsc{A-Nash}: Is it the case that $\pi(\vec{\sigma})\models \phi$, for all $\vec{\sigma} \in \NE(\LTLGame)$~?
\end{quote}
We write $(\LTLGame,\phi)\in\text{\textsc{E-Nash}}$ to denote that $(\LTLGame,\phi)$ is an instance of \textsc{E-Nash}, \textit{i.e.}, given a game $\LTLGame$ and a \LTL formula $\phi$, the answer to \textsc{E-Nash} problem is a ``yes''; and, similarly for \textsc{A-Nash}.

Because we are working on a bisimulation-invariant setting, we can ensure something even stronger: that for any two games $\LTLGame$ and $\LTLGame'$, whose underlying CGSs are $\CGModel$ and $\CGModel'$, respectively, we know that if $\CGModel$ is bisimilar to $\CGModel'$, then $(\LTLGame,\phi)\in\text{\textsc{E-Nash}}$ if and only if $(\LTLGame',\phi)\in\text{\textsc{E-Nash}}$, for all \LTL formulae~$\phi$; and, similarly for \textsc{A-Nash}, as desired. 

In order to solve \textsc{E-Nash} and \textsc{A-Nash} via \textsc{Non-Emptiness}, one could use the following result, whose proof is a simple adaptation of the same result for iterated Boolean games~\cite{GutierrezHW15} and for multi-player games with \LTL goals modelled using SRML~\cite{GHW17-aij}, which was first presented in~\cite{Gao0W17}. 

\begin{lemma}
	\label{lem:nashtononemptiness}
	Let $G$ be a game and $\phi$ be an \LTL formula. There is a game $H$ of linear size in $G$, such that 
	$
	\NE(H)\neq\emptyset \ \text{if and only if} \ \exists \vec{\sigma}\in \NE(G) . \ \pi(\vec{\sigma})\models\phi \ .
%	\NE(H)\neq\emptyset \ \text{if and only if} (G,\phi)
	$
\end{lemma}

However, since we have Algorithm~\ref{algo:NEviaParity} at our
disposal, an easier -- and more direct -- solution can be obtained. To
solve \textsc{E-Nash} we can modify line~12 of
Algorithm~\ref{algo:NEviaParity} to include the restriction that such
an algorithm, which now receives $\phi$ as a parameter, returns
``Yes'' in line~13 if and only if~$\phi$ is satisfied in some run in
the set of Nash equilibrium witnesses. The new line~12 is ``{\bf if}
$\Language(\bigtimes_{i \in W}(\mthname{S}_{i}) \times
\mthname{S}_{\phi} ) \neq \emptyset$'', where $\mthname{S}_{\phi}$ is
the DSW automaton representing~$\phi$. All complexities remain the
same; the modified algorithm for \textsc{E-Nash} is denoted as
Algorithm~\ref{algo:NEviaParity}'. We can then use
Algorithm~\ref{algo:NEviaParity}' to solve \textsc{A-Nash}, also as
described in~\cite{Gao0W17}: essentially, we can check whether
Algorithm~\ref{algo:NEviaParity}'($\LTLGame,\neg\phi$) returns ``No''
in line~16. If it does, then no Nash equilibrium of $\LTLGame$
satisfies~$\neg\phi$, either because no Nash equilibrium exists at all
(thus, \textsc{A-Nash} is vacuously true) or because all Nash
equilibria of~$\LTLGame$ satisfy~$\phi$, then solving \textsc{A-Nash}
positively. Note that in this case, since \textsc{A-Nash} is solved
positively when the algorithm returns ``No'' in line~16, then no
specific Nash equilibrium strategy profile is synthesised, as
expected. However, if the algorithm returns ``Yes'', that is, the case
when the answer to \textsc{A-Nash} problem with $(\LTLGame,\phi)$
instance is negative, then a strategy profile is synthesised from
Algorithm~\ref{algo:NEviaParity}' which corresponds to a
counter-example for $(\LTLGame,\phi)\in\text{\textsc{A-Nash}}$. It
should be easy to see that implementing \textsc{E-Nash} and
\textsc{A-Nash} is straightforward from
Algorithm~\ref{algo:NEviaParity}. Also, as already known, it is also
easy to see that Algorithm~\ref{algo:NEviaParity}' solves
\textsc{Non-Emptiness} if and only if
$(\LTLGame,\top)\in\text{\textsc{E-Nash}}$.

\section{Implementation}\label{sec:implementation}
%implementation as a separate file: implementation.tex

%%%
% This is the body of Implementation section
%%%

\newcommand{\ra}[1]{\renewcommand{\arraystretch}{#1}}
\setlength{\tabcolsep}{5pt}

We have implemented the decision procedures presented in this
paper. Our implementation uses SRML~\cite{HoekLW06} as a modelling
language. SRML is based on the {Reactive Modules}
language~\cite{AlurH99b} which is used in a number of verification
tools, including \PRISM~\cite{KwiatkowskaNP09} and
\MOCHA~\cite{AlurHMQRT98}.
The tool that implements our algorithms is called \EVE\ (for
\emph{E}quilibrium \emph{V}erification
\emph{E}nvironment)~\cite{GNPW18}.  \EVE\ is the {\em first and only
  tool} able to analyse the linear temporal logic properties that hold
in equilibrium in a concurrent, reactive, and multi-agent system
within a bisimulation-invariant framework. It is also the only tool
that supports all of the following combined features: a high-level
description language using SRML, general-sum multi-player games with
\LTL goals, bisimulation-invariant strategies, and perfect recall. It
is also the only tool for Nash equilibrium analysis that relies on a
procedure based on the solution of parity games, which has allowed us
to solve the (rational) synthesis problem for individual players in
the system using very powerful techniques originally developed to
solve the synthesis problem from (linear-time) temporal logic
specifications.

To the best of our knowledge, there are only two other tools that can
be used to reason about temporal logic equilibrium properties of
concurrent/multi-agent systems: \PRALINE~\cite{Brenguier13} and
\MCMAS~\cite{CermakLMM14,CLMM18}.

\PRALINE allows one to compute a Nash equilibrium in a game played in
a concurrent game structure~\cite{Brenguier13}. The underlying
technique uses alternating B{\"u}chi automata and relies on the
solution of a two-player zero-sum game called the `suspect
game'~\cite{BouyerBMU15}. \PRALINE can be used to analyse games with
different kinds of players goals ({\em e.g.}, reachability, safety,
and others), but does not permit \LTL goals, and 
does not compute bisimulation-invariant strategies.

\MCMAS is a model checking tool for multi-agent
systems~\cite{LomuscioQR17}. Since it can be used to model check
Strategy Logic (SL~\cite{MogaveroMPV14}) formulae~\cite{CLMM18},
and SL can express the existence of a Nash equilibrium, one can model
a multi-agent system in \MCMAS and check for the existence of a Nash
equilibrium in such a system using SL. However, \MCMAS only supports
SL with memoryless strategies (while our implementation does not have
this restriction) and, as \PRALINE, does not compute
bisimulation-invariant strategies either.%
%\footnote{\textsf{UPPAAL}\cite{DavidLLMP15} is another tool that can be used to analyse equilibrium behaviour in a system~\cite{DavidJLMT15,DBLP:journals/corr/abs-1202-4506}. However, \textsf{UPPAAL} differs from \textsf{EVE} is various ways: {\em e.g.}, it works in a quantitative setting, uses statistical model checking, and computes approximate Nash equilibria.}

From the many differences between \PRALINE, \MCMAS, and \EVE 
(and their associated underlying reasoning and verification techniques), 
one of the most important ones is bisimulation-invariance, a feature needed 
to be able to do verification and synthesis, 
{\em e.g.}, when using symbolic methods with OBDDs or some model-minimisation techniques. 
Not being bisimulation-invariant also means that in some cases \PRALINE, \MCMAS, and \EVE 
would deliver completely different answers. For instance, unlike \EVE, with \PRALINE and \MCMAS it may be 
the case that for two bisimilar systems \PRALINE and \MCMAS would compute a Nash equilibrium in 
one of them and none in the other. A particular instance is the ``motivating example'' in~\cite{GHPW17}. 
Since the two systems there are bisimilar, \EVE is able to compute a bisimulation-invariant 
Nash equilibrium in both systems, while \PRALINE and \MCMAS, both of which are not using bisimulation-invariant model of strategies, cannot. The experiment supporting this claim is reported in Section \ref{sec:bisim_exp} along with the performance results.
Indeed, even in cases where all tools are able to compute a Nash equilibrium, \EVE outperforms 
the other two tools as the size of the input system grows, despite the fact that the model of 
strategies we use in our procedure is {\em richer} in the sense that it takes into account more information 
of the underlying game.%
\footnote{As mentioned before, not all games can be tested in all tools since, for instance, \PRALINE does not support \LTL objectives, but only goals expressed directly as B{\"u}chi conditions.}

\subsection{Tool Description}
\paragraph{Modelling Language} Systems in \EVE are specified with the \textit{Simple Reactive Modules Language} (SRML~\cite{HoekLW06}), that can be used to model non-deterministic systems. Each system component (agent/player) in SRML is represented as a {\em module}, which consists of an \textit{interface} that defines the name of the module and lists a non-empty set of Boolean variables controlled by the module, and a set of \textit{guarded commands}, which define the choices available to the module at each state. There are two kinds of guarded commands: \textbf{init}, used for initialising the variables, and \textbf{update}, used for updating variables subsequently. 
% we refer to~\cite{HoekLW06} for further details on the semantics of SRML. 

A guarded command has two parts: a ``condition" part (the ``guard") and an ``action" part. The ``guard" determines whether a guarded command can be executed or not given the current state, while the ``action" part defines how to update the value of (some of) the variables controlled by a corresponding module. Intuitively, $\varphi \leadsto \alpha$ can be read as ``if the condition $\varphi$ is satisfied, then \textit{one} of the choices available to the module is to execute $\alpha$". Note that the value of $\varphi$ being {true} does not guarantee the execution of $\alpha$, but only that it is \textit{enabled} for execution, and thus \textit{may be chosen}. If no guarded command of a module is enabled in some state, then that module has no choice and the values of the variables controlled by it remain unchanged in the next state.

Formally, an SRML module $m_i$ is defined as a triple $m_i = (\Phi_i,I_i,U_i)$, where $\Phi_i \subseteq \Phi$ is the finite set of Boolean variables controlled by $m_i$, $I_i$ a finite set of \textbf{init} guarded commands, such that for all $g \in I_i$, we have $ctr(g) \subseteq \Phi_i$, and $U_i$ a finite set of \textbf{update} guarded commands, such that for all $g \in U_i$, we have $ctr(g) \subseteq \Phi_i$. A guarded command $g$ over a set of variables~$\Phi$ is an expression
\begin{displaymath}
g : \quad \varphi \leadsto x'_1 := \psi_1;\dots;x'_k := \psi_k
\end{displaymath}
where the guard $\varphi$ is a propositional logic formula over $\Phi$, each $x_i$ is a member of $\Phi$ and $\psi_i$ is a propositional logic formula over $\Phi$. Let $guard(g)$ denote the guard of $g$, thus, in the above rule, we have $guard(g) = \varphi$. It is required that no variable $x_i$ appears on the left hand side of more than one assignment statements in the same guarded command, hence no issue on the (potentially) conflicting updates arises. The variables $x_1,\dots,x_k$ are controlled variables in $g \in U_i$ and we denote this set by $ctr(g)$. If no guarded command of a module is enabled, then the values of all variables in $ctr(g)$ are unchanged. A set of guarded commands is said to be \textit{disjoint} if their controlled variables are mutually disjoint.
To make it clearer, here is an example of a guarded command:
	\[ \underbrace{(p \wedge q)}_\text{guard} \leadsto \underbrace{p' := \top; q':=\bot}_\text{action}\]
The guard is the propositional logic formula $ (p \wedge q) $, so this guarded command will be enabled if both \textit{p} and \textit{q} are true. If the guarded command is chosen (to be executed), then in the next time-step, variable \textit{p} will be assigned {true} and variable \textit{q} will be assigned {false}.

\begin{figure}
	\begin{center}
		\begin{tabular}{l}
			$ \textbf{module } toggle \textbf{ controls } x $ \\ 
			$ \quad \textbf{init} $ \\
			$ \quad :: \top \leadsto x':= \top; $ \\
			$ \quad :: \top \leadsto x':= \bot; $ \\
			$ \quad \textbf{update} $ \\
			$ \quad :: \lnot x \leadsto x':= \top; $ \\
			$ \quad :: x \leadsto x':= \bot; $ \\
		\end{tabular}
	\end{center}
	\caption{Example of module toggle in SRML.}
	\label{fig:toggle}
\end{figure}

Figure \ref{fig:toggle} shows a module named $ toggle $ that controls a Boolean variable named $ x $. There are two \textbf{init} guarded commands and two \textbf{update} guarded commands. The \textbf{init} guarded commands define two choices for the initialisation of variable $ x $: {true} or {false}. The first \textbf{update} guarded command says that if $ x $ has the value of {true}, then the corresponding choice is to assign it to {false}, while the second command says that if $ x $ has the value of {false}, then it can be assigned to {true}. Intuitively, the module would choose (in a non-deterministic manner) an initial value for $ x $, and then on subsequent rounds toggles this value. In this particular example, the \textbf{init} commands are non-deterministic, while the \textbf{update} commands are deterministic.
We refer to~\cite{GHW17-aij} for further details on the semantics of SRML. In particular, in Figure~12 of~\cite{GHW17-aij}, we detail how to build a Kripke structure that models the behaviour of an SRML system. 
In addition, we associate each module with a goal, which is specified as an \LTL formula. 

\begin{figure}[t]
%	\vspace{-10pt}
	\includegraphics[width=\textwidth]{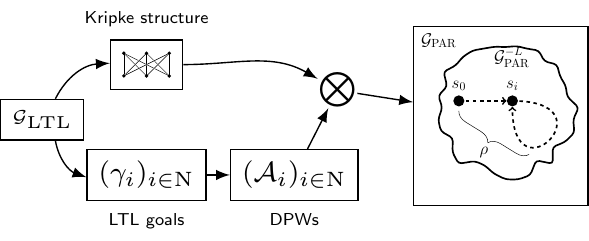}
	\caption{High-level workflow of \textsf{EVE}.}
	\label{fig:flow}
%	\vspace{-10pt}
\end{figure}

At this point, readers might notice that the way SRML modules are
defined leads to the possibility of having multiple initial states --
which appears to contradict the definition of CMGS. However, this is
not a problem, since we can always add an extra ``pre''-initial state
whose outgoing edges are labelled according to init guarded commands,
and use it as the ``real'' initial state.

\paragraph{\bf Automated Temporal Equilibrium Analysis} Once a multi-agent system is modelled in SRML, it can be seen as a multi-player game in which players (the modules) use strategies to resolve the non-deterministic choices in the system. 
%Over these games we can then ask, {\em e.g.}, whether they have any Nash equilibrium, that is, the \textsc{Non-Emptiness} problem.\footnote{Solving \textsc{E-Nash} and \textsc{A-Nash} is done using the reductions presented in \cite{Gao0W17}.} 
\textsf{EVE} uses Algorithm~\ref{algo:NEviaParity} to solve \textsc{Non-Emptiness}. The main idea behind this algorithm is illustrated in Figure~\ref{fig:flow}. The general flow of the implementation is as follows. 
Let $ \mathcal{G}_{\text{LTL}} $ be a game, modelled using SRML, with a set of players/modules $ \text{N}=\{1,\dots,n\} $ and \LTL goals $ \Gamma=\{\gamma_1,\dots,\gamma_n\} $, one for each player. Using $ \mathcal{G}_{\text{LTL}} $ we construct an associated concurrent game with parity goals $ \mathcal{G}_{\text{PAR}} $ in order to shift reasoning on the set of Nash equilibria of $ \mathcal{G}_{\text{LTL}} $ into the set of Nash equilibria of $ \mathcal{G}_{\text{PAR}} $. 
The basic idea of this construction is, firstly, to transform all LTL goals in $ \mathcal{G}_{\text{LTL}} $ into deterministic parity word (DPW) automata. To do this, we use \textsf{LTL2BA} tool \cite{GO01,ltl2ba} to transform the formulae into nondeterministic B\"{u}chi word (NBW) automata. From NBWs, we construct the associated deterministic parity word (DPW) automata via construction described in \cite{Piterman07}. Secondly, to perform a product construction of the Kripke structure that represents $ \mathcal{G}_{\text{LTL}} $ with the collection of DPWs in which the set of Nash equilibria of the input game is preserved. With $ \mathcal{G}_{\text{PAR}} $ in our hands, we can then reason about Nash equilibria by solving a collection of parity games. To solve these parity games, we use \textsf{PGSolver} tool \cite{Friedmann10thepgsolver,pgsolver}. 
%As shown in \cite{GutierrezHW15-concur}, the existence of Nash equilibria in LTL games can be characterized in terms of punishment strategies, an idea underlying the algorithm that \textsf{EVE} uses. Intuitively, \textit{punishment strategies} are strategies that prevent a player~\textit{i} to achieve its goal~$\gamma_i$, thus eliminating any incentive of~$i$ to deviate. 
%%In particular, in a parity game, punishment strategies can be characterised by \textit{punishment regions}, that is, states in which the existence of a punishment strategy against a potential ``deviator''~\textit{i} is guaranteed. Computing these regions, for every $i\in \text{N}$, can be done by computing a winning strategy against each~$i$ in a suitable two-player zero-sum (parity) game. 
\textsf{EVE} then iterates through all possible set of ``winners'' $ W \subseteq \Ag $ (Algorithm~\ref{algo:NEviaParity} line 4) and computes a punishment region~$ \text{Pun}_j(\mathcal{G}_{\text{PAR}}) $ for each~$ j \in L = \text{N} \backslash W $, with which a reduced parity game $ \mathcal{G}_{\text{PAR}}^{-L} = \bigcap_{j \in L} \text{Pun}_j(\mathcal{G}_{\text{PAR}}) $ is built. Notice that for each player $ j $, $ \Pun_{j}(\Game_{\Par})  $ need only computed once and can be stored, thus resulting in a more efficient running time. Lastly, \textsf{EVE} checks whether there exists a path $ \rho $ in $ \mathcal{G}_{\text{PAR}}^{-L} $ that satisfies the goals of each~$ i \in W $. To do this, we translate $ \mathcal{G}_{\text{PAR}}^{-L} $ into a deterministic Streett automata, whose language is empty if and only if so is the set of Nash equilibria of $ \mathcal{G}_{\text{PAR}} $. For \textsc{E-Nash} problem, we simply need to find a run in the witness returned when we check for \textsc{Non-Emptiness}; this can be done via automata intersection\footnote{For \textsc{A-Nash} is straightforward, since it is the dual of \textsc{E-Nash}.}.
%
%This algorithm runs in doubly exponential time, matching the optimal upper bound of the problem \cite{Mogavero:2014:RSM:2656934.2631917}. We obtain a doubly exponential blowup when converting LTL goals to DPWs. Computing punishment regions takes exponential time in the number of players and parity game priorities, while checking Streett automata emptyness can be done in polynomial time, thus resulting in an overall 2EXPTIME algorithm.
%\figseqgam \label{fig:seq}
%
%\paragraph{Implementation and Usage} 

\textsf{EVE} was developed in Python and available online
from~\cite{eve18}. \textsf{EVE} takes as input a concurrent and
multi-agent system described in SRML code, with player goals and a
property~$ \phi $ to be checked specified in \LTL. For
\textsc{Non-Emptiness}, \textsf{EVE} returns ``YES" (along with a set
of winning players~$W$) if the set of Nash equilibria in the system is
not empty, and returns ``NO" otherwise. For \textsc{E-Nash}
(\textsc{A-Nash}), \textsf{EVE} returns ``YES" if $ \phi $ holds on
\textit{some} (\textit{every}) Nash equilibrium of the system, and
``NO" otherwise.

In the next subsection, we present some case studies to evaluate the performance of \EVE. The case studies are based on distributed and concurrent systems that can naturally be modelled as multi-agent systems. We note, however, that such case studies bear no special relevance to multi-agent systems research. Instead, our only purpose is to use such case studies and multi-agent systems to evaluate \EVE's {\em performance}, rather than to solve problems of particular relevance in the AI or multi-agent systems literatures. Nevertheless, one could easily see that the case studies are based on systems that one can imagine to be found in many AI systems nowadays. 

\subsection{Case Studies}\label{sec:case_studies}
In this section, we present two examples from the literature of
concurrent and distributed systems to illustrate the practical usage
of \textsf{EVE}. Among other things, these two examples differ in the
way they are modelled as a concurrent game. While the first one is
played in an arena implicitly given by the specification of the
players in the game (as done in~\cite{GHW17-aij}), the second one is
played on a graph, {\em e.g.}, as done in~\cite{AlurHK02} with the use
of concurrent game structures. Both of these models of games
(modelling approaches) can be used within our tool. We will also use
these two examples to evaluate \textsf{EVE}'s practical performance
and compare it against \MCMAS and \PRALINE in
Section~\ref{sec:exp_conclusions}. Furthermore, since \textsf{PRALINE}
and \textsf{MCMAS} use different modelling languages -- ISPL in the
case of MCMAS -- we need to translate the examples modelled in SRML
into \PRALINE's input language and ISPL. Given the high-level nature
of SRML, the translation might introduce exponential blowup. However,
we argue that this is not a problem from the comparison point of view,
since the exponential blowup is also unavoidable when building Kripke
structures from SRML games.

\begin{figure}[t]
	\begin{minipage}{0.5\textwidth}
		\centering
		%		\vspace{-25pt}
		\includegraphics[scale=0.8]{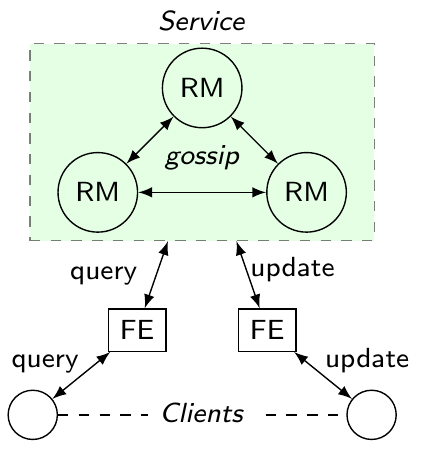}
%		\vspace{-10pt}
		\caption{Gossip framework structure.}
		\label{fig:gossip_arch}
		%		\vspace{-15pt}
	\end{minipage}
	\begin{minipage}{0.5\textwidth}
		%		\vspace{-25pt}
		\includegraphics[scale=0.8]{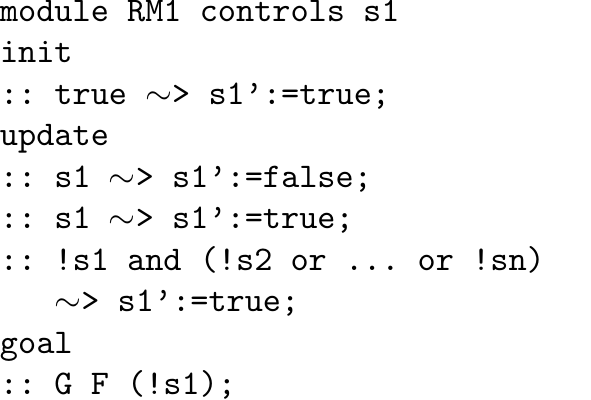}
		%		\vspace{-10pt}
		\caption{SRML machine readable code for module~$ \text{RM}_1 $ as written in \textsf{EVE}'s input code.}
		\label{fig:server_srml}
		%		\vspace{-5pt}
	\end{minipage}\hspace{0.2cm}
%	\vspace{-10pt}
\end{figure}

%\subsection*{Gossip Protocol} 
\paragraph{Gossip protocols} These are a class of networking and communication protocols that mimic the way social networks disseminate information. They have been used to solve problems in many large-scale distributed systems, such as \textit{peer-to-peer} and \textit{cloud} computing systems. Ladin \textit{et al.}~\cite{Ladin:1992:PHA:138873.138877} developed a framework to provide high availability services via replication which is based on the gossip approach first introduced in~\cite{Fischer:1982:SSA:588111.588124,Wuu:1984:ESR:800222.806750}. The main feature of this framework is the use of \textit{replica managers} (RMs) which exchange ``gossip" messages periodically in order to keep the data updated. The architecture of such an approach is shown in Figure~\ref{fig:gossip_arch}. 

We can model each RM as a module in SRML as follows: (1) When in \textit{servicing mode}, an RM can choose either to keep in servicing mode or to switch to gossiping mode; (2) If it is in gossiping mode and there is at least another RM also in gossiping mode\footnote{The core of the protocol involves (at least) pairwise interactions periodically.}, since the information during gossip exchange is of (small) bounded size, it goes back to servicing mode in the subsequent step. We then set the goal of each RM to be able to gossip infinitely often. As shown in Figure~\ref{fig:server_srml}, the module \texttt{RM1} controls a variable: \texttt{s1}. Its value being true signifies that \texttt{RM1} is in servicing mode; otherwise, it is in gossiping mode. Behaviour (1) is reflected in the first and second update commands, while behaviour (2) is reflected in the third update command. The goal of \texttt{RM1} is specified with the \LTL formula \textbf{GF} $ \lnot $ \hspace{-2pt} \texttt{s1}, which expresses that \texttt{RM1}'s goal is to gossip infinitely often: ``always'' (\textbf{G}) ``eventually'' (\textbf{F}) gossip ($ \lnot $\hspace{-2pt} \texttt{s1}).

Observe that with all RMs rationally pursuing their goals, they will adopt any strategy which induces a run where each RM can gossip (with at least one other RM) infinitely often. In fact, this kind of game-like modelling gives rise to a powerful characteristic: on \textit{all} runs that are sustained by a Nash equilibrium, the distributed system is guaranteed to have two crucial \textit{non-starvation}/\textit{liveness} properties: RMs can gossip infinitely often and clients can be served infinitely often. Indeed, these properties are verified in the experiments; with \textsc{E-Nash}: no Nash equilibrium sustains ``all RMs forever gossiping"; and with \textsc{A-Nash}: in all Nash equilibria at least one of the RM is in servicing mode infinitely often. We also notice that each RM is modelled as a non-deterministic open system: non-determinism is used in the first two updated commands, as they have the same guard \texttt{s1} and therefore will be both enabled at the same time; and the system is open since each module's state space and choices depend on the states of other modules, as reflected by the third updated command. 

\begin{figure}[t]
%	\vspace{-1cm}
	\centering
	\includegraphics{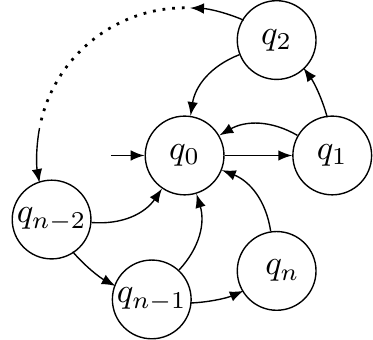}
	\caption{Gifford's protocol modelled as a game.}
	\label{fig:giff}
%	\vspace{-0.5cm}
\end{figure}

\paragraph{Replica Control Protocol} Consensus is a key issue in distributed computing and multi-agent systems. An important application domain is in maintaining data consistency. Gifford~\cite{Gifford:1979:WVR:800215.806583} proposed a quorum-based voting protocol to ensure data consistency by not allowing more than one processes to read/write a data item concurrently. To do this, each copy of a replicated item is assigned a vote.

We can model a (modified version of) Gifford's protocol as a game as follows. The set of players $ \text{N}=\{1,\dots,n\} $ in the game is arranged in a request queue represented by the sequence of states $ q_1,\dots,q_n $, where $ q_i $ means that player \textit{i} is requesting to read/write the data item. At state~$ q_i $, other players in $ \text{N}\backslash \{i\} $ then can vote whether to allow player~\textit{i} to read/write. If the majority of players in \text{N} vote ``yes", then the transition goes to $ q_0 $, \textit{i.e.}, player~\textit{i} is allowed to read/write, and otherwise it goes to $ q_{i+1} $\footnote{We assume arithmetic modulo $ (|\text{N}|+1) $ in this example.}. The voting process then restarts from $ q_1 $. The protocol's structure is shown in Figure~\ref{fig:giff}. Notice that at the last state, $ q_n $, there is only one outgoing arrow to $ q_0 $.
As in the previous example, the goal of each player~$i$ is to visit $ q_0 $ right after $ q_i $ infinitely often, so that the desired behaviour of the system is sustained on all Nash equilibria of the system: a data item is not concurrently accessed by two different processes and the data is updated in \textit{every} round. The associated temporal properties are automatically verified in the experiments in Section~\ref{sec:exp_conclusions}. Specifically, the temporal properties we check are as follows. With \textsc{E-Nash}: there is no Nash equlibrium in which the data is never updated; and, with \textsc{A-Nash}: on all Nash equilibria, for each player, its request will be granted infinitely often. Also, in this example, we define a module, called ``Environment", which is used to represent the underlying concurrent game structure, shown in Figure~\ref{fig:giff}, where the game is played. 

\subsection{Experiment I}\label{sec:exp_conclusions}
\setlength{\tabcolsep}{2pt}

\begin{table}[!t]
%	\vspace{-10pt}
	\centering
	\caption{Gossip Protocol experiment results.}
	\label{tab:experiment_gossip}
	\ra{1.1}
	\begin{tabular}{@{}rrrrrrrrrrrrrrr@{}}
		\toprule
		\multirow{2}{*}{P} & \multirow{2}{*}{S} & \multirow{2}{*}{E} & & \multicolumn{3}{c}{\textbf{\EVE}} & & \multicolumn{3}{c}{\textbf{\PRALINE}} &  & \multicolumn{3}{c}{\textbf{\MCMAS}} \\ \cmidrule{5-7} \cmidrule{9-11} \cmidrule{13-15}
		&&&& {$ \nu $ (s)} & {$ \epsilon $ (s)}	& $ \alpha $ (s) && {$ \nu $ (s)} & $ \epsilon $ (s) & $ \alpha $ (s) && $ {\nu} $ (s) & $ {\epsilon} $ (s) &  $ {\alpha} $ (s) \\
		%		&&&&&&&&&&&&&&\\
		\midrule
		2	& 4		& 9		&&0.02  			&0.24 			&0.08 			&&0.02		&1.71 	&1.73	&& \textbf{0.01}& \textbf{0.01}	&\textbf{0.01}\\
		3	& 8		& 27	&&0.09 				&0.43 			&0.26 			&&0.33		&26.74 	&27.85	&& \textbf{0.02}& \textbf{0.06}	&\textbf{0.06}\\
		4	& 16	& 81	&&\textbf{0.42} 	&\textbf{3.51} 	&\textbf{1.41} 	&&0.76		&547.97 &548.82	&& 760.65		& 3257.56		&3272.57\\
		5	& 32	& 243	&&\textbf{2.30} 	&\textbf{35.80} &\textbf{25.77} &&10.06		&TO 	&TO 	&& TO			& TO			&TO		\\
		6	& 64	& 729	&&\textbf{16.63} 	&\textbf{633.68}&\textbf{336.42}&&255.02	&TO 	&TO 	&& TO			& TO			&TO		\\
		7	& 128	& 2187	&&\textbf{203.05} 	&TO 			&TO				&&5156.48	&TO		&TO 	&& TO			& TO			&TO		\\
		8	& 256	& 6561	&&\textbf{4697.49} 	&TO 			&TO				&&TO		&TO 	&TO 	&& TO			& TO			&TO		\\
		\bottomrule
	\end{tabular}
	\centering
	\caption{Replica control experiment results.}
	\label{tab:experiment_replica}
	\ra{1.1}
	\begin{tabular}{@{}rrrrrrrrrrrrrrr@{}}
		\toprule
		\multirow{2}{*}{P} & \multirow{2}{*}{S} & \multirow{2}{*}{E} & & \multicolumn{3}{c}{\textbf{\EVE}} & & \multicolumn{3}{c}{\textbf{\PRALINE}} &  & \multicolumn{3}{c}{\textbf{\MCMAS}} \\ \cmidrule{5-7} \cmidrule{9-11} \cmidrule{13-15}
		&&&& {$ \nu $ (s)} & {$ \epsilon $ (s)}	& $ \alpha $ (s) && {$ \nu $ (s)} & $ \epsilon $ (s) & $ \alpha $ (s) && $ {\nu} $ (s) & $ {\epsilon} $ (s) &  $ {\alpha} $ (s) \\
		%		&&&&&&&&&&&&&&\\
		\midrule
		2& 3	& 8		&&0.04  			&0.11 			&0.10 			&&0.05		&0.64 		&0.74 		&&\textbf{0.01}	&\textbf{0.01}	&\textbf{0.02}	\\
		3& 4	& 20	&&0.11 				&1.53 			&0.22 			&&0.12		&4.96 		&5.46 		&&\textbf{0.02}	&\textbf{0.06}	&\textbf{0.11}	\\
		4& 5	& 48	&&\textbf{0.34} 	&\textbf{1.73} 	&\textbf{0.68} 	&&0.56		&65.50		&67.45 		&&1.99 			&4.15			&11.28	\\
		5& 6	& 112	&&\textbf{1.43} 	&\textbf{2.66} 	&\textbf{2.91} 	&&6.86		&1546.90 	&1554.80	&&1728.73 		&6590.53 		&TO		 \\
		6& 7	& 256	&&\textbf{5.87} 	&\textbf{13.69} &\textbf{16.03}	&&94.39		&TO			&TO 		&&TO 			&TO				&TO		\\
		7& 8	& 576	&&\textbf{32.84} 	&\textbf{76.50} &\textbf{102.12}&&2159.88	&TO			&TO 		&&TO 			&TO				&TO		\\
		8& 9	& 1280	&&\textbf{166.60} 	&\textbf{485.99}&\textbf{746.55}&&TO		&TO			&TO 		&&TO 			&TO				&TO		\\

		\bottomrule
	\end{tabular}
%	\vspace{-10pt}
\end{table}

In order to evaluate the practical performance of our tool and approach (against \textsf{MCMAS} and \textsf{PRALINE}), we present results on the temporal equilibrium analysis for the examples in Section~\ref{sec:case_studies}. We ran the tools on the two examples with different numbers of players (``P"), states (``S"), and edges (``E"). The experiments were obtained on a PC with Intel i5-4690S CPU 3.20 GHz machine with 8 GB of RAM running Linux kernel version 4.12.14-300.fc26.x86\_64. We report the running time\footnote{To carry out a fairer comparison (since \textsf{PRALINE} does not accept \LTL goals), we added to \textsf{PRALINE}'s running time the time needed to convert \LTL games into its input.} for solving \textsc{Non-Emptiness} (``$ \nu $"), \textsc{E-Nash} (``$ \epsilon $"), and \textsc{A-Nash} (``$ \alpha $"). For the last two problems, since there is no direct support in \textsf{PRALINE} and \textsf{MCMAS}, we used the reduction of \textsc{E/A-Nash} to \textsc{Non-Emptiness} presented in \cite{Gao0W17}. Intuitively, the reduction is as follows: given a game \textit{G} and formula $ \phi $, we construct a new game \textit{H} with two additional agents, say $ n+1 $ and $ n+2 $, with goals $ \gamma_{n+1} = \phi \vee (p \leftrightarrow q) $ and $ \gamma_{n+2} = \phi \vee \lnot (p \leftrightarrow q) $, where $ \Phi_{n+1} = \{ p \} $ and $ \Phi_{n+2} = \{ q \} $, $ p $ and $ q $ are fresh Boolean variables. This means that it is the case $ \NE(H) \neq \emptyset $ if and only if there exists a Nash equilibrium run in $ G $ satisfying $ \phi $.

From the experiment results shown in Table \ref{tab:experiment_gossip} and \ref{tab:experiment_replica}, we observe that, in general, \textsf{EVE} has the best performance, followed by \textsf{PRALINE} and \textsf{MCMAS}. Although \textsf{PRALINE} performed better than \textsf{MCMAS}, both struggled (timed-out\footnote{Time-out was fixed to be 7200 seconds.}) with inputs with more than 100 edges, while \textsf{EVE} could handle up to 6000 edges (for \textsc{Non-Emptiness}).

\subsection{Experiment II}\label{sec:bisim_exp}

\begin{figure}[t]
	%	\vspace{-1cm}
	\[
	\scalebox{1}{
		\begin{tikzpicture}[scale=1]
		\arraycolsep=1.4pt\def\arraystretch{0.5}
		\tikzstyle{every ellipse node}=[draw,inner xsep=3.5em,inner ysep=1.2em,fill=black!15!white,draw=black!15!white]
		\tikzstyle{every circle node}=[fill=white,draw,minimum size=1.6em,inner sep=0pt]
		
		% \draw[use as bounding box,draw=black] (-2,-2) rectangle (6.5,6);	  
		
		\draw(0,0) node(0){}
		++ 	( 0:.9)	  node[label=-90:{\footnotesize$\bar p\bar q$},circle](v0){$s_0$}	
		++	(  40:4)  node[label=-90:{\footnotesize$\bar p\bar q$},circle](v1){$s_1$}
%		++	(  0:4)  node[label=-90:{\footnotesize$\bar p\bar q$},circle](v11){$s_1'$}
		+( 0:5)  node[label=-90:{\footnotesize$     p\bar q$},circle](v2){$s_2$}
		+(-27:5.7)  node[label=-90:{\footnotesize$\bar p     q$},circle](v3){$s_3$}
		;
		\draw(0,.9) ++(-30:10)	node[label=-90:{\footnotesize$\bar p\bar q$},circle](v4){$s_4$};
		
		\draw(0,0) ++(0:4)	node[label=-90:{\footnotesize$\bar p\bar q$},circle](v11){$s_1'$};
		
		\draw[-latex] (v0) --node[pos=.5,fill=white](){\footnotesize$\begin{array}{l}b,a,a\\b,a,a'\\ a,b,b \\ a,b,b' \end{array}$} (v1); 
		
		\draw[-latex] (v1) --node[pos=.75,fill=white](){\footnotesize$\begin{array}{l}b,\ast,a \\ a,\ast,b \end{array}$} (v2);
		
		\draw[-latex] (v1) --node[pos=.74,fill=white](){\footnotesize$\begin{array}{l}\ast,b,a' \\ \ast, a, b'\end{array}$} (v3);
		
		\draw[-latex] (v1) --node[pos=.6,fill=white](){\footnotesize$\begin{array}{l}a,\ast,a\\ b,\ast,b \\ \ast,a,a' \\ \ast,b,b'\end{array}$} (v4);
		
		\draw[-latex] (v0) to[bend right=30] node[pos=.5,fill=white](){\footnotesize$\begin{array}{l}a,a,\ast \\ b,b,\ast \end{array}$} (v4);
		
		\draw[-latex] (v0) -- node[pos=.49,fill=white](){\footnotesize$\begin{array}{l}a,b,a \\ a,b,a' \\ b,a,b \\ b,a,b' \end{array}$} (v11);
		
		\draw[-latex] (v11) -- node[pos=.74,fill=white](){\footnotesize$\begin{array}{l}b,\ast,a \\ a,\ast,b \end{array}$} (v2);
		
		\draw[-latex] (v11) -- node[pos=.74,fill=white](){\footnotesize$\begin{array}{l} \ast,b,a' \\ \ast,a,b' \end{array}$} (v3);
		
		\draw[-latex] (v11) to[bend right=20] node[pos=.4,fill=white](){\footnotesize$\begin{array}{l}a,\ast,a \\ b,\ast,b \\ \ast,a,a' \\ \ast,b,b' \end{array}$} (v4);
		
		\draw[-latex] (0) -- (v0);
		
		% loops
		\draw[-latex] (v2.70-90) .. controls +(60-90:4em) and +(120-90:4em) .. node[pos=.5,fill=white,right,xshift=.5ex](){\footnotesize$\ast,\ast,\ast$} (v2.110-90);
		
		\draw[-latex] (v3.70-90) .. controls +(60-90:4em) and +(120-90:4em) .. node[pos=.5,fill=white,right,xshift=.5ex](){\footnotesize$\ast,\ast,\ast$} (v3.110-90);
		
		\draw[-latex] (v4.70-90) .. controls +(60-90:4em) and +(120-90:4em) .. node[pos=.5,fill=white,right,xshift=.5ex](){\footnotesize$\ast,\ast,\ast$} (v4.110-90);

%		 \draw[-latex] (v2.70) .. controls +(60:4em) and +(120:4em) .. node[pos=.5,yshift=-.1em,above](){$y$} (v2.110);	
		% 
%		 \draw[-latex] (v0) -- node[pos=.45,yshift=0em,above](){$y$}  (v2);
		% \draw[-latex] (v1) to[bend left=45] node[pos=.5,yshift=.1em,below](){$\bar y$} (v0.-30);
		% 
		% \draw[-latex] (v1) to[bend right=30] node[pos=.5,above](){$y$} (v2);
		% \draw[-latex] (v2) to[bend right=30] node[pos=.5,above](){$\bar y$} (v1);
		\end{tikzpicture}
	}
	\]
	\caption{A 3-player game with Nash equilibrium.}
	\label{fig:bisim1}
	%	\vspace{-0.5cm}
\end{figure}
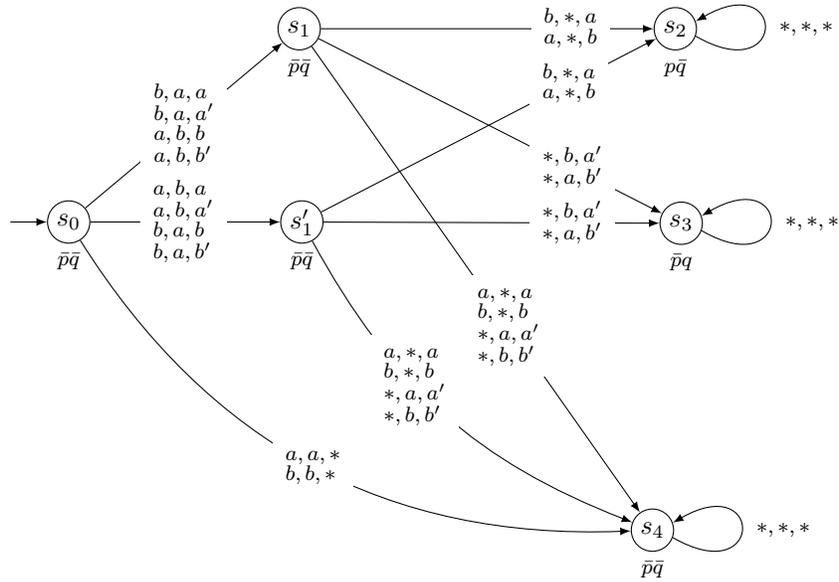

This experiment is taken from the motivating examples in~\cite{GHPW17}. Suppose the systems shown in Figure \ref{fig:bisim1} and \ref{fig:bisim2} represents a 3-player game, where each transition is labelled by the actions $ x, y, z $ of player 1, 2, and 3, respectively, an asterisk $ \ast $ being a wildcard. The goals of the players can be represented by the \LTL formulae $ \gamma_1 = \sometime p, \gamma_2 = \sometime q, \text{and}~\gamma_3 = \always \lnot (p \vee q) $. The system in Figure \ref{fig:bisim1} has a Nash equilibrium, whereas no (non-bisimulation-invariant strategies) Nash equilibria exists in the (bisimilar) system in Figure \ref{fig:bisim2}.

In this experiment, we extended the number of states by adding more layers to the game structures used there in order to test the practical performance of \EVE, \MCMAS, and \PRALINE. The experiments were performed on a PC with Intel i7-4702MQ CPU 2.20GHz machine with 12GB of RAM running Linux kernel version 4.14.16-300.fc26.x86\_64. We divided the test cases based on the number of Kripke states and edges; then, for each case, we report (i) the total running time\footnote{Similarly to Experiment I (Section \ref{sec:exp_conclusions}), we added to \textsf{PRALINE}'s running time the time needed to convert \LTL games into its input to carry out a fairer comparison.} (``time'') and (ii) whether the tools find any Nash equilibria (``NE'').
%, and (iii) \textit{discounted execution time} (``disc.\ time''). Discounted execution time is the amount of time used after $ \mathcal{G}_{\text{PAR}} $ has been built until the tool terminates and outputs the result. This is to enable a comparison between \EVE and \PRALINE, since the latter only accepts B{\"u}chi goals (while \EVE accepts \LTL goals). 

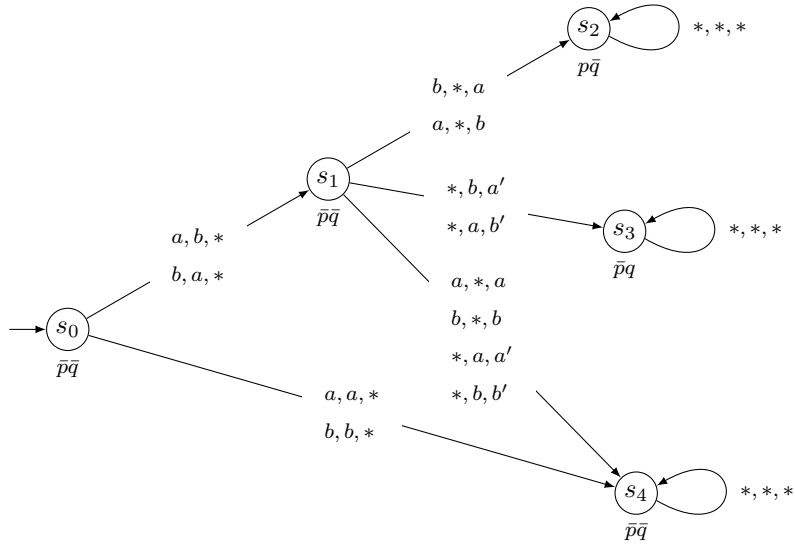
\begin{figure}[t]
	%	\vspace{-1cm}
	\[
	\scalebox{1}{
		\begin{tikzpicture}[scale=1]
		
		\tikzstyle{every ellipse node}=[draw,inner xsep=3.5em,inner ysep=1.2em,fill=black!15!white,draw=black!15!white]
		\tikzstyle{every circle node}=[fill=white,draw,minimum size=1.6em,inner sep=0pt]
		
		% \draw[use as bounding box,draw=black] (-2,-2) rectangle (6.5,6);	  
		
		\draw(0,0) node(0){}
		++ 	( 0:.9)	  node[label=-90:{\footnotesize$\bar p\bar q$},circle](v0){$s_0$}	
		++	(  30:4)  node[label=-90:{\footnotesize$\bar p\bar q$},circle](v1){$s_1$}
		+( 30:4)  node[label=-90:{\footnotesize$     p\bar q$},circle](v2){$s_2$}
		+(-10:4)  node[label=-90:{\footnotesize$\bar p     q$},circle](v3){$s_3$}
		;
		\draw(0,.9) ++(-20:9)	node[label=-90:{\footnotesize$\bar p\bar q$},circle](v4){$s_4$};
		
		\draw[-latex] (v0) --node[pos=.5,fill=white](){\footnotesize$\begin{array}{l}a,b,\ast\\b,a,\ast \end{array}$} (v1); 
		
		\draw[-latex] (v1) --node[pos=.5,fill=white](){\footnotesize$\begin{array}{l}b,\ast,a\\a,\ast,b\end{array}$} (v2);
		
		\draw[-latex] (v1) --node[pos=.5,fill=white](){\footnotesize$\begin{array}{l}\ast,b,a'\\ \ast,a,b'\end{array}$} (v3);
		
		\draw[-latex] (v1) --node[pos=.5,fill=white](){\footnotesize$\begin{array}{l}a,\ast,a\\ b,\ast,b \\ \ast,a,a' \\ \ast,b,b'\end{array}$} (v4);
		
		\draw[-latex] (v0) --node[pos=.5,fill=white](){\footnotesize$\begin{array}{l}a,a,\ast \\ b,b,\ast \end{array}$} (v4);
		
		\draw[-latex] (0) -- (v0);
		
		% loops
		\draw[-latex] (v2.70-90) .. controls +(60-90:4em) and +(120-90:4em) .. node[pos=.5,fill=white,right,xshift=.5ex](){\footnotesize$\ast,\ast,\ast$} (v2.110-90);
		
		\draw[-latex] (v3.70-90) .. controls +(60-90:4em) and +(120-90:4em) .. node[pos=.5,fill=white,right,xshift=.5ex](){\footnotesize$\ast,\ast,\ast$} (v3.110-90);
		
		\draw[-latex] (v4.70-90) .. controls +(60-90:4em) and +(120-90:4em) .. node[pos=.5,fill=white,right,xshift=.5ex](){\footnotesize$\ast,\ast,\ast$} (v4.110-90);

		% \draw[-latex] (v2.70) .. controls +(60:4em) and +(120:4em) .. node[pos=.5,yshift=-.1em,above](){$y$} (v2.110);	
		% 
		% \draw[-latex] (v0) -- node[pos=.45,yshift=0em,above](){$y$}  (v2);
		% \draw[-latex] (v1) to[bend left=45] node[pos=.5,yshift=.1em,below](){$\bar y$} (v0.-30);
		% 
		% \draw[-latex] (v1) to[bend right=30] node[pos=.5,above](){$y$} (v2);
		% \draw[-latex] (v2) to[bend right=30] node[pos=.5,above](){$\bar y$} (v1);
		\end{tikzpicture}
	}
	\]
	\caption{A 3-player game without (non-bisimulation-invariant strategies) Nash equilibria.}
	\label{fig:bisim2}
	%	\vspace{-0.5cm}
\end{figure}

\begin{table}[!t]
	\centering
	\caption{Example with no Nash equilibrium.}
	\label{tab:none}
	\ra{1.1}
	\begin{tabular}{@{}rrcrcrcrcrc@{}}
		\toprule
		\multirow{2}{*}{states} & \multirow{2}{*}{edges} & & \multicolumn{2}{c}{\textbf{\MCMAS}} & & \multicolumn{2}{c}{\textbf{\EVE}} &  & \multicolumn{2}{c}{\textbf{\PRALINE}} \\ \cmidrule{4-5} \cmidrule{7-8} \cmidrule{10-11}
		&		&& \multirow{1}{*}{time (s)} &  \multirow{1}{*}{NE}	&& \multirow{1}{*}{time (s)} & \multirow{1}{*}{NE}	&& \multirow{1}{*}{\shortstack{time (s)}} &  \multirow{1}{*}{NE} \\
		\midrule
		5	& 80	&& \textbf{0.04}& No	&& 0.75				& Yes		&& 0.77 	& No\\
		8	& 128	&& \textbf{0.24}& No	&& 2.99				& Yes		&& 2.06 	& No\\
		11	& 176	&& 6.28			& No	&& \textbf{3.86}			& Yes		&& 4.42 	& No\\
		14	& 224	&& 273.14		& No	&& \textbf{7.46}	& Yes		&& 8.53 	& No\\
		17	& 272	&& TO			& --	&& \textbf{13.31}	& Yes		&& 15.33 	& No\\
		\vdots	& \vdots&& \vdots		& \vdots&& \vdots 		& \vdots 	&& \vdots	& \vdots\\
		50	& 800	&& TO			& --	&& \textbf{655.80}	& Yes 		&& 789.77 	& No\\
		\bottomrule
	\end{tabular}
	\centering
	\caption{Example with Nash equilibria}
	\label{tab:ne}
	\ra{1.1}
	\begin{tabular}{@{}rrcrcrcrcrc@{}}
		\toprule
		\multirow{2}{*}{states} & \multirow{2}{*}{edges} & & \multicolumn{2}{c}{\textbf{\MCMAS}} & & \multicolumn{2}{c}{\textbf{\EVE}} &  & \multicolumn{2}{c}{\textbf{\PRALINE}} \\ \cmidrule{4-5} \cmidrule{7-8} \cmidrule{10-11}
		&		&& \multirow{1}{*}{time (s)} &  \multirow{1}{*}{NE}	&& \multirow{1}{*}{time (s)} & \multirow{1}{*}{NE}	&& \multirow{1}{*}{\shortstack{time (s)}} &  \multirow{1}{*}{NE} \\
		\midrule
		6	& 96	&& \textbf{0.02}	& Yes	&& 1.09				& Yes	&& 1.19 	& Yes\\
		9	& 144	&& \textbf{0.77}	& Yes	&& 3.36				& Yes	&& 3.76 	& Yes\\
		12	& 192	&& 65.31			& No	&& \textbf{7.45}	& Yes	&& 8.89 	& Yes\\
		15	& 240	&& TO				& --	&& \textbf{15.52}	& Yes	&& 17.72 	& Yes\\
		18	& 288	&& TO				& --	&& \textbf{30.06}	& Yes	&& 30.53	& Yes\\
		\vdots	& \vdots&& \vdots			& \vdots&& \vdots 		& \vdots&& \vdots	& \vdots\\
		51	& 816	&& TO				& --	&& \textbf{1314.47}	& Yes 	&& 1563.79 	& Yes\\
		\bottomrule
	\end{tabular}
\end{table}

Table~\ref{tab:none} shows the results of the experiments on the example in which the model of strategies that depends only on the run (sequence of states) of the game (run-based strategies~\cite{GHPW17}) cannot sustain any Nash equilibria, a model of strategies that is not invariant under bisimilarity. Indeed, since \MCMAS and \PRALINE use this model of strategies, both did not find any Nash equilibria in the game, as shown in Table~\ref{tab:none}. \EVE, which uses a model of strategies that not only depends on the run of the game but also on the actions of players (computation-based~\cite{GHPW17}), found a Nash equilibrium in the game. We can also see that \EVE outperformed \MCMAS on games with 14 or more states. In fact, \MCMAS timed-out\footnote{We fixed the time-out value to be 3600 seconds (1 hour).} on games with 17 states or more, while \EVE kept working efficiently for games of bigger size. We can also observe that \PRALINE performed almost as efficiently as \EVE in this experiment, although \EVE performed better in both small and large instances of these games. 

In Table~\ref{tab:ne}, we used the example in which Nash equilibria is sustained in run-based strategies. As shown in the table, \MCMAS found Nash equilibria in games with 6 and 9 states. However, since \MCMAS uses imperfect recall, when the third layer was added (case with 12 states in Table~\ref{tab:ne}) to the game, it could not find any Nash equilibria. Regarding running times, \EVE outperformed \MCMAS from the game with 12 states and beyond, where \MCMAS timed-out on games with 15 or more states. As for \PRALINE, it performed comparably to \EVE in this experiment, but again, \EVE performed better in all instances. 

\subsection{Experiment III}\label{sec:exp_grid}
%In this experiment, we have two agents inhabiting a grid world with dimensions $ n \times n $. Initially, the agents are located at opposing corners of the grid; specifically, agent 1 is located at the  top-left corner (coordinate $ (0,0) $) and agent 2 at the bottom-right corner $ (n-1,n-1) $. The agents are each able to move around the grid in directions \textit{north}, \textit{south}, \textit{east}, and \textit{west}. The goal of each agent is to reach the opposite corner, that is, agent 1's goal is to reach position~$ (n-1,n-1) $, and agent 2's goal is to reach position~$ (0,0) $. A number of obstacles are also placed (uniformly) randomly on the grid. The agents are not allowed to move into a coordinate occupied by an obstacle, the other agent, or outside the grid world. We used a binary encoding to represent the spatial information of the grid world which includes the grid coordinates, as well as the obstacles and the agents locations.

This experiment is based on the example previously presented in
Section \ref{secn:prelim}. For this particular experiment, we assume
that initially the agents are located at opposing corners of the grid;
specifically, agent 1 is located at the  top-left corner (coordinate $
(0,0) $) and agent 2 at the bottom-right corner $ (n-1,n-1) $. A
number of obstacles are also placed (uniformly) randomly on the
grid. We use a binary encoding to represent the spatial information of
the grid world which includes the grid coordinates, as well as the
obstacles and the agents locations. For instance, to encode a position
of an agent 1 in $ 4 \times 4 $ grid, we need 4 Boolean variables
arranged as a tuple $pos_1 = \tuple{ x_0^1,x_1^1,y_0^1,y_1^1 } $. An
instance of such a tuple $ pos_1 = \tuple{0,1,1,0} $ means that agent
$ 1 $ is at $ (2,1) $. For each time step and $ i \in \{1,2\} $, the
update guarded command set $ U_i $ is such a way that agent $ i $ can
only move horizontally and vertically, 1 step at a time. Furthermore,
the commands in $ U_i $ respect the legality of movement,
\textit{i.e.}, agent $ i $ cannot move out of bound or into an
obstacle. The goal of each agent can be expressed by the \LTL formulae
\[\gamma_1 = \sometime ( \bigwedge_{i \in \{0,\dots,n-1\}} x_i^1
  \wedge \bigwedge_{i \in \{0,\dots,n-1\}} y_i^1 ) \] and \[ \gamma_2
  = \sometime ( \bigwedge_{i \in \{0,\dots,n-1\}} \lnot x_i^2 \wedge
  \bigwedge_{i \in \{0,\dots,n-1\}} \lnot y_i^2 ).\] A safety
specification (no more than one agent occupying the same position at
the same time) can be expressed by the following \LTL formula: \[\phi = \always \lnot ( \bigwedge_{i \in \{0,\dots,n-1\}} (x_i^1 \leftrightarrow x_i^2) \wedge \bigwedge_{i \in \{0,\dots,n-1\}} (y_i^1 \leftrightarrow y_i^2) ).\]

%\GridEx

%To make it clearer, consider the example shown in Figure~\ref{fig:GridEx}; a (grey) filled square depicts an obstacle. Agent 1, depicted by $ \blacksquare $, can move north to $ (2,0) $, south to $ (2,2) $, east to $ (3,1) $, and west to $ (1,1) $. Whereas agent 2, depicted by $ \bigcirc $, can only move north to $ (0,1) $ and south to $ (0,3) $ (she cannot move west because it is outside the world, nor east because there is an obstacle.)

%In this experiment we make the following assumptions: (1) at each timestep, each agent has to make a move, that is, she cannot stay at the same position for two consecutive timesteps, and she can only move at most one step; (2) the goal of each agent is, as stated previously, to eventually reach the opposite corner of her initial position. Furthermore, for \textsc{E-Nash}, the property to be checked is ``two agents never occupying the same coordinate at the same time", in other words, two agents never crash into each other.

\begin{table}[!t]
	\flushleft
	\caption{Grid world experiment results.}
	\label{tab:grid}
	\setlength{\tabcolsep}{9pt}
	\ra{1.1}
	\begin{tabular}{@{}rcrrrr@{}}
		\toprule
		{Size} && \# Obs & \multicolumn{1}{c}{KS} & \multicolumn{1}{c}{KE} & \multicolumn{1}{c}{GS} \\
		%		& & & & & & &	& & &  \\
		\midrule
		3	&& 3	& $ 15(13,18) $		& $ 44(32,72) $			& $ 60(53,73) $	\\
		4	&& 6	& $ 40(32,52) $		& $ 150(98,200) $		& $ 156(121,209) $\\
		5	&& 10	& $ 94(61,125) $	& $ 398(242,512) $		& $ 376(453,741) $ \\
		6	&& 15	& $ 155(113,185) $	& $ 655(450,800) $		& $ 619(453,741) $ \\
		7	&& 21	& $ 228(181,290) $	& $ 994(800,1250) $		& $ 909(725,1161) $	\\
		8	&& 28	& $ 491(394,666) $	& $ 2297(1922,2888) $	& $ 1963(1577,2665) $	\\
		9	&& 36	& $ 564(269,765) $	& $ 2687(1352,3698) $	& $ 2256(1077,3061) $	\\
		10	&& 45	& $ 916(730,1258) $	& $ 4780(3528,6498) $	& $ 3657(2921,5033) $	\\
		\bottomrule
	\end{tabular}
	
	\setlength{\tabcolsep}{5pt}
	\begin{tabular}{@{}rrrr@{}}
		\toprule
		{Size} & \multicolumn{1}{c}{GE} & \multicolumn{1}{c}{$ \nu $ (s)} & \multicolumn{1}{c}{$ \epsilon $ (s)} \\
		%		& & & & & & &	& & &  \\
		\midrule
		3	& $ 173(129,289) $ 		& $ 0.44(0.19,1.14) $		& $ 1.21(0.5,2.63) $ 	\\
		4	& $ 595(379,801) $ 		& $ 0.98(0.63,1.16) $ 		& $ 1.57(1.01,2.24) $\\
		5	& $ 1591(969,2049) $ 	& $ 4.73(2.62,6.22) $ 		& $ 22.51(18.22,26.25) $\\
		6	& $ 2622(1801,3201) $ 	& $ 9.53(7.13,11.49) $		& $ 32.32(26.05,37.35) $\\
		7	& $ 3969(3161,5001) $ 	& $ 17.69(13.81,21.58) $	& $ 48.90(39.70,59.50) $\\
		8	& $ 9190(7689,11553) $ 	& $ 50.91(38.38,72.49) $	& $ 121.33(95.03,167.25) $\\
		9	& $ 10748(5409,14793) $	& $ 100.94(45.81,137.91) $	& $ 6002.80(5477.63,6374.26) $\\
		10	& $ 19102(14113,25993) $& $ 211.30(152.74,311.43) $	& $ 6871.16(6340.64,7650.87) $\\
		\bottomrule
	\end{tabular}
\end{table}

\GridPlot

The experiment was obtained on a PC with Intel i5-4690S CPU 3.20 GHz machine with 8 GB of RAM running Linux kernel version 4.12.14-300.fc26.x86\_64. We varied the size of the grid world (``size") from $ 3 \times 3 $ to $ 10 \times 10 $, each with a fixed number of obstacles (``\# Obs"), randomly distributed on the grid. We report the number of Kripke states (``KS"), Kripke edges (``KE"), $ \ParGame $ states (``GS"), $ \ParGame $ edges (``GE"), \textsc{Non-Emptiness} execution time (``$ \nu $"), and \textsc{E-Nash} execution time (``$ \epsilon $"). We ran the experiment for five replications, and report the average (\textit{ave}), minimum (\textit{min}), and maximum (\textit{max}) times from the replications. The results are reported in Table \ref{tab:grid}, with the following format: \textit{ave}(\textit{min},\,\textit{max}).

From the experiment results, we see that \EVE works well for \textsc{Non-Emptiness} up until size~10. From the plots in Figure \ref{fig:GridPlot}, we can clearly see that the values of each variable, except for $ \epsilon $, grow exponentially. For $ \epsilon $ (\textsc{E-Nash}), however, it seems to grow faster than the rest. Specifically, it is clearly visible in transitions between numbers that have different size of bit representation, {\em i.e.}, 4 to 5 and 8 to 9\footnote{Since the grid coordinate index starts at 0, the ``actual" transitions are 3 to 4 and 7 to 8.}. These jumps correspond to the time used to build deterministic parity automata on words from \LTL properties to be checked in \textsc{E-Nash}, which is essentially, bit-for-bit comparisons between the position of agent 1 and 2.

From the experiments shown in this section it is also clear that the bottleneck in the performance is the translation of \LTL\ goals and the high-level description of the game into the underlying parity game. Once an explicit parity game is constructed, then the performance improves radically. This result is perfectly consistent with what the theoretical complexity of the decision procedure predicts: our algorithm works in doubly-exponential time in the size of the goals of the players, while it is only singly-exponential in the size of the SRML specification. These two exponential-time reductions are in fact optimal, so there is no hope that they can be improved, at least in theory. On the other hand, the actual subroutine that finds a Nash equilibrium and computes players' strategies from the parity games representation of the problem is rather efficient in theory -- but still not known to be in polynomial time using the best algorithms to solve parity games. Then, it is clear that a natural way to make rational verification a feasible problem, in theory, is to look at cases where goals and/or game representations are simpler. Such study is conducted in \cite{GNPW19-ijcai}, where several positive results on the complexity of solving the rational verification problem are obtained.

%\newpage
\section{Concluding Remarks and Related Work}\label{secn:conc}

This paper contains a complete study, from theory to implementation, of the temporal equilibrium analysis of multi-agent AI systems formally modelled as multi-player games. The two main contributions of the paper are: (1) a novel and optimal decision procedure, based on the solution of parity games, that can be used to solve both the rational verification and the automated synthesis problems for multi-player games; and (2) a complete implementation of the general game-theoretic modelling and reasoning framework -- with full support of goals expressed as \LTL\ formulae and high-level game descriptions in SRML -- which is available online. Our work builds on several previous results in the computer science (synthesis and verification) and AI literatures (multi-agent systems). Relevant related literature will be discussed next. 

\paragraph{\bf Equilibrium Analysis in Multi-Agent Systems}
Rational verification was proposed as an complementary verification
methodology to conventional methods, such as model checking. A
legitimate question is, then, when is rational verification an
appropriate verification approach? A possible answer is given next.
The verification problem~\cite{CGP02}, as conventionally formulated,
is concerned with checking that some property, usually defined using a
modal or a temporal logic~\cite{Emerson90}, holds on some or on every
computation run of a system. In a game-theoretic setting, this can be
a very strong requirement -- and in some cases even inappropriate --
since only some computations of the system will arise (be sustained)
as the result of agents in the system choosing strategies in
equilibrium, that is, due to strategic and rational play. It was
precisely this concern that 
motivated the rational
  verification approach~\cite{GHW17-aij,WooldridgeGHMPT16}. In rational verification, we ask if a given
temporal property holds on some or every computation run that can be
sustained by agents choosing Nash equilibrium strategies. Rational
verification can be reduced to the \textsc{Non-Emptiness} problem, as
stated in this paper; cf.,~\cite{Gao0W17}. As a consequence, along
with the polynomial transformations in~\cite{Gao0W17}, our results
provide a complete framework (theory, algorithms, and implementation)
for automated temporal equilibrium analysis, specifically, to do
rational synthesis and formal verification of logic-based multi-agent
systems.  The framework, in particular, provides a concrete and
algorithmic solution to the rational synthesis problem as studied
in~\cite{FismanKL10}, where the Boolean case (iterated games where
players control Boolean variables, whose valuations define sequences
of states in the game, {\em i.e.}, the plays in the game) was given an
interesting automata-theoretic solution via (an extension of) Strategy
Logic~\cite{ChatterjeeHP10}.

\paragraph{\bf Automata and logic}
In computer science, a common technique to reason about Nash equilibria in multi-player games %(representing a concurrent and reactive system) 
is using alternating parity \emph{automata on infinite trees} (APTs~\cite{Loding12}). This approach is used to do rational synthesis~\cite{FismanKL10,KupfermanPV16}; equilibrium checking and rational verification~\cite{WooldridgeGHMPT16,GutierrezHW15,GHW17-aij}; and model checking of logics for strategic reasoning capable to specify the existence of a Nash equilibrium in concurrent game structures~\cite{AlurHK02}, both in two-player games~\cite{ChatterjeeHP10,FinkbeinerS10} and in multi-player games~\cite{LaroussinieM15,MogaveroMPV14}. In cases where players' goals are simpler than general \LTL formulae, \emph{e.g.}, for reachability or safety goals, alternating B{\"u}chi automata can be used instead~\cite{BouyerBMU15}. \emph{Our technique is different from all these automata-based approaches, and in some cases more general}, as it can be used to handle either a more complex model of strategies or a more complex type of goals, and delivers an immediate procedure to synthesise individual strategies for players in the game, while being amenable to implementation.

\paragraph{\bf Tools and algorithms}
In theory, the kind of equilibrium analysis that can be done using
\MCMAS~\cite{CermakLMM14,CermakLM15,CermakLMM18} and
\PRALINE~\cite{Brenguier13,BouyerBMU15} rely on the automata-based
approach. However, the algorithms that are actually implemented have a
different flavour. \MCMAS uses a procedure for SL which works as a
\emph{labelling algorithm} since it only considers memoryless
strategies~\cite{CermakLMM18}. On the other hand, \PRALINE, which
works for B{\"u}chi definable objectives, uses a procedure based on
the \emph{``suspect game''}~\cite{BouyerBMU15}. Despite some
similarities between our construction and the suspect game, introduced
in~\cite{BouyerBMU15}, the two procedures are substantially
different. Unlike our procedure, the suspect game is a standard
two-player zero-sum turn-based game $\mthname{H}(\Game, \pi)$,
constructed from a game $\Game$ and a possible path~$\pi$, in which
one of the players (``Eve'') has a winning strategy if, and only if,
$\pi$ can be sustained by a Nash equilibrium in~$\Game$. The overall
procedure in~\cite{BouyerBMU15} relies on the construction of such a
game, whose size (space complexity) is exponential in the number of
agents~\cite[Section 4.3]{BouyerBMU15}. Instead, our procedure solves,
independently, a collection of parity games that avoids an exponential
use of space but may require to be executed exponentially many
times. Key to the correctness of our approach is that we deal with
parity conditions, which are prefix-independent, ensuring that
punishment strategies do not depend on the history of the
game. Regarding similarities, our procedure also checks for the
existence of a path sustained by a Nash Equilibrium, but our algorithm
does this for every subset $W \subseteq \Ag$ of agents, if
needed. Doing this (\emph{i.e.}, trading exponential space for
exponential time), at every call of this subroutine, our algorithm
avoids building an exponentially sized game, like $\mthname{H}$. On
the other hand, from a practical point of view, avoiding the
construction of such an exponential sized game leads to better
performance (running times), even in cases where no Nash equilibrium
exists, when our subroutine is necessarily called exponentially many
times.
In addition to all of the above, neither the algorithm used for \MCMAS nor the one used for \PRALINE computes pure Nash equilibria in a bisimulation-invariant framework, as our procedure does.  
While \MCMAS and \PRALINE are the two closest tools to \EVE, 
they are not the only available options to reason about games. 
For instance, PRISM-games~\cite{KwiatkowskaPW16}, EAGLE~\cite{ToumiGW15}, 
and \textsf{UPPAAL}~\cite{DavidLLMP15} are other interesting tools 
to reason about games. PRISM-games allows one to do strategy synthesis
for turn-based stochastic games as well as model checking for
long-run, average, and ratio rewards properties. Only until very recently, 
PRISM-games had no support of equilibrium reasoning, but see~\cite{KNPS19}.
EAGLE is a tool specifically designed to
reason about pure Nash equilibria in multi-player games. EAGLE considers
games where goals are given as \CTL formulae and allows one to check if
a given strategy profile is a Nash equilibrium of a given multi-agent system. 
This decision problem, called \textsc{Membership} within the rational verification
framework~\cite{WooldridgeGHMPT16}, is, theoretically, simpler than
\textsc{Non-Emptiness}: while the former can be solved in EXPTIME (for
branching-time goals expressed using \CTL
formulae~\cite{GHW17-apal}), the latter is 2EXPTIME-complete
for \LTL goals, and even 2EXPTIME-hard for \CTL goals and
nondeterministic strategies~\cite{GHW17-apal}.
\textsf{UPPAAL} is another tool that can be used to analyse equilibrium behaviour in a system~\cite{DavidJLMT15,DBLP:journals/corr/abs-1202-4506}. However, \textsf{UPPAAL} differs from \textsf{EVE} in various critical ways: \emph{e.g.}, it works in a quantitative setting, uses statistical model checking, and most importantly, computes approximate Nash equilibria of a game. 

\paragraph{\bf The Role of Bisimilarity}
One crucial aspect of our approach to rational verification and synthesis is the role
of
\emph{bisimilarity}~\cite{Milner80,HM85,NicolaV95,GlabbeekW96}.
Bisimulation is the most important type of behavioural equivalence
relation considered in computer science, and in particular two
bisimilar systems will satisfy the same temporal logic properties.  In
our setting, it is highly desirable that properties which hold in
equilibrium are sustained across all bisimilar systems to
$P_1,\ldots,P_n$. That is, that for every (temporal logic) property
$\phi$ and every system component $P'_i$ modelled as an agent in a
multi-player game, if $P'_i$ is bisimilar to
$P_i\in\{P_1,\ldots,P_n\}$, then $\phi$ is satisfied in equilibrium -- that is, 
on a run induced by some Nash equilibrium of the game -- by 
$P_1,\ldots, P_i, \ldots P_n$ if and only if is also satisfied in
equilibrium by $P_1,\ldots, P'_i, \ldots,P_n$, the system in which $P_i$
is replaced by $P'_i$, that is, across all bisimilar systems
to~$P_1,\ldots,P_n$. This property is called \emph{invariance under
bisimilarity}. 
% and has been widely used for decades for the semantic
%analysis (\emph{e.g.}, for modular and compositional reasoning) and
%formal verification (\emph{e.g.}, for temporal logic model checking) of
%concurrent and distributed systems. 
Unfortunately, as shown
in~\cite{GutierrezHW15-concur,GHPW17}, the satisfaction of temporal
logic properties in equilibrium is not invariant under bisimilarity,
thus posing a challenge for the modular and compositional reasoning of
concurrent systems, since individual system components in a concurrent
system cannot be replaced by (behaviourally equivalent) bisimilar
ones, while preserving the temporal logic properties that the overall
multi-agent system satisfies in equilibrium. This is also a problem
from a synthesis point of view. Indeed, a strategy for a system
component $P_i$ may not be a valid strategy for a bisimilar system
component~$P'_i$. As a consequence, the problem of building strategies
for individual processes in the concurrent system
$P_1,\ldots, P_i, \ldots P_n$ may not, in general, be the same as
building strategies for a bisimilar system
$P_1,\ldots, P'_i, \ldots P_n$, again, deterring any hope of being
able to do modular reasoning on concurrent and multi-agent systems.
These problems were first identified in~\cite{GutierrezHW15-concur}
and further studied in~\cite{GHPW17}. However, no algorithmic
solutions to these two problems were presented in
either~\cite{GutierrezHW15-concur} or~\cite{GHPW17}. 
%Instead, the
%focus in \cite{GutierrezHW15-concur,GHPW17} was on classifying classes
%of games and investigating models of strategies where different kinds
%of properties (not necessarily temporal logic properties) could be
%preserved in equilibrium by bisimilarity.
%
Specifically, in this paper, bisimilarity was exploited in two ways. 
Firstly, our construction of punishment strategies (used in the characterisation of 
Nash equilibrium given by Theorem~\ref{thm:NEchar}) assumes that players 
have access to the history of choices that other players in the game have 
made. As shown in \cite{GHPW17,GHPW19}, with a model of strategies 
where this is not the case, the preservation of Nash equilibria in the game, 
as well as of temporal logic properties in equilibrium, may not be guaranteed. 
Secondly, our implementation in \EVE\ guarantees that any two games
whose underlying CGSs are bisimilar, and therefore should be regarded 
as observationally equivalent from a concurrency point of view,  
will produce the same answers to the 
rational verification and automated synthesis problems. 
It is also worth noting that even though bisimilarity is probably the most widely used behavioural equivalence in concurrency, in the context of multi-agent systems other relations may be preferred, for instance, equivalence relations that take a detailed account of the independent interactions and behaviour of individual components in a multi-agent system.
In such a setting, ``alternating'' relations with natural ATL$^*$ characterisations have been studied~\cite{AHKV98}.
Alternating bisimulation is very similar to bisimilarity on labelled transition systems~\cite{Milner80,HM85}, only that when defined on CGSs, instead of action profiles (directions) taken as possible transitions, one allows individual player's actions, which must be matched in the bisimulation game. Because of this, it immediately follows that any alternating bisimulation as defined in~\cite{AHKV98} is also a bisimilarity as defined here. Despite having a different formal definition, a simple observation can be made: Nash equilibria are not preserved by the alternating (bisimulation) equivalence relations in \cite{AHKV98} either, which discourages the use of these even stronger equivalence relations for multi-agent systems. 
In fact, as discussed in~\cite{Benthem02}, the ``right'' notion of equivalence for games (which can be indirectly used as an observationally equivalence between multi-agent systems) and their game theoretic solution concepts is, undoubtedly, an important and interesting topic of debate, which deserves to be investigated further.

%\noindent 
%\paragraph{\bf Parity games and bisimulation-invariance}
\paragraph{\bf Some features of our framework}
Unlike other approaches to rational synthesis and temporal equilibrium analysis, \emph{e.g.}~\cite{CermakLMM18,BouyerBMU15,FismanKL10,GHW17-aij}, we employ parity games~\cite{EmersonJ91}, which are an intuitively \emph{simple verification model} with an abundant associated set of algorithmic solutions~\cite{FriedmannL09}. 
In particular, strategies in our framework, as in \cite{GHW17-aij}, 
can depend on players' actions, leading to a much richer game-theoretic setting where 
Nash equilibrium is invariant under bisimilarity~\cite{GHPW17,GHPW19}, a
desirable property for concurrent and reactive
systems~\cite{Milner80,HM85,NicolaV95,GlabbeekW96}. 
%Bisimulation invariance, in turn, enables the use of standard verification
%techniques for temporal logics when reasoning about (pure Nash)
%equilibria. 
Our reasoning and verification approach applies to
multi-player games that are concurrent and
synchronous, %(at each round all players make their choices independently and at the same time),
with perfect recall and perfect information, and which can be
represented in a high-level, succinct manner using
SRML~\cite{HoekLW06}.
In addition, the technique developed in this paper, and its associated
implementation, considers games with \LTL goals, deterministic and pure
strategies, and dichotomous preferences. In particular, strategies in
these games are assumed to be able to see all past players' actions. 
%,
%leading to a setting where Nash equilibrium is invariant under
%bisimilarity~\cite{GHPW17}. This invariance property, in turn, enables
%the use of standard verification techniques for temporal logics when
%reasoning about (Nash) equilibria. In addition, the games are
%concurrent and synchronous (at each round all players make their
%choices independently and at the same time), with perfect information,
%and represented using the Simple Reactive Modules Language
%(SRML~\cite{HoekLW06}). 
We do not consider mixed or nondeterministic
strategies, or goals given by branching-time formulae. We also do not
allow for quantitative or probabilistic systems, \emph{e.g.}, such as
stochastic games or similar game models. We note, however, that some
of these aspects of our reasoning framework have been placed to avoid
undesirable computational properties. For instance, it is known that
checking for the existence of a Nash equilibrium in multi-player games
like the ones we consider is an undecidable problem if either
imperfect information or (various kinds of) quantitative/probabilistic
information is allowed~\cite{GutierrezPW18,UmmelsW11}.

\paragraph{\bf Future Work}
This paper gives a solution to the temporal equilibrium problem 
(both automated synthesis and formal verification) in a noncooperative setting. 
In future work, we plan to investigate the cooperative games setting~\cite{AgotnesHW09}. 
The paper also solves the problem in practice for perfect information games. 
We also plan to investigate if our main algorithms can be extended to 
decidable classes of imperfect information games, for instance, as those 
studied to model the behaviour of multi-agent systems 
in~\cite{GutierrezPW18,BelardinelliLMR17,AminofMM14,BerthonMM17}. 
Whenever possible, such studies will be complemented with 
practical implementations in \EVE. Finally, extensions to epistemic systems
and quantitative information in the context of multi-agent systems 
may be another avenue for further applications~\cite{HerzigLMS16,BelardinelliL09}, 
as well as settings with more complex preference
relations~\cite{GHW17-apal,FismanKL10,GMPRW17,AlmagorKP18}, which would provide 
a strictly stronger modelling power. 

\paragraph{Acknowledgements}	
The authors gratefully acknowledge the financial support of the ERC
Advanced Investigator Grant 291528 (``RACE'') at Oxford.
Giuseppe Perelli conducted this research partially while being member of the University of Oxford, working on the aforementioned grant, and now supported in part by European Research Council under the European Union’s Horizon 2020 Programme through the ERC Advanced Investigator Grant 834228 (``WhiteMech'').
Muhammad Najib was supported by the Indonesia Endowment Fund for Education (LPDP) while working on this research at the University of Oxford, and now by the European Research Council (ERC) under the European Union’s Horizon 2020 research and innovation programme (grant agreement no 759969).
Part of this paper, focussing on the \EVE\ system, has been presented at ATVA'18~\cite{GNPW18}.

\bibliography{refs-all,refs-aij18}

\begin{thebibliography}{10}

\bibitem{eve18}
{EVE}: A tool for temporal equilibrium analysis.
\newblock \url{https://github.com/eve-mas/eve-parity}.
\newblock Accessed: 09-09-2019.

\bibitem{pgsolver}
{PGS}olver.
\newblock \url{https://github.com/tcsprojects/pgsolver}.
\newblock Accessed: 09-09-2019.

\bibitem{AbrahamAH11}
I.~Abraham, L.~Alvisi, and J.~Y. Halpern.
\newblock Distributed computing meets game theory: combining insights from two
  fields.
\newblock {\em {SIGACT} News}, 42(2):69--76, 2011.

\bibitem{AgotnesHW09}
T.~{\AA}gotnes, W.~van~der Hoek, and M.~Wooldridge.
\newblock Reasoning about coalitional games.
\newblock {\em Artificial Intelligence}, 173(1):45--79, 2009.

\bibitem{AlmagorKP18}
S.~Almagor, O.~Kupferman, and G.~Perelli.
\newblock Synthesis of controllable nash equilibria in quantitative objective
  game.
\newblock In {\em Proceedings of the Twenty-Seventh International Joint
  Conference on Artificial Intelligence, {IJCAI} 2018, July 13-19, 2018,
  Stockholm, Sweden}, pages 35--41, 2018.

\bibitem{AlurH99b}
R.~Alur and T.~A. Henzinger.
\newblock Reactive modules.
\newblock {\em Formal Methods in System Design}, 15(1):7--48, 1999.

\bibitem{AlurHK02}
R.~Alur, T.~A. Henzinger, and O.~Kupferman.
\newblock Alternating-time temporal logic.
\newblock {\em Journal of the {ACM}}, 49(5):672--713, 2002.

\bibitem{AHKV98}
R.~Alur, T.~A. Henzinger, O.~Kupferman, and M.~Y. Vardi.
\newblock Alternating refinement relations.
\newblock In {\em {CONCUR}}, volume 1466 of {\em LNCS}, pages 163--178.
  Springer, 1998.

\bibitem{AlurHMQRT98}
R.~Alur, T.~A. Henzinger, F.~Y.~C. Mang, S.~Qadeer, S.~K. Rajamani, and
  S.~Tasiran.
\newblock {MOCHA:} modularity in model checking.
\newblock In {\em {CAV}}, volume 1427 of {\em LNCS}, pages 521--525. Springer,
  1998.

\bibitem{AminofMM14}
B.~Aminof, F.~Mogavero, and A.~Murano.
\newblock Synthesis of hierarchical systems.
\newblock {\em Science of Computer Programming}, 83:56--79, 2014.

\bibitem{BelardinelliL09}
F.~Belardinelli and A.~Lomuscio.
\newblock Quantified epistemic logics for reasoning about knowledge in
  multi-agent systems.
\newblock {\em Artificial Intelligence}, 173(9-10):982--1013, 2009.

\bibitem{BelardinelliLMR17}
F.~Belardinelli, A.~Lomuscio, A.~Murano, and S.~Rubin.
\newblock Verification of multi-agent systems with imperfect information and
  public actions.
\newblock In {\em {AAMAS}}, pages 1268--1276. {ACM}, 2017.

\bibitem{BerthonMM17}
R.~Berthon, B.~Maubert, and A.~Murano.
\newblock Decidability results for {ATL*} with imperfect information and
  perfect recall.
\newblock In {\em {AAMAS}}, pages 1250--1258. {ACM}, 2017.

\bibitem{BouyerBMU15}
P.~Bouyer, R.~Brenguier, N.~Markey, and M.~Ummels.
\newblock Pure {N}ash equilibria in concurrent deterministic games.
\newblock {\em Logical Methods in Computer Science}, 11(2):1--72, 2015.

\bibitem{BrafmanD08}
R.~I. Brafman and C.~Domshlak.
\newblock From one to many: Planning for loosely coupled multi-agent systems.
\newblock In {\em Proceedings of the Eighteenth International Conference on
  Automated Planning and Scheduling, {ICAPS} 2008, Sydney, Australia, September
  14-18, 2008}, pages 28--35, 2008.

\bibitem{Brenguier13}
R.~Brenguier.
\newblock {PRALINE:} {A} tool for computing {N}ash equilibria in concurrent
  games.
\newblock In {\em {CAV}}, volume 8044 of {\em LNCS}, pages 890--895. Springer,
  2013.

\bibitem{DBLP:journals/corr/abs-1202-4506}
P.~E. Bulychev, A.~David, K.~G. Larsen, A.~Legay, and M.~Mikucionis.
\newblock Computing {N}ash equilibrium in wireless ad hoc networks: {A}
  simulation-based approach.
\newblock In {\em Proceedings Second International Workshop on Interactions,
  Games and Protocols, {IWIGP}}, volume~78 of {\em {EPTCS}}, pages 1--14, 2012.

\bibitem{CaludeJKLS17}
C.~S. Calude, S.~Jain, B.~Khoussainov, W.~Li, and F.~Stephan.
\newblock Deciding parity games in quasipolynomial time.
\newblock In {\em STOC}, pages 252--263. ACM, 2017.

\bibitem{CermakLMM14}
P.~Cerm{\'{a}}k, A.~Lomuscio, F.~Mogavero, and A.~Murano.
\newblock {MCMAS-SLK:} {A} model checker for the verification of strategy logic
  specifications.
\newblock In {\em CAV}, volume 8559 of {\em LNCS}, pages 525--532. Springer,
  2014.

\bibitem{CLMM18}
P.~Cerm{\'{a}}k, A.~Lomuscio, F.~Mogavero, and A.~Murano.
\newblock {Practical Verification of Multi-Agent Systems Against SLK
  Specifications}.
\newblock {\em {Information and Computation}}, {261}({Part}):{588--614},
  {2018}.

\bibitem{CermakLMM18}
P.~Cerm{\'{a}}k, A.~Lomuscio, F.~Mogavero, and A.~Murano.
\newblock Practical verification of multi-agent systems against {SLK}
  specifications.
\newblock {\em Information and Computation}, 261(Part):588--614, 2018.

\bibitem{CermakLM15}
P.~Cerm{\'{a}}k, A.~Lomuscio, and A.~Murano.
\newblock Verifying and synthesising multi-agent systems against one-goal
  strategy logic specifications.
\newblock In {\em {AAAI}}, pages 2038--2044. {AAAI} Press, 2015.

\bibitem{ChatterjeeHP10}
K.~Chatterjee, T.~A. Henzinger, and N.~Piterman.
\newblock Strategy logic.
\newblock {\em Information and Computation}, 208(6):677--693, 2010.

\bibitem{CGP02}
E.~M. Clarke, O.~Grumberg, and D.~A. Peled.
\newblock {\em Model Checking}.
\newblock {MIT} Press, Cambridge, MA, USA, 2002.

\bibitem{DavidJLMT15}
A.~David, P.~G. Jensen, K.~G. Larsen, M.~Mikucionis, and J.~H. Taankvist.
\newblock Uppaal stratego.
\newblock In {\em {TACAS}}, volume 9035 of {\em LNCS}, pages 206--211.
  Springer, 2015.

\bibitem{DavidLLMP15}
A.~David, K.~G. Larsen, A.~Legay, M.~Mikucionis, and D.~B. Poulsen.
\newblock Uppaal {SMC} tutorial.
\newblock {\em {STTT}}, 17(4):397--415, 2015.

\bibitem{NicolaV95}
R.~{De Nicola} and F.~W. Vaandrager.
\newblock Three logics for branching bisimulation.
\newblock {\em Journal of the {ACM}}, 42(2):458--487, 1995.

\bibitem{Emerson90}
E.~A. Emerson.
\newblock Temporal and modal logic.
\newblock In {\em Handbook of Theoretical Computer Science, Volume {B:} Formal
  Models and Sematics {(B)}}, pages 995--1072. {MIT} Press, Cambridge, MA, USA,
  1990.

\bibitem{EmersonJ91}
E.~A. Emerson and C.~S. Jutla.
\newblock Tree automata, mu-calculus and determinacy.
\newblock In {\em {FOCS}}, pages 368--377. {IEEE}, 1991.

\bibitem{FinkbeinerS10}
B.~Finkbeiner and S.~Schewe.
\newblock Coordination logic.
\newblock In {\em CSL}, volume 6247 of {\em LNCS}, pages 305--319. Springer,
  2010.

\bibitem{Fischer:1982:SSA:588111.588124}
M.~J. Fischer and A.~Michael.
\newblock Sacrificing serializability to attain high availability of data in an
  unreliable network.
\newblock In {\em PODS}, pages 70--75, New York, NY, USA, 1982. ACM.

\bibitem{FismanKL10}
D.~Fisman, O.~Kupferman, and Y.~Lustig.
\newblock Rational synthesis.
\newblock In {\em {TACAS}}, volume 6015 of {\em LNCS}, pages 190--204.
  Springer, 2010.

\bibitem{FriedmannL09}
O.~Friedmann and M.~Lange.
\newblock Solving parity games in practice.
\newblock In {\em {ATVA}}, volume 5799 of {\em LNCS}, pages 182--196. Springer,
  2009.

\bibitem{Friedmann10thepgsolver}
O.~Friedmann and M.~Lange.
\newblock The pgsolver collection of parity game solvers -- version 3, 2010.

\bibitem{Gao0W17}
T.~Gao, J.~Gutierrez, and M.~Wooldridge.
\newblock Iterated {B}oolean games for rational verification.
\newblock In {\em {AAMAS}}, pages 705--713. {ACM}, 2017.

\bibitem{GO01}
P.~Gastin and D.~Oddoux.
\newblock Fast {LTL} to {B}{\"{u}}chi automata translation.
\newblock In {\em CAV}, pages 53--65. Springer, 2001.

\bibitem{Gifford:1979:WVR:800215.806583}
D.~K. Gifford.
\newblock Weighted voting for replicated data.
\newblock In {\em SOSP}, pages 150--162, New York, NY, USA, 1979. ACM.

\bibitem{GHPW17}
J.~Gutierrez, P.~Harrenstein, G.~Perelli, and M.~Wooldridge.
\newblock Nash equilibrium and bisimulation invariance.
\newblock In {\em CONCUR}, volume~85 of {\em LIPIcs}, pages 17:1--17:16.
  Schloss Dagstuhl, 2017.

\bibitem{GHPW19}
J.~Gutierrez, P.~Harrenstein, G.~Perelli, and M.~J. Wooldridge.
\newblock Nash equilibrium and bisimulation invariance.
\newblock {\em Logical Methods in Computer Science}, 15(3), 2019.

\bibitem{GutierrezHW15-concur}
J.~Gutierrez, P.~Harrenstein, and M.~Wooldridge.
\newblock Expresiveness and complexity results for strategic reasoning.
\newblock In {\em CONCUR}, volume~42 of {\em LIPIcs}, pages 268--282. Schloss
  Dagstuhl, 2015.

\bibitem{GutierrezHW15}
J.~Gutierrez, P.~Harrenstein, and M.~Wooldridge.
\newblock Iterated {B}oolean games.
\newblock {\em Information and Computation}, 242:53--79, 2015.

\bibitem{GHW17-aij}
J.~Gutierrez, P.~Harrenstein, and M.~Wooldridge.
\newblock From model checking to equilibrium checking: Reactive modules for
  rational verification.
\newblock {\em Artificial Intelligence}, 248:123--157, 2017.

\bibitem{GHW17-apal}
J.~Gutierrez, P.~Harrenstein, and M.~Wooldridge.
\newblock Reasoning about equilibria in game-like concurrent systems.
\newblock {\em Annals of Pure Applied Logic}, 168(2):373--403, 2017.

\bibitem{GMPRW17}
J.~Gutierrez, A.~Murano, G.~Perelli, S.~Rubin, and M.~J. Wooldridge.
\newblock Nash equilibria in concurrent games with lexicographic preferences.
\newblock In {\em {IJCAI}}, pages 1067--1073. ijcai.org, 2017.

\bibitem{GNPW18}
J.~Gutierrez, M.~Najib, G.~Perelli, and M.~Wooldridge.
\newblock Eve: A tool for temporal equilibrium analysis.
\newblock In {\em ATVA}, Vol 11138 of LNCS, pages 551--557, Cham, 2018.
  Springer.

\bibitem{GNPW19-ijcai}
J.~Gutierrez, M.~Najib, G.~Perelli, and M.~J. Wooldridge.
\newblock On computational tractability for rational verification.
\newblock In {\em {IJCAI}}, pages 329--335. ijcai.org, 2019.

\bibitem{GutierrezPW18}
J.~Gutierrez, G.~Perelli, and M.~Wooldridge.
\newblock Imperfect information in reactive modules games.
\newblock {\em Information and Computation}, 261(Part):650--675, 2018.

\bibitem{Halpern08}
J.~Y. Halpern.
\newblock Beyond nash equilibrium: Solution concepts for the 21st century.
\newblock In {\em {KR}}, pages 6--15. {AAAI} Press, 2008.

\bibitem{HM85}
M.~Hennessy and R.~Milner.
\newblock Algebraic laws for nondeterminism and concurrency.
\newblock {\em Journal of the {ACM}}, 32(1):137--161, 1985.

\bibitem{HerzigLMS16}
A.~Herzig, E.~Lorini, F.~Maffre, and F.~Schwarzentruber.
\newblock Epistemic {B}oolean games based on a logic of visibility and control.
\newblock In {\em {IJCAI}}, pages 1116--1122. {IJCAI/AAAI} Press, 2016.

\bibitem{Jurdzinski98}
M.~Jurdzinski.
\newblock Deciding the winner in parity games is in {UP} $\cap$ co-up.
\newblock {\em Information Processing Letters}, 68(3):119--124, 1998.

\bibitem{Kupferman18}
O.~Kupferman.
\newblock Automata theory and model checking.
\newblock In {\em Handbook of Model Checking}, pages 107--151. Springer
  International Publishing, 2018.

\bibitem{KupfermanPV16}
O.~Kupferman, G.~Perelli, and M.~Y. Vardi.
\newblock Synthesis with rational environments.
\newblock {\em Annals of Mathematics and Artificial Intelligence}, 78(1):3--20,
  2016.

\bibitem{KNPS19}
M.~Kwiatkowska, G.~Norman, D.~Parker, and G.~Santos.
\newblock Equilibria-based probabilistic model checking for concurrent
  stochastic games.
\newblock In {\em {FM}}, volume 11800 of {\em LNCS}, pages 298--315. Springer,
  2019.

\bibitem{KwiatkowskaPW16}
M.~Kwiatkowska, D.~Parker, and C.~Wiltsche.
\newblock Prism-games 2.0: {A} tool for multi-objective strategy synthesis for
  stochastic games.
\newblock In {\em TACAS}, volume 9636 of {\em LNCS}, pages 560--566. Springer,
  2016.

\bibitem{KwiatkowskaNP09}
M.~Z. Kwiatkowska, G.~Norman, and D.~Parker.
\newblock {PRISM:} probabilistic model checking for performance and reliability
  analysis.
\newblock {\em {SIGMETRICS} Performance Evaluation Review}, 36(4):40--45, 2009.

\bibitem{Ladin:1992:PHA:138873.138877}
R.~Ladin, B.~Liskov, L.~Shrira, and S.~Ghemawat.
\newblock Providing high availability using lazy replication.
\newblock {\em ACM Transactions on Computer Systems}, 10(4):360--391, Nov.
  1992.

\bibitem{LaroussinieM15}
F.~Laroussinie and N.~Markey.
\newblock Augmenting {ATL} with strategy contexts.
\newblock {\em Information and Computation}, 245:98--123, 2015.

\bibitem{Loding12}
C.~L{\"{o}}ding.
\newblock Basics on tree automata.
\newblock In {\em Modern Applications of Automata Theory}, pages 79--110.
  Indian Institute of Science, Bangalore, India, 2012.

\bibitem{LomuscioQR17}
A.~Lomuscio, H.~Qu, and F.~Raimondi.
\newblock {MCMAS:} an open-source model checker for the verification of
  multi-agent systems.
\newblock {\em {STTT}}, 19(1):9--30, 2017.

\bibitem{ltl2ba}
{LTL 2 BA}: fast translation from {LTL} formulae to {B}{\"{u}}chi automata.
\newblock \url{http://www.lsv.fr/~gastin/ltl2ba/}.
\newblock Accessed: 09-09-2019.

\bibitem{Milner80}
R.~Milner.
\newblock {\em A Calculus of Communicating Systems}, volume~92 of {\em LNCS}.
\newblock Springer, 1980.

\bibitem{milner:89a}
R.~Milner.
\newblock {\em Communication and Concurrency}.
\newblock Prentice Hall, 1989.

\bibitem{MogaveroMPV14}
F.~Mogavero, A.~Murano, G.~Perelli, and M.~Y. Vardi.
\newblock Reasoning about strategies: On the model-checking problem.
\newblock {\em {ACM} Transactions on Computational Logic}, 15(4):34:1--34:47,
  2014.

\bibitem{OR94}
M.~J. Osborne and A.~Rubinstein.
\newblock {\em A Course in Game Theory}.
\newblock MIT Press, 1994.

\bibitem{PP04}
D.~Perrin and J.~Pin.
\newblock {\em {Infinite Words.}}
\newblock {Pure and Applied Mathematics.} Elsevier, 2004.

\bibitem{Piterman07}
N.~Piterman.
\newblock From nondeterministic {B}{\"{u}}chi and {S}treett automata to
  deterministic parity automata.
\newblock {\em Logical Methods in Computer Science}, 3(3):1--21, 2007.

\bibitem{pnueli:77a}
A.~Pnueli.
\newblock The temporal logic of programs.
\newblock In {\em FOCS}, pages 46--57. IEEE, 1977.

\bibitem{shoham:2008a}
Y.~Shoham and K.~{Leyton-Brown}.
\newblock {\em Multiagent Systems: Algorithmic, Game-Theoretic, and Logical
  Foundations}.
\newblock CUP, 2008.

\bibitem{SistlaVW87}
A.~P. Sistla, M.~Y. Vardi, and P.~Wolper.
\newblock The complementation problem for {B}{\"{u}}chi automata with
  appplications to temporal logic.
\newblock {\em Theoretical Computer Science}, 49:217--237, 1987.

\bibitem{ToumiGW15}
A.~Toumi, J.~Gutierrez, and M.~Wooldridge.
\newblock A tool for the automated verification of {N}ash equilibria in
  concurrent games.
\newblock In {\em ICTAC}, volume 9399 of {\em LNCS}, pages 583--594. Springer,
  2015.

\bibitem{UmmelsW11}
M.~Ummels and D.~Wojtczak.
\newblock The complexity of {N}ash equilibria in stochastic multiplayer games.
\newblock {\em Logical Methods in Computer Science}, 7(3):1--45, 2011.

\bibitem{Benthem02}
J.~van Benthem.
\newblock Extensive games as process models.
\newblock {\em Journal of Logic, Language and Information}, 11(3):289--313,
  2002.

\bibitem{HoekLW06}
W.~van~der Hoek, A.~Lomuscio, and M.~Wooldridge.
\newblock On the complexity of practical {ATL} model checking.
\newblock In {\em {AAMAS}}, pages 201--208. {ACM}, 2006.

\bibitem{GlabbeekW96}
R.~J. van Glabbeek and W.~P. Weijland.
\newblock Branching time and abstraction in bisimulation semantics.
\newblock {\em Journal of the {ACM}}, 43(3):555--600, 1996.

\bibitem{Woo01}
M.~Wooldridge.
\newblock {\em Introduction to Multiagent Systems}.
\newblock Wiley, Chichester, UK, 2001.

\bibitem{WooldridgeGHMPT16}
M.~Wooldridge, J.~Gutierrez, P.~Harrenstein, E.~Marchioni, G.~Perelli, and
  A.~Toumi.
\newblock Rational verification: From model checking to equilibrium checking.
\newblock In {\em {AAAI}}, pages 4184--4191, 2016.

\bibitem{Wuu:1984:ESR:800222.806750}
G.~T. Wuu and A.~J. Bernstein.
\newblock Efficient solutions to the replicated log and dictionary problems.
\newblock In {\em PODC}, pages 233--242, New York, NY, USA, 1984. ACM.

\bibitem{Zie98}
W.~Zielonka.
\newblock Infinite games on finitely coloured graphs with applications to
  automata on infinite trees.
\newblock {\em Theoretical Computer Science}, 200(1-2):135--183, 1998.

\end{thebibliography}

%\newpage
%\appendix
%\input{eve-mcmas-exp}

\end{document}